\documentclass[aps,pra, onecolumn,14 pts,a4paper]{revtex4}
 \pdfoutput=1

\usepackage{amssymb,amsmath,amsfonts,graphicx}
\usepackage{bm}
\usepackage{epsfig,color,times}
 \usepackage{bm,color}
\usepackage{graphicx}
\usepackage{amssymb}
\usepackage{epstopdf}

\begin{document}

\title{Ultimate Statistical Physics: fluorescence of a single atom}

\author{Yves Pomeau$^1$, Martine Le Berre$^2$ and Jean Ginibre$^3$}
\affiliation{$^1$  Department of Mathematics, U. of Arizona,Tucson, AZ 85721, USA
\\$^2$ Institut des Sciences Mol\'eculaires d'Orsay ISMO
- CNRS, Universit\'e Paris-Sud, Bat. 210, 91405 Orsay Cedex, France.
\\$^3$ Laboratoire de Physique Th\'eorique, Universit\'e Paris-Sud, Bat. 210, 91405 Orsay Cedex, France.
}

\date{\today }

\begin{abstract}

We discuss the statistics of emission of photons by a single atom or ion illuminated by a laser beam at the frequency of quasi-resonance between two energy levels, a situation that corresponds to real experiments. We extend this to the case of two laser beams resonant with the energy differences between two excited levels and the ground state (three level atom in V-configuration). We use a novel approach of this type of problem by considering  Kolmogorov equation for the probability distribution of the atomic state which takes into account first the deterministic evolution of this state under the effect of the incoming laser beam and the random emission of photons during the spontaneous decay of the excited state(s) to the ground state. This approach yields solvable equations in the two level atom case.  For the three level atom case we set the problem and define clearly its frame. The results obtained are valid both in the opposite limits of rare and of frequent spontaneous decay, compared to the period of the optical Rabi oscillations due to the interaction between the resonant excitation and the atomic levels. Our analysis gives access to various statistical properties of the fluorescence light, including one showing that its fluctuations in time are not invariants under time reversal. This result puts in evidence the fundamentally irreversible character of quantum measurements, represented here by the emission of photons of fluorescence. 

\end{abstract}

\maketitle
\section{Introduction}

Statistical concepts are deeply embedded into quantum mechanics and its understanding. The relationship between  quantum fluctuations and other kinds of randomness in physics remains a difficult question. For classical (= non quantum) systems one of the sources of randomness is the large number of degrees of freedom of a fluid in a big container for instance. Chaotic dynamics is another source of randomness in classical systems. As an example of quantum fluctuations in a simple system we consider below  the fluorescence of a single two-level atom illuminated by a monochromatic light beam at the frequency of resonance or quasi-resonance between the two quantum levels.  We hope that this could put some light on fundamental issues of quantum statistics, and  motivate new experiments.
We 
explain
 that by observing this fluorescence one can show
 how irreversible dynamics is linked to the measurement process, or, equivalently, to so-called quantum jumps.  Later on we extend this to a three-level system. 
In the absence of spontaneous decay of the excited level(s) by the random emission of photons, such systems present regular oscillations, called optical Rabi oscillations. 
This naming is based on the similarity of the relevant dynamical equations with the ones of a spin in an uniform  magnetic field and a periodic electromagnetic field at Larmor frequency. The optical Rabi oscillations (just called Rabi oscillations later on) of a single atom get a statistical underpinning when one takes into account the possibility of spontaneous decay, namely the possibility of connecting the atom with the outside space where electromagnetic waves with all possible directions can propagate. This spontaneous decay, or quantum jump, occurs typically with a very short duration but at random times. 
We represent by a Kolmogorov equation this randomness together with the reversible dynamics of the Rabi oscillations. This equation is for the probability distribution of the state of the atom and it can be solved in various instances. The solutions give access to physical quantities like the spectrum of the fluorescence light, and also to higher order time correlations allowing to display the fundamentally irreversible character of fluorescence: by analyzing the fluctuations as explained in reference \cite{ref.1} one can show that the time dependent fluctuations are different when looked at in the forward and backward time direction. This lack of asymmetry under time reversal is absent in the fluctuations of a system at equilibrium, as was shown by Onsager.

Perhaps it is not without interest to explain how this study began. The idea was to apply the concept of breaking of time reversal symmetry to spectra of light and more generally of electromagnetic waves. Likely, by far, the most studied spectrum of radiation is the cosmological background, which is experimentally found to be an equilibrium spectrum at a well defined temperature, but with various fluctuations seemingly above the level expected for such a black-body radiation. The claim that this radiation is at equilibrium relies on the comparison between the dependence of the power received as a function of the frequency and Planck's spectrum.  This is only one feature of equilibrium. Another one, perhaps more fundamental, is the invariance of any observation with respect to time reversal symmetry: some correlation functions are exactly zero for equilibrium systems. As it is unclear if such correlations could be measured on the cosmological background because of its faintness, we turned to other kinds of radiation where measurements of time reversal symmetry could be at least potentially doable. This made one of the motivations of this paper.

Another important motivation is to apply the classical Kolmogorov approach to the description of the random emission of photons by a single atom,  a topic extensively studied in quantum optics. In the next section we argue in favor of the relevance of such a classical approach to describe the atomic fluorescence and we emphasize  the novelty of our point of view as compared with that of previous works. 
 As in  most of the quantum optics studies, we start from a Schr\"{o}dinger description of the atom-field interaction, but while in quantum optics the main goal is to analyse the evolution of the density matrix of the system composed of the atom, the exciting field and eventually the ``reservoir``, here we propose and study a Kolmogorov-type kinetic equation for the evolution of a probability $p$ on the set of density matrices describing the state of the atom. That probability evolves 
as follows: $i)$ in the coherent regime, namely between two successive photon emissions, the probability is transported by the Schr\"{o}dinger flow describing the Rabi oscillations; $ii)$ when a photon is emitted $p$ undergoes an instantaneous jump. The description of fluorescence by quantum jumps is standard in the subject. The previous procedure  describes the quantum trajectories followed by the atom in the course of its history with  the successive emission of photons. 
The conservation of the total probability is automatically ensured in the Kolmogorov approach, where in addition the trace normalisation of the density matrices under consideration is ensured by  the very definition of the model. In practice  the probability bears on the coordinates of the generic density matrix. In the two-level atom case the set of coordinates eventually reduces to a single angle $\theta$.

 In a number of quantum optical studies of fluorescence, the starting point is a Lindblad type master equations \cite{lindblad} for the density matrix, as used for instance in the dressed atom picture of the radiative cascade\cite{cohen}.   That equation preserves  the trace normalisation condition. However  the latter may be violated by the use of various approximations made in the course of the analysis.  Some of our results disagree with those of previous works. This concerns for instance the probability distribution of successive times of spontaneous emission of photons, see subsection \ref{sec:between}. We suspect that such a discrepancy might come from the problem of non-conservation of the trace of the  density matrix in the Lindblad-type studies.
 
In  other quantum optics studies of the fluorescence some authors start from the optical Bloch equations which describe mean values of  the atomic operators in presence of a classical strong driving field \cite{scully}. In order
to derive variances and correlations a random operator is phenomenologically added to the original Bloch equations which become Bloch-Langevin equations. From these equations one may derive the evolution of the mean value of any operator, while maintaining the commutation relations for all times. But again that procedure  produces a difficulty with the trace normalisation condition of the density matrix, since that condition is preserved only in the mean \cite{knight}. Another treatment of the problem consists in introducing a random process on the set of pure states, namely wave-functions  of the system, which eventually leads to a Lindblad-type equation in the mean \cite{dalibard92}.

 This paper is organized as follows.
 In section \ref{sec:Kolm} we present the framework of our Kolmogorov
approach to the study of fluorescence. We emphasize the originality of using Kolmogorov equation to describe the random emission of photons by the atom compared to other studies  which start from a Lindblad-like equation for the density matrix.  
Section \ref{Keq2level} is the
main part of the paper, it is devoted to the two-level atom case.  
Firstly we derive and solve Kolmogorov equation for the exact resonant  atom-laser interaction and also for a detuned laser  (= not exactly resonant with the frequency of the atomic transition). We show that the detuned laser case gives formally the same equation as the resonant laser case, a remarkable result. As an application we  first calculate the spectrum of the fluorescence (at exact resonance) which displays the well known triplet structure for strong laser intensity (small damping), and a single peak in the opposite case. In the latter case (large damping) our statistical approach gives a spectral width increasing with damping, or equivalently an average time interval between successive photons decreasing with damping, which  disagrees qualitatively with previous work \cite{cohen}-\cite{Arimondo}, see  our Fig.\ref{fig:Kol2}.  Secondly we show the fundamental irreversibility of the photon  emission process. Choosing appropriate correlation functions to test this irreversibility, we prove that this property may be quantified. Quite naturally the irreversibility increases with the damping rate, which definitely shows that irreversibility is due to the spontaneous emission, as expected.
In section \ref{3levelJG} we outline the premise of the derivation of
Kolmogorov equation in the three-level case. Our aim is to interpret the
intermittent fluorescence observed when two lasers  illuminate a
three-level atom (or ion) in the so-called V-configuration.  We show that by choosing an appropriate set of variables to describe the phase space,  the deterministic dynamics  reduce to a rotation at constant angular velocity along a circle of definite radius, this radius changing after each quantum jump. In Appendix C  we make step toward the representation  of the  shelving process  \cite{nagourney}.

 \section{Kolmogorov equation in general}
  \label{sec:Kolm}
 We shall introduce below several examples of what we call Kolmogorov equations. This kind of equation describes the evolution of the probability distribution of a system (in our case a single atom or ion) under the effect of two processes: first a deterministic dynamics changes the parameters of the system. Let $\Theta(t)$ be this set of time dependent parameters with the time derivative ${\partial_{t}{\Theta}} = v(\Theta) $, a function of $\Theta$. For this deterministic dynamics the conservation of probability yields the equation of evolution of the probability distribution $p(\Theta, t)$, 
\begin{equation}
\partial_{t}{p}(\Theta, t) + \partial_{\Theta}(v(\Theta)p(\Theta, t))= 0
\textrm{,}
\label{eq:dotprobb}
\end{equation}
 where $\partial_{t}$ is here and elsewhere for the derivative with respect to time,  and $\partial_{\Theta}$ for the derivative with respect to $\Theta$. This is actually a gradient in general because $\Theta$ has more than one component, but this only complicates the writing in an unessential way.  Moreover the probability distribution is positive and normalized. 
 
 Kolmogorov equations add to this equation a right-hand side representing instantaneous transitions occurring at random instants of time. 
 This effect of the transitions (or jumps) is represented by a positive valued function $\Gamma (\Theta' \vert \Theta)$. During a small interval of time $ \mathrm{d}t$, if the system is in state  $\Theta$, it  jumps with probability $\Gamma (\Theta' \vert \Theta)  \mathrm{d}t$ to state $\Theta'$, the duration of a jump being much shorter than the time scale of the deterministic dynamics. Kolmogorov equation describes both the deterministic dynamics and the jump process  and reads \cite{kolmo},
 \begin{equation}
\partial_{t}{p}(\Theta, t) + \partial_{\theta}(v(\Theta)p(\Theta, t)) = \int \mathrm{d}\Theta_1 \Gamma (\Theta\vert \Theta_{1}) p(\Theta_1, t) -  p(\Theta, t) \int \mathrm{d}\Theta' \Gamma (\Theta' \vert \Theta)
\textrm{.}
\label{eq:dotproba}
\end{equation}
In the right-hand side the first positive term describes the increase of probability of the $\Theta$-state due to jumps from other states to $\Theta$. The second term represents the loss of probability because of jumps from  $\Theta$ to any other state $ \Theta'$. We assume that the jump probability $\Gamma (\Theta' \vert \Theta)$ is smooth enough near $ \Theta =  \Theta'$ for not having to care for jumps from  $\Theta$ to almost identical states (as it may happen for instance because of the grazing two-body collisions in Boltzmann kinetic theory with a smooth pair potential).  Indeed the very existence of the probability transition $ \Gamma $ implies that we are considering a Markov process where the transition rate depends on the present state of the system only. 
 
 By integration over $ \Theta$ one finds that the $L^1$-norm $ \int \mathrm{d}\Theta \ p(\Theta, t)$ is constant (if it converges, as we assume it).  Below we shall derive an explicit Kolmogorov equation for a two-level atom subject both to external electric field at  quasi-resonance frequency between the two levels (the pump field) and to spontaneous decay by radiation (the process of fluorescence). We shall also outline the derivation of  Kolmogorov equation in the three-level case with two pump fields at the two frequencies defined by the resonances between the ground state and the two excited states.  Our ultimate goal is to find a way to display the irreversible character of quantum transitions as a result of measurements of the emitted light. 
 
 The validity of this approach requires the existence of two widely different time scales. The long time scale is associated to the deterministic dynamics, and so to the left-hand side of Kolmogorov equation. From time to time there is a jump from one state to another state, a jump that is instantaneous at the time scale of the deterministic dynamics. This splitting into two time scales, and consequently the use of a Kolmogorov equation is particularly well justified in the case of fluorescence of a single atom: the deterministic dynamics is due to the interaction with the external pump field, whereas the quick random jumps are due to the emission of a photon when the excited state decays to the ground state. This decay occurs rarely but with a time scale of the order of the period of the emitted photon, a time much shorter than the two other typical times of the problem, namely the average interval between two spontaneous jumps from the excited state and the period of the Rabi oscillations due to the interaction with classical external pump field.

 The physical results we shall derive from this Kolmogorov approach will rest on explicit calculations of time dependent correlations of the fluctuations of the system around its steady state, given by a steady solution of Kolmogorov equation. Although this is standard let us recall how to derive such time dependent correlations. Let $F(\Theta, t)$ and  $G(\Theta, t)$ be two functions of the state of the system at time $t$. We are interested in the computation of the time dependent correlation $<F(\Theta, t) G(\Theta, t')>$ where the average is done on the steady state $P_{st}(\Theta)$, the time independent solution of equation (\ref{eq:dotproba}). When $t = t'$ the correlation $<F(\Theta, t) G(\Theta, t)>$ is time independent and is simply
 $$<F(\Theta, t) G(\Theta, t)> =  \int \mathrm{d}\Theta P_{st}(\Theta) F(\Theta) G(\Theta) \textrm{.}$$
 The time dependent correlation is found as follows. Let $P( \Theta,t \vert  \Theta_{0})$ be the solution of Kolmogorov equation (\ref{eq:dotproba}) with variables $(\Theta, t)$ and  initial condition $P(\Theta, t = 0) = \delta(\Theta_0 -  \Theta)$, $ \delta(.)$ being Dirac delta function. The time dependent correlation reads
  \begin{equation}
  <F(\Theta, 0) G(\Theta, t)> =  \int \mathrm{d}\Theta_0 P_{st}(\Theta_0)  \int \mathrm{d}\Theta P( \Theta, t  \vert  \Theta_0)  F(\Theta_0) G(\Theta) 
\textrm{.}
\label{eq:dotprob1}
\end{equation}

 As  sketched in the introduction, our approach of single atom fluorescence by using a Kolmogorov equation is not the one used in the literature. 
Quantum optics studies generally use 
the density matrix of  the interacting atom-field system.
 The density matrix is used instead of wave-functions in order to be able to consider arbitrary superposition of quantum states.  The  standard procedure consists in deriving and studying a Lindblad-type equation for a single
 density matrix of  the interacting atom-field system (see equation (2.1) of \cite{cohen}),
 \begin{equation} 
   \partial_{t}{\rho} =-\frac{i}{\hbar}[H,\rho] -\frac{\gamma}{2}(S^{+}S^{-}\rho + \rho S^{+}S^{-}) +\gamma S^{-}\rho S^{+}
   \textrm{,}
\label{eq:lindblad}
\end{equation}
which describes the non-unitary evolution of the density matrix $\rho$. In equation (\ref{eq:lindblad})
 $H$ is the Hamiltonian describing atom and laser interacting together, $\gamma$ is the Einstein coefficient associated with spontaneous emission, and $S^{+}$ and $S^{-}$ are the raising and lowering atomic operators, $\vert e><g \vert$ and $\vert g><e \vert $ respectively.
  That equation is trace-preserving. Our point view is significantly different: we also use the density matrix description but instead of using an evolution equation for a single density matrix,  we put a probability $p(\rho_{at},t)$ on the set  of density matrices  $\rho_{at}$ of the atom, and we derive and study an evolution equation for that probability, an equation of the form
  \begin{equation}
  \partial_{t}{p} =M(p,\rho_{at})
\label{eq:pl}
\end{equation}   
 where the r.h. s. has to be determined.
   In other terms we represent the evolution of an ensemble of points (matrix densities) by putting a probability on that ensemble, whereas the standard quantum optics studies consider eventually the evolution of a single point of that ensemble.

 We argue  that the Kolmogorov approach  is  the relevant one for such a study. Because of the quantum jumps the atom follows a trajectory with successive discontinuities. Since in addition the jumps occur at random times, the trajectory has a random character, which justifies putting a probability on the set of density matrices which are the generic points of that trajectory.
 In contrast with this, the Lindblad equation is an (ordinary) differential equation for the density matrix, and is expected to yield  a unique smooth solution for a given initial condition.  Such a smooth solution is inadequate to describe a trajectory with jumps and can describe at best the evolution of the mean value of the trajectory of the atom, or in mathematical terms, of its matrix density.  In other words a deterministic theory of the density matrix evolution is inappropriate as soon as the fluctuations of this density matrix (due to the quantum jumps) are of order one with respect to density matrix solution. In such a situation a statistical approach is necessary.

In addition to this fondamental difference between Kolmogorov and Lindblad-type descriptions,  the following point seems to be at the origin of some discrepancies between our results and previous ones. In the Lindblad description of fluorescence the decay of the excited state is generally represented by the addition of a damping term to a Schr\"{o}dinger-like evolution equation for the state of the atom, with no accretive term (i.e. no gain term). This is likely to be a fair representation if the decay process is much 
 slower than the time evolution of the atom under the effect of the external light beam. However whenever the decay rate becomes large,  it seems to us that the accretive term cannot be neglected. Then one has to consider the full radiative cascade from one multiplicity to the lower one, in other words the fluctuations of the state of the system  at the time scale of its deterministic dynamics 
 cannot be ignored  anymore. 
 This point is illustrated by a result obtained in \cite{cohen} by  the 
Schr\"{o}dinger-plus-damping treatment 
namely the fact that, in the limit of a strong damping by emission of photons, the predicted time interval between two successive photons  {\it{increases}} as the damping term {\it{increases}} when the period of the Rabi oscillations is kept constant. On the contrary, in  our Kolmogorov-inspired theory this time interval, as expected, {\it{decreases}} as the damping term {\it{increases}}. 
 
This Kolmogorov equation takes into account accretive and dissipative effects,  which are described by the first and second term respectively in the right-hand side of equation(\ref{eq:dotproba})), thereby conserving strictly the total probability. Those  accretive and dissipative terms should play the same role as the third and second term in the r.h.s. of equation (\ref{eq:lindblad}).
 
Moreover our preliminary work on the three level system allows us to preserve the amplitude of the spectator level after a transition (for instance level 2 after a transition 1-0), a condition which is not a priori  easily implementable in a relaxation model.

  \subsection{Kolmogorov equation and irreversibility}
  As shown in reference \cite{ref.1} an analysis of the transition probabilities allows to decide if a Markov process yields or not a reversible dynamics in the sense of this reference. The result proved for finite time discrete Markov process can be extended to the case of continuous time and a continuum of states, the one relevant for Kolmogorov equation (\ref{eq:dotproba}). We shall use also the slightly simpler case of discrete representations and continuous time.  To ease the writing let first consider the case of a finite number of states indexed by letters like $i, j, k$, then give the formulae  with continuous variables.
    
We assume that random jumps (from a state to another state) occur without memory other than the knowledge of the present state.  It follows that the probability $P_i(t)$ of being in state $i$ at time step $t$ (taken as discrete) obeys  Markov equation of the form,
  \begin{equation}
  P_i(t) = \sum_j M_{ij}  P_j(t - 1)
\textrm{,}
\label{eq:dotprobM1}
\end{equation}
  where the transition probabilities $M_{ij}$ are positive constants. They are subject to constraints related to probability theory, namely to the fact that for any  $P_j(t - 1)$ such that the total probability is one, namely that $ \sum_j P_j(t - 1) = 1$, then  $ \sum_j P_j(t ) = 1$. This requires  
  \begin{equation}
    \sum_i M_{ij} = 1 \textrm{.}
    \label{eq:dotprobM1.1}
\end{equation}
Therefore the $M_{ij}$ are in between $0$ and $1$. A shown in reference \cite{ref.1}, the fluctuations around the steady  state make a distinction between the forward and backward direction of time if some conditions are met by the coefficients $M_{ij}$. By considering the short time evolution of Kolmogorov like equations one can derive from the conditions for the discrete Markov process constraints for the reversibility of the fluctuations near the steady solution of a Kolmogorov equation. 
  
  There are more than one condition for the $M_{ij}$ to yield time reversible fluctuations. We shall derive one requiring the knowledge of the steady solution of equation (\ref{eq:dotprobM1}) because this steady solution is known explicitly in our case. Let $P_{st}$ be the probability distribution in the steady state reached by the discrete Markov process of equation (\ref{eq:dotprobM1}). It is a solution of   
    \begin{equation}
  P_{i, st} = \sum_j M_{ij}  P_{j, st}
 \textrm{.}
\label{eq:dotprobM2}
\end{equation}
such that $\sum_i P_{i, st} = 1$.  We shall assume this solution to be unique, although it is easy to find examples where it is not. This is not what happens in the case of  Kolmogorov equation we shall study. 

The fluctuations are time reversible if no analysis of time dependent fluctuations can show a difference between the forward and backward direction of time. In the discrete Markov process, this is equivalent to say that time correlation like $<A(i, t) B(j, t + T)>$ are equal to  $<B(i, t) A(j, t + T)>$ where $A$ and $B$ are two different functions of the state of the system. This property can be formulated by means of the time autocorrelation function $Q(i, j; T)$ satisfying the same Markov equation as $P(i, t)$:  
  \begin{equation}
 Q(i, j; T) = \sum_k M_{jk} Q(i, k; T - 1)
 \textrm{.}
\label{eq:dotprobM3}
\end{equation}
with the initial condition $Q(i, j; T=0) = \delta_{ij}$ where $\delta_{ij}$ is the usual Kronecker delta. The time correlations have the following expression:
 \begin{equation}
 <A(i, t) B(j, t + T)> = \sum_{k, i} P_{i, st}  A(i) B(k) Q(i, k; T)
 \textrm{.}
\label{eq:dotprobM4}
\end{equation}
Not surprisingly the time reversal invariance of the fluctuations implies a relation between the transition probabilities from $i$ to $j$ and from $j$ to $i$. In the present case, this is a straightforward consequence of equation (\ref{eq:dotprobM4}) and involves the steady distribution of the Markov process. It reads: 
 \begin{equation}
 P_{i, st} M_{ik} =  P_{k, st} M_{ki} 
 \textrm{.}
\label{eq:dotprobM5}
\end{equation}
Such conditions are called sometime condition of detailed balance in the statistical physics literature. 

The present work has to do with Kolmogorov equation with continuous time and states. The previous considerations have been developed for discrete time and states for the purpose of a more streamlined presentation. It can be adapted to this situation quite simply. First one can substitute for the discrete time step a small time shift during which the probability distribution changes a bit. Let us consider the following equation for a Markov process with a continuous time but still a discrete set of states: 
 \begin{equation}
  {\partial_{t}{P}}(i, t) = \sum_j N_{ij}  P(j, t) - P(i, t)  \sum_j N_{ji} 
\textrm{.}
\label{eq:dotprobM6}
\end{equation}
The quantities $N_{ij}$ are the positive rates of transition per unit time from state $j$ to $i$. Integrating formally this equation from time $t$ to time $t + \Delta t$ with $\Delta t$ small with respect to the inverse transition rates, one derives  
the following equation for a discrete  Markov process: 
\begin{equation}
P (i, t + \Delta t )  =  P (i, t ) +  \Delta t  \left( \sum_j N_{ij}  P(j, t) - P(i, t) \sum_j N_{ji} \right)
\textrm{.}
\label{eq:dotprobM7}
\end{equation}
This equation  can be mapped into a form identical to the general Markov equation (\ref{eq:dotprobM1}) by defining the coefficients $M_{ij}$ as 
\begin{equation}
M_{ij}  =  \delta_{ij}  +    \Delta t (N_{ij}  - \sum_k N_{ki})
\textrm{.}
\label{eq:dotprobM8}
\end{equation}
Notice that the condition of conservation of probability is satisfied exactly by the transition probabilities defined in this way. This yields the condition to be satisfied by the coefficients $N_{ij}$ for the continuous Markov process to be time reversible in the sense defined before. From equation (\ref{eq:dotprobM5}) this condition reads:
 \begin{equation}
 P_{i, st} (N_{ik}  - \sum_l N_{li}) =  P_{k, st} (N_{ki}  - \sum_l N_{lk}) 
 \textrm{.}
\label{eq:dotprobM9}
\end{equation}
Notice that the Kronecker delta does not contribute to the writing of this condition. We shall use it in the case of Kolmogorov equation where the state variable belongs to a continuous  set. In this case one replaces the index $i, j, k, ...$ by variables $\Theta, \Theta', ...$, discrete sums by integrals and the coefficients  $N_{ij}$ by the coefficients $\Gamma( \Theta' \vert \Theta)$, as they appear in the writing of Kolmogorov equation (\ref{eq:dotproba}). The condition of reversibility reads now: 
 \begin{equation}
 P_{st} (\Theta) \left(\Gamma( \Theta'  \vert  \Theta) -  \int \mathrm{d}\Theta'' \Gamma(\Theta''  \vert  \Theta)\right) =   P_{st} (\Theta')  \left(\Gamma( \Theta \vert \Theta') -  \int \mathrm{d}\Theta'' \Gamma( \Theta''  \vert \Theta')\right)  
 \textrm{.}
\label{eq:dotprobirr}
\end{equation}
This is sufficient to show the irreversibility of the fluctuations described by Kolmogorov equation we shall deal with, because it is not satisfied in our case. The reason is rather straightforward: we shall introduce functions $\Gamma(\Theta'  \vert \Theta)$ which are delta-Dirac functions of one of their argument, the one with a prime. Let us write this function as $\Gamma( \Theta' \vert \Theta) = g(\Theta) \delta(\Theta')$ with $g(.)$ smooth function of its argument. The steady distribution $P_{st} (\Theta)$ is not a smooth function of its argument, but its singularity at $\Theta = 0$ is a jump, a singularity weaker than a Dirac delta. Therefore the most singular piece on the right-hand side of equation(\ref{eq:dotprobirr}) is a Dirac delta of argument $\Theta$, although it is a Dirac delta with argument $\Theta'$ on the left-hand side. Therefore equation(\ref{eq:dotprobirr}) is not satisfied for the type of Kolmogorov equation we shall deal with and there the fluctuations near the steady state are not time reversible in the sense of ref.\cite{ref.1}. This implies that some time correlations are not zero, although they should be so for a time reversible dynamics.  In section \ref{sec.irr} we give examples of correlations showing this asymmetry with respect to time reversal and which could be measured out of fluctuations of the fluorescence of a single atom (or ion).

Let us remark that the irreversibility in the sense just explained could be a sensitive criterion to test theories. As shown in reference  \cite{ref.1}  the Langevin equation of various systems describes time dependent fluctuations having the property of time reversal symmetry because of a rather subtle balance between the damping term (fundamentally irreversible) and the noise source. Therefore it could well be that the addition of a Langevin noise, as done often, to the Bloch equations with damping yields at the end time reversible fluctuations, of the fluorescence light for instance. As shown below, Kolmogorov model shows instead irreversible time dependent fluctuations of the intensity of fluorescence, rooted in the structure of the probability transitions which do not satisfy what is called sometimes the property of detailed balance.

  \section{Kolmogorov equation for a two-level system}
    \label{Keq2level} 
 Let us consider a two-level quantum system illuminated by a laser beam at resonance (practically this system is a single atom or ion in a Paul's trap). This system does two things: first starting from the ground state it will get to the excited state and back, following Rabi oscillations. Moreover it will decay spontaneously from the excited to the ground state by random emission of photons at the transition frequency: this is the fluorescence process. Although this has been much studied over the years, it does not seem to have been looked at as an out-of-equilibrium system such that irreversibility, in the precise meaning explained in section \ref{sec.irr}, could be put in evidence, for instance by the statistics of the fluctuations of fluorescence. This statistics requires a statistical theory of the process of emission-absorption of the light, including in particular the knowledge of the steady state as well as how it fluctuates all the time around this state.  
 This knowledge of the fluctuations requires to introduce a Kolmogorov-type equation for this two-level system, an equation for the evolution of the probability distribution of its quantum states, namely  an equation for the evolution of the density matrix $\rho_{at}$ of the atom.

The Kolmogorov equation is classical ( = non quantum) in the sense that different 
  sequences of quantum jumps yield quantum states of the atom which cannot interfere between themselves because they bear a random phase.  Therefore the contribution of each trajectory adds to the one of the others in the sense of classical probability, without quantum interference between different 
  sequences. From the point of view of interpretation of quantum mechanics, probabilities are added because the various trajectories belong to different Universes in the sense of Everett  \cite{everett}. This is explained as follows. Let $(...t_j, t_{j+1}, ...)$ be the set of times of emission of a fluorescence photon with $t_j < t_{j+1}$.  Together with the  coherent evolution between jumps, this sequence of times  defines what we shall call  a trajectory of the atom.  The quantity measuring the properties of the whole system, atom and radiated photons is the full density matrix which can be written symbolically like 
  $\rho( At,  F; At', F')$ 
 where $At$ is for the degrees of freedom of the atom (in the present case this will be a finite discrete set representing the indices of the quantum levels involved) and where $F$ is for the degrees of freedom of the field of the emitted photons. The primes are representing another choice of the variables, either for the degree of freedom of the atom or for the field. The rules of quantum mechanics are such that the state of the atom is given by the reduced density matrix derived from the overall density matrix by tracing it over the degrees of freedom of the field,
  \begin{equation} 
  \rho_{ at} = \Sigma_{F} \ \rho( At, F;  At' , F) \textrm{.}
  \label{eq:atF}
\end{equation}
 Consider the contribution to the reduced density matrix arising from different 
 sequences of quantum jumps. Those contributions correspond to different set of field coordinates $F$, because there is no overlap of the fields emitted by the atom at different times: the emitted fields propagate at the speed of light out of the atom and have no overlap if the emissions are at different times. Therefore the contributions of different histories to this density matrix add to each other in the ordinary classical sense. In particular, no contribution to the off-diagonal part of the density matrix may arise from the addition of  contributions of different trajectories which have no correlation in the quantum sense. This explains why one can add contributions to the reduced density matrix with a probability distribution in the ordinary sense for each contribution. For this problem this average with a probability distribution in the classical sense is the translation in this specific case of the summation over the degrees of freedom of the field, as done in the general formula defining the reduced density matrix.
 
 As regards stimulated emission and absorption in the coherent regime, it
  has been shown by Lee \cite{Lee} that the interaction between the coherent laser beam and a single atom has also a fluctuating component because of the randomness of the stimulated absorption and emission of photons. Such a randomness introduce fluctuations in the Rabi oscillations. Those fluctuations are washed out by averaging on a large number of atoms undergoing Rabi oscillations in a pump field. For times of observation sufficiently long to include many typical time scales of the process we shall study, we assume that this kind of fluctuation is also washed out by long time averages of the fluorescence of a single atom or ion. 

 \subsection{Deterministic regime: Rabi oscillations} 
 \label{Kdeterm}

We consider here the deterministic regime which begins at the instant when a photon is emitted by the atom, and goes on until a new photon is emitted. Let us consider the two coupled equations for the quantum amplitudes of the ground state $\vert 0>$ and the excited state $\vert 1>$ (in Dirac notation) of a two-level atom, of energy  $E_{0}$ and $E_{1}$ respectively ($E_{0} < E_{1}$).   
In absence of external field the wave function reduces to
\begin{equation}
\vert \Psi > = c_{0}(t) e^{-iE_{0 }t/\hbar} \vert 0 > + c_{1}(t) e^{-iE_{1 }t/\hbar} \vert 1 > 
\textrm{.}
\label{eq:wavef}
\end{equation}
with constant complex coefficients $c_{0,1}$. In presence of a quasi-resonant classical field $\mathcal{E} \cos(\omega_{L}t)$, of frequency $\omega_{L}$ close to the difference of  Bohr frequencies of the atom, $ \omega_{0}= (E_{1}-E_{0})/\hbar$, and assuming that all other atomic levels play no role, the decomposition (\ref{eq:wavef}) of the wave-function is still valid but with coefficients depending on time.  The interaction representation gets rid of the free evolution by introducing the coefficients $a_{0}=c_{0}$ and $a_{1}=c_{1}e^{i\omega_{L}t}$ which  are associated to the slow evolution of the state vector.  In absence of any relaxation by fluorescence the dynamics is purely deterministic and the two complex amplitudes
  obey the two coupled differential equations 
\begin{equation}
{\partial_{t}{a}}_0 = - i \frac{\Omega }{2} e^{i (t \delta  -\xi) } a_1
\textrm{.}
\label{eq:dot0}
\end{equation}
and 
\begin{equation}
{\partial_{t}{a}}_1 = - i \frac{\Omega}{2} e^{- i(t \delta  -\xi) }   a_0
\textrm{.}
\label{eq:dot1}
\end{equation}
where  $\delta=\omega_{L}-\omega_{0}$ is the detuning between the laser and the frequency difference of the two atomic levels, $\xi$  is the phase of the matrix element of the electric dipole moment $d = e <0\vert x \vert 1>$ of the atom
and  $\Omega$ is the Rabi frequency of the atom illuminated by  the laser,  proportional to the amplitude $\mathcal{E}$ of the incoming pump field, and to the modulus of $d$,
\begin{equation}
\Omega=- \frac{\vert d \vert \mathcal{E}}{\hbar}
\textrm{.}
\label{eq:omegR}
\end{equation}
 The dynamics described by equations(\ref{eq:dot0}) and (\ref{eq:dot1}) is unitary because the sum $(a_1  a^*_1 + a_0  a^*_0)$ is constant in time, $a^*$ being the complex conjugate of $a$.  We consider the case of real dipolar moment, or $\xi=0$, for simplicity. A complex dipolar moment changes nothing essential.
 Note that using a quantum description of the laser field, as in the dressed atom picture,  one get the same equations as (\ref{eq:dot0})-(\ref{eq:dot1}) for the amplitudes of the compound atom-laser system\cite{cohen}. This occurs  because starting from  a given state $\vert 0,n >$ or $ \vert 1,n-1>$ ($n$ being the laser photon number), the  system stays in the same manifold  in absence of quantum jumps. Its wave function  is  given by 
\begin{equation}
\vert \Psi > = a_{0}(t)  \vert 0,n > + a_{1}(t) \vert 1,n-1 > 
\textrm{,}
\label{eq:wavef2}
\end{equation} 
with amplitudes still obeying equations(\ref{eq:dot0}) and (\ref{eq:dot1}).
 
 \subsubsection{ Laser at exact resonance}
 
In addition to the real dipole condition, we  consider first the simplest case of laser at exact resonance with the atomic  transition, that gives $\delta=0$. Therefore we have :
  \begin{equation} 
 \delta=\xi=0
\textrm{.}
\label{eq:param}
\end{equation} 
 
Let us introduce the scaled time  $\tilde{t}= \Omega t /2 $,  the spinor $A$ and Pauli matrix $\sigma_{1}$ defined as,
\begin{equation}
A= 
 \left( \begin{array}{ccc}
a_{0} \\
a_{1}
\end{array}
\right)
\textrm{,}
\label{eq:dot1cc}
\end{equation}
and
\begin{equation}
\sigma_{1}= 
 \left( \begin{array}{ccc}
0 & 1 \\
1 &0
\end{array}
\right)
\textrm{.}
\label{eq:pauli}
\end{equation}
The system (\ref{eq:dot0})-(\ref{eq:dot1}) can be written as $\partial_{\tilde{t}} A= - i \sigma_{1} A$. The formal solution  $A(t)=\exp(-i \tilde{t} \sigma_{1}) A(0)$ can be written as
\begin{equation}
A(\tilde{t}) =(\cos \tilde{t} - i \sigma_{1} \sin \tilde{t}) A(0)
\textrm{.}
\label{eq:dot10}
\end{equation}

Because the deterministic regime begins just after the emission of a photon by transition from state $\vert 1>$ to $\vert 0>$, the amplitude of state $\vert 1>$  vanishes exactly after this emission. Therefore the post-decay initial conditions are $\vert a_{0}(0)\vert =1$ and $\vert a_{1}(0) \vert =0$,  and equation (\ref{eq:dot10}) becomes
\begin{equation}
A(\tilde{t})= e^{i\phi}
 \left( \begin{array}{ccc}
 \cos\tilde{t} \\
 i   \sin\tilde{t}
\end{array}
\right)
\textrm{.}
\label{eq:dot1d}
\end{equation}
Equation(\ref{eq:dot1d}) shows that the two amplitudes evolve in quadrature between two jumps. We have introduced
 the constant phase $\phi$ while it plays no role in this section devoted to the derivation of Kolmogorov equation and can be ignored for the moment. However this parameter  will play an important role when deriving the  statistical properties of the light emitted by the atom  because  this phase is associated to the random times of emission, therefore $\phi$ changes after each quantum jump (see section \ref{sec:Fluctlight} below for a discussion of this point).
Returning to the original variable and introducing the angle variable
\begin{equation}
\theta(t) =  \frac{\Omega }{2} t
\textrm{,}
\label{eq:thetat}
\end{equation}
we obtain 
 \begin{equation}
a_{0}(t)= i  e^{i\phi} \cos\theta (t);    \;\; \; a_{1}(t)= e^{i\phi} \sin\theta (t)
\textrm{,}
\label{eq:thetatt}
\end{equation}
which describes the Rabi nutation of the atom between the fundamental state and the excited state. Equation (\ref{eq:thetat}) shows  that the amplitudes $a_{0}$ and $a_{1}$ remain in quadrature during the deterministic motion, an important result which characterizes the resonant case only. This explains why we could  derive Kolmogorov equation in the resonant case from the dynamics of the complex amplitudes. 
Therefore in the deterministic regime  without damping the  dynamical equation for the probability $p(\theta,t)$ is
\begin{equation}
\partial_{t} p + \frac{\Omega}{2} \partial_{\theta} p =0
\textrm{.}
\label{eq:pleft}
\end{equation}
Equation (\ref{eq:pleft}) conserves the norm defined in equation (\ref{eq:1}). We point out the peculiar status of the resonant situation, for which the quantum flow can be studied directly by using the dynamics of the amplitudes $ a_{0,1}$, as done here.

Let us notice that other variables may be used to describe the trajectory in another phase space. For exemple, taking the variables ($r,\theta$)  used in the next subsection and defined in appendix A,  which are quadratic combinations of the amplitudes $ a_{0}$ and $a_{1}$, it can be shown that the quantum flow rotates in the plane ($r\cos2\theta, r\sin2\theta$)
 at constant angular velocity $\Omega/2$ along a circle of radius unity. With these variables the probability density $p$  is such that $p(r,\theta+\Omega t,t)=\delta_{D}(r-1) p(\theta,0)$, where $\delta_{D}$ is the Dirac distribution (not to be confused with the detuning  $\delta$). Moreover we show below that the probability is  still carried by equation (\ref{eq:pleft})  where the variable $\theta$ has not the same meaning (it differs by a factor $2$ because the functions $u,v$ are quadratic with respect to the amplitudes).

\subsubsection{Laser detuned from resonance}

When the laser is not at exact resonance with the atomic transition, the solution of equations (\ref{eq:dot0})-(\ref{eq:dot1})  can be expressed in terms of two  time dependent phases, $\theta(t)=\Omega t/2$ and $\Psi(t)= t \delta /2 $, see Appendix A, equations (\ref{eq:apa0})-(\ref{eq:apa1}).  But it appears that when we use the description of the motion in terms of the amplitudes $a_{0,1}$, the trajectory is complicated because their relative phase evolves with time, contrary to the resonant case where the amplitudes are in quadrature all along the deterministic trajectory. 

 We propose in Appendix A  to use a pertinent set of variables ($r,\theta$) such that the situation becomes very much like the one of the resonant case. These variables are the modulus and the phase of a complex function $u+iv$ where $u$ and $v$ are quadratic functions of new amplitudes $c_{0,1}$ defined in terms of $a_{0,1}$ by equation (\ref{eq:RB}). In the phase space ($u=r \cos2\theta, v=r \sin2\theta$) we show that the quantum flow displays a rotation at constant angular velocity $\omega=\Omega_{\delta}/2$, where
\begin{equation}
\Omega_{\delta}=\sqrt{\Omega^{2} +\delta^{2}}
\textrm{,}
\label{eq:omegeff}
\end{equation}
 is the effective Rabi frequency in presence of detuning. We show that the rotation takes place along a circle of radius 
\begin{equation}
r_{\delta}=\frac{\Omega}{\Omega_{\delta}}
\textrm{,}
\label{eq:rdet}
\end{equation}
 smaller than unity. At exact resonance the trajectory is a circle of radius unity in this phase space, as written above. After each emission of one photon, the trajectory starts again  on the circle from the same point $\theta=0$. Therefore one may forget the inside and outside of the circle and restrict to the circle of radius $r_{\delta}$ given by equation (\ref{eq:rdet}). On this trajectory one may put a probability density $p(r,\theta,t)$ which obeys the partial differential equation having the same form as (\ref{eq:pleft}),
\begin{equation}
\partial_{t} p+ \frac{\Omega_{\delta}}{2} \partial_{\theta} p =0
\textrm{.}
\label{eq:pdet}
\end{equation} 
 
\subsection{Rabi oscillations and damping}
\label{sec:2level-damping}

The dynamics of the system results from two physical phenomena,  first the Rabi oscillation described by the simple equation (\ref{eq:thetat}) which is an oscillation between the quantum amplitudes of the two states, secondly there is spontaneous decay (not described by the equations above) of the state  $\vert1>$ toward the ground state $\vert0>$. The rate of transition is proportional to $\gamma$, a quantity with the physical dimension of an inverse time which is proportional to the square of $d$, the atomic dipolar moment, and to $\omega_{L}^{3}$. Any statistical theory is based on a probability distribution. In the present case, we  could consider the probability distribution of  the values of $\cos(\theta)$ or $\sin(\theta)$, both in the interval $[-1,1]$, or more simply of the variable $\theta$,  all  measuring the probability of the atom to be in the excited or the ground state. Let $p(\theta, t)$ be the probability distribution describing this system, a wrapped  probability distribution because it is a periodic function of $\theta$, of period $\pi$ instead of $2\pi$ because a change of sign in front of  both $a_0$ and $a_1$ does not change the state of the system. The probability  $p(\theta, t)$ is normalized by the constraint 
\begin{equation}
 \int_{-\pi/2}^{\pi/2} \mathrm{d}{\theta} \ p(\theta, t) =1
\textrm{.}
\label{eq:1}
\end{equation}
As explained before (beginning of the present section) we consider probabilities with their classical (non quantum) meaning. This relies on the assumption that the various quantum trajectories (or histories) are independent, namely that for different values of the angle $\theta$ at a {\emph{given time}} no quantum correlations exist. Such correlations would forbid one to consider trajectories with different $\theta$ at a given time as statistically independent. This statistical independence in the quantum sense is there because what we call different histories are different set of times of emission of fluorescence photons. 

Another issue must also be discussed: why is the photon emission treated as a quantum jump? It implies that this process is much faster than any other physical process in the problem, so that its time duration is so short that it can be seen as instantaneous to make valid Kolmogorov equation. The typical time scale for the duration of a "quantum jump" is monitored by atomic processes. Roughly speaking it is about of the order of magnitude it takes for the emitted photon to leave the close neighborhood of an atom. This time is almost  independent of the weak interaction with the laser field for instance. Therefore the only relevant time scale for such an atomic process is the period of the electromagnetic wave emitted by the transition from the excited to the ground state. In the situation of the kind of experiment we are thinking to \cite{exp}, this period of the electromagnetic  waves is much shorter than the interval between two emissions of photon (or, equivalently than the lifetime of the excited state) and much shorter too than the period of Rabi's oscillations. Therefore it is legitimate to consider this "duration of the quantum jump" as negligibly small compared to any other relevant time scale.

Let us derive the equation of motion of the probability $p(\theta, t)$, taking into account the joint effect of Rabi oscillations 
and the spontaneous decay toward the ground state.  The probability distribution is expected to be a periodic function of the variable $\theta$, of period $\pi$. Let us consider the domain $-\pi/2 < \theta < \pi/2$. 

\subsubsection{Laser at resonance}
The probability of the event ``the atom decays  at time $t$ from the excited state  toward the ground state'' is  proportional to  the square modulus of the amplitude of the excited state, namely to  $\sin^2(\theta)$ 
times the decay rate $\gamma$, a data of the problem with the physical dimension of an inverse time. 
Let us write now explicitly the right-hand side of equation (\ref{eq:dotproba}). The  probability  $\Gamma( \theta;\theta')$ for the atom  to make a quantum jump from the state $\theta$ towards the state $\theta'$  is proportional to $\delta_{p}(\theta') $ because any jump lands on $\theta'=0$ in  the interval  $[-\pi/2, \pi/2]$
and this probability is proportional to  $ \gamma \sin^{2 }\theta $ because it comes from the state $a_1$ with the squared amplitude $\sin^{2}\theta$. Therefore
\begin{equation}
\Gamma( \theta \vert  \theta')=  \gamma \ \sin^{2 }\theta \ \delta_{p}(\theta')  
\textrm{.}
\label{eq:Gamma}
\end{equation}
Although this does not make any difference in the interval  $[-\pi/2, \pi/2]$, we shall replace sometime Dirac function $\delta(\theta)$ by $\delta( \sin(\theta))$ to make more obvious the periodicity with respect to $\theta$ of a given expression.     
  Using the above expression for the jump probability one obtains the following Kolmogorov equation for the dynamics of  the $\pi$-periodic function $p(\theta, t)$, 
\begin{equation}
{\partial_{t}{p}} + \frac{\Omega}{2} \partial_{\theta} p = \gamma \left(  \delta(\sin(\theta))  \left(  \int_{-\pi/2 }^{\pi/2} \mathrm{d}{\theta'}\ p(\theta', t)  \sin^2(\theta') \right) - p(\theta, t)  \sin^2(\theta) \right)
\textrm{.}
\label{eq:p}
\end{equation}

\subsubsection{Laser detuned from resonance}

In presence of a detuning, we have shown in the previous subsection that the trajectory in the phase space ($r,\theta$) is confined on the circle of radius  $r_{\delta}$ given by (\ref{eq:rdet}). After each emission of a photon, in the plane $u=r\cos2\theta, v=r\sin2\theta$ schematized in Fig.\ref{fig:frontier}-(b), the quantum flow  starts from the same point $u=r, v=0$, and rotate with angular velocity $\Omega_{\delta}/2$ until the next emission. To derive the right hand side of Kolmogorov equation, we must express the transition function $\Gamma(\theta,\theta')$ in this space.  We start naturally from the expression,
\begin{equation}
\Gamma( \theta \vert  \theta') =  \gamma \vert a_{1}(\theta) \vert^{2} \; \delta_{p}(\theta')  
\textrm{.}
\label{eq:Gammadet}
\end{equation}
The  amplitude $a_{1}(\theta)$ is obtained via the amplitudes $c_{0,1}$ expressed in terms of the functions ($u,v$) as detailed in Appendix A. We get an expression $\vert a_{1}\vert^{2}$ which is surprisingly proportional to $\sin^{2}(\theta)$, that gives a transfert function (\ref{eq:Gammadet}) which becomes (\ref{eq:Gamma}) for zero detuning. Finally the probability $p(\theta,t)$ obeys the following equation
\begin{equation}
{\partial_{t}{p}} +  \frac{\Omega_{\delta}}{2}  \partial_{\theta} p =  \gamma \left(   \delta_{p}(\theta)  \left(  \int_{-\pi/2 }^{\pi/2} \mathrm{d}{\theta'}\ p(\theta', t) f(\theta') \right) - p(\theta, t)  f(\theta)  \right)
\textrm{,}
\label{eq:Kolmdet}
\end{equation}
where
\begin{equation}
f(\theta)=  \gamma (\frac{\Omega}{\Omega_\delta})^{2} \sin^{2}\theta
\textrm{.}
\label{eq:fdet}
\end{equation}

In summary we have proved that Kolmogorov equation for the quasi-resonant case and exact resonant case are  formally identical. They only differ by two coefficients, one concerns the Rabi frequency in the left hand side, the other is the factor $ \gamma (\frac{\Omega}{\Omega_\delta})^{2} $ which is in front of $\sin^{2}\theta$ in the right hand side. When rescaling the time, as done below, these coefficients disappear. In the following we focus on equation (\ref{eq:p}).

 \subsection{Solution of Kolmogorov equation}
 \label{sec:exactsol} 

Solutions of  equation(\ref{eq:p}) keep constant the $L^1$ - norm $ \int_{0_{-}}^{\pi} \mathrm{d}{\theta}  \  p(\theta, t) $ for any periodic distribution of the variable $\theta$. 
 The left-hand side of this equation describes the Rabi oscillation which amounts to an uniform drift in time of the angle $\theta$. The right-hand side represents the effect of the spontaneous decay of the excited state toward the ground state, it has a gain term for the ground state $\theta=0$ and a loss term for any other value of $\theta$. 
 
 Below we look at various questions related to this Kolmogorov equation. We give first its explicit solution and derive some interesting properties, then its steady solution reached from arbitrary initial conditions. Afterwards we look at various limiting cases namely the one of large and of small damping. Lastly we explain how to obtain various physical quantities out of solutions of this Kolmogorov equation, in particular for the spectral properties of the emitted light.

 \subsubsection{ Derivation of the probability density and average value of $\sin^{2}(\theta)$. }
 \label{sec:exactsol} 

Let us introduce the auxiliary function
\begin{equation}
 b(t) =  \int_{-\pi/2}^{\pi/2} \mathrm{d}{\theta'}\ p(\theta', t)   \sin^2(\theta') \textrm{.}
\label{eq:b(t)}
\end{equation}
  The equation we are trying to solve becomes
\begin{equation}
{\partial_{t}{p}} + \frac{\Omega}{2} \partial_{\theta} p + \gamma \sin^2(\theta) p = \gamma \delta(\sin(\theta)) b(t)  
\textrm{.}
\label{eq:pgamqsoltransf}
\end{equation}
Let us take $ \frac{2}{\Omega}$ as unit of time and introduce the dimensionless parameter $\gamma' =  \frac{2 \gamma}{\Omega}$. equation(\ref{eq:pgamqsoltransf}) becomes
\begin{equation}
{\partial_{t}{p}} + \partial_{\theta} p = g(\theta, t) - f(\theta) p
\textrm{.}
\label{eq:JG1}
\end{equation}
with 
\begin{equation}
 g(\theta, t) =  \gamma'  \delta(\sin(\theta)) b(t),
 \label{eq:g}
\end{equation}
where $b(t)$ is still given by equation(\ref{eq:b(t)})
  and 
  \begin{equation}
  f(\theta) =  \gamma' \sin^2(\theta).
  \label{eq:f(x)}
\end{equation}
 Let us derive from equation(\ref{eq:JG1}) the differential equation satisfied by the function $h(\theta, t) = p(\theta + t, t)$ which takes the form
 \begin{equation}
\partial_{t} h(\theta,t)= g(\theta + t, t) - f(\theta+t)) h(\theta,t)
\textrm{.}
\label{eq:JG2}
\end{equation}
This can be solved as an initial value problem as follows. Take $s(\theta, t) = h (\theta, t) e^{ \int_{0}^{t} \mathrm{d}{t'}\ f(\theta+t')}$. Therefore $s(\theta, t = 0) = h(\theta, t = 0)$. The auxiliary function $s(\theta, t)$ is a solution of 
$$ \partial_{t} s = g(\theta + t, t) e^{\int_{0}^{t} \mathrm{d}{t'}\ f(\theta+t')}\textrm{.}$$
This has the solution 
$$s(\theta, t) = s(\theta, t = 0) + \int_{0}^{t} \mathrm{d}{t'}\  g(\theta + t', t') e^{\int_{0}^{t'} \mathrm{d}{t''}\ f(\theta+t'')}\textrm{.}$$
The equivalent result for the function $h(\theta, t) $ reads
$$h(\theta, t) = h(\theta, 0) e^{- \int_{0}^{t} \mathrm{d}{t'}\  f(\theta + t')} +  \int_{0}^{t} \mathrm{d}{t'}\  g(\theta + t', t') exp{\left(\int_{0}^{t'} \mathrm{d}{t''}\  f(\theta + t'') - \int_{0}^{t} \mathrm{d}{t''}\  f(\theta + t'')\right)}\textrm{.}$$
Tracing back the path from this explicit solution to the original equation, one finds the general solution of equation(\ref{eq:JG1}): 
 \begin{equation}
 p(\theta, t) =  p(\theta - t, 0)\alpha(\theta, t) + \int_{0}^{t} \mathrm{d}{t'}\alpha(\theta, t') g(\theta - t', t - t')
\textrm{.}
\label{eq:JG3}
\end{equation}
where 
 \begin{equation}
 \alpha(\theta, t) =e^{\left(- \int_{0}^{t} \mathrm{d}{t'}\  f(\theta - t')\right)},
 \label{eq:alpha}
\end{equation}
 and $g(.)$ is given by equation(\ref{eq:g}). 
Multiplying both sides of this equation by $f(\theta) = \gamma' \sin^2(\theta)$ and integrating the result over one period for $\theta$, one finds the following Fredholm integral equation for $b(t)$,
 \begin{equation}
 b(t) = m(t) +  \int_{0}^{t} \mathrm{d}{t'}\ b( t') l(t-t') 
\textrm{.}
\label{eq:JG4}
\end{equation}
In (\ref{eq:JG4}) we have
 \begin{equation}
  m(t) =  \int_{T}  \mathrm{d}{\theta}\ p(\theta, 0) f(\theta+t) \alpha(\theta+t, t) \textrm{,}
  \label{eq:m(t)}
\end{equation}
where $T$ stands for the period  of the function $f$ (here $T=(0, \pi)$),
and 
 \begin{equation}
 l(t)=f(t)\alpha(t)
  \textrm{,}
   \label{eq:l(t)}
\end{equation}
where $\alpha(t)$ is the reduction to $\theta=t$ of the function of two variables $\alpha(\theta,t)$ (hopefully no confusion will arise from the use of the same notation, $\alpha$ for $\alpha(\theta,t)$ and $\alpha(t, t) = \alpha(t)$) : 
 \begin{equation}
 \alpha(t) =\alpha(t,t)=e^{- \int_{0}^{t} \mathrm{d}{t'}\  f( t')}
  \textrm{.}
   \label{eq:alphat}
\end{equation}
Note the relation
 \begin{equation} 
 l(t)= - \partial_{t} \alpha (t)
 \textrm{.}
\label{eq:JG9}
\end{equation}

\begin{figure}
\centerline{
 \includegraphics[height=1.5in]{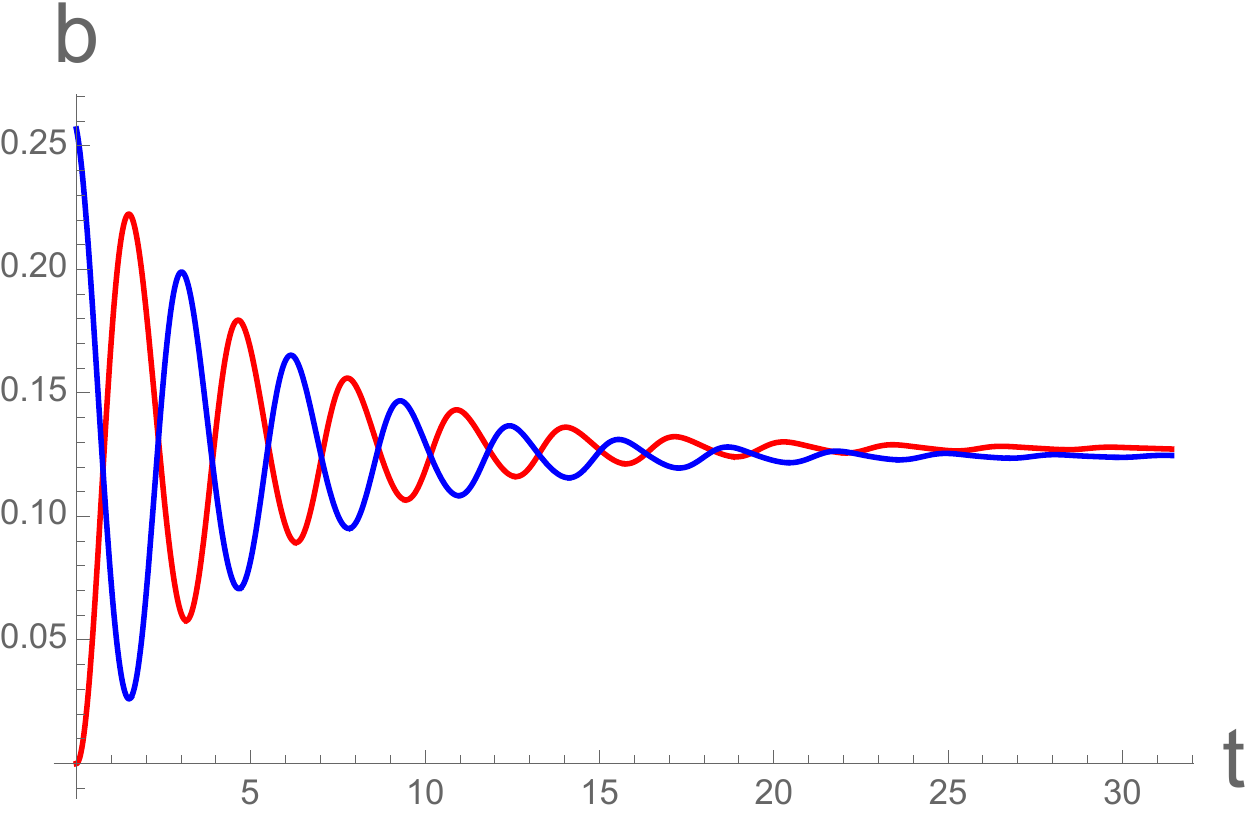}  
  }  
\caption{Solution of equation (\ref{eq:JG4}) in the small damping range $\gamma=1/7.8$. The  initial condition is given by equation (\ref{eq:icp}) with  $\theta_{0}=0$ (red curve)  and $\theta_{0}=\pi/2$ (blue curve).  
 }
\label{fig:b}
\end{figure}

For $m(t)$ given,  $b(t)$ can be derived from equation(\ref{eq:JG4}) either by iterations or by Laplace transforming both sides.  In the numerics  we use the iteration method which gives the result drawn in Fig.\ref{fig:b} where the initial condition for the probability is chosen as
 \begin{equation}
p(\theta, t = 0)=\delta(\theta-\theta_{0} ) \textrm{,}
  \label{eq:icp}
\end{equation}
This choice is made in view of the derivation of correlation functions, see  subsection \ref{sec:Fluctlight}. It follows that any mean value calculated by using the probability (\ref{eq:JG3}) with initial condition (\ref{eq:icp}), is actually a conditional average and should depend on the parameter $\theta_{0}$ and should be labelled  $b(t,\theta_{0})$  as done below. Figure \ref{fig:b} shows this function $b(t)$ for two values of $\theta_{0}$, both converging at large time towards the stationary value $<b_{st}>$ (see below and appendix B for a sketch of proof of this property).
With initial condition (\ref{eq:icp}) the  probability  $p(\theta,t)$ in (\ref{eq:JG3}) becomes the conditional probability,
 \begin{equation}
p(\theta, t \vert \theta_{0},0)=\delta(\theta-t -\theta_{0}) \alpha(\theta_{0},t) + \int_{0}^{t} \mathrm{d}{t'}\alpha(\theta, t') g(\theta - t', t - t')
\textrm{,}
  \label{eq:condp}
\end{equation}
where $g(\theta,t)$ is given by equation(\ref{eq:g}).

Formally Laplace transform method is very simple: 
Let $\Phi_L(z)$ be Laplace transform of a function $\Phi(t)$ of time, defined as
 \begin{equation}
 \Phi_L(z) = \int_{0}^{ \infty} \mathrm{d}{t}\ \Phi(t) e^{- zt}
\textrm{.}
\label{eq:JG5}
\end{equation}
To ensure the convergence of the integral in this definition of Laplace's transform, the real part of $z$ must be big enough, depending on how $\Phi(t)$ behaves at infinity (assuming it is a smooth function otherwise). The Fredholm equation (\ref{eq:JG4}) has a simple solution in Laplace transform $ b_L(z) = m_L(z)/(1 - l_L(z))$, or using equation (\ref{eq:JG9}) 
  \begin{equation}
 b_L(z) =- \frac{m_L(z)} {z \alpha_L(z)} 
\textrm{,}
\label{eq:JG6}
\end{equation}
but this result is not very useful because the inverse Laplace transform requires to know the singularities of the right-hand side, a difficult task.

\subsubsection {Properties of Kolmogorov solution}
\label{sec:propK}
Several properties of the solution  (\ref{eq:JG3})-(\ref{eq:JG4}) of Kolmogorov equation deserves to be mentioned. They are valid for any periodic function $f(.)$ which is positive and bounded.
An important result  is
 \begin{equation}
M(t)=B(t)
\textrm{,}
\label{eq:JG11}
\end{equation}
 where $M(t)$ is  the integral of $m(t)$, 
 \begin{equation}
M(t)= \int_{0}^{ t} \mathrm{d}{t'}\  m(t')
\textrm{,}
\label{eq:JG7}
\end{equation}
and 
 \begin{equation} 
 B(t)= \int_{0}^{ t} \mathrm{d}{t'}\  b(t') \alpha(t-t')
\textrm{.}
\label{eq:JG8}
\end{equation}
 The proof goes as follows. 
Using equation (\ref{eq:JG9}), equation (\ref{eq:b(t)}) becomes 
 \begin{equation} 
m(t)=b(t)+ \int_{0}^{t} \mathrm{d}{t'}\  b( t') \partial_{t}\alpha(t-t')=  \partial_{t}B (t)
 \textrm{.}
\label{eq:JG10}
\end{equation}
Integration of (\ref{eq:JG10}) with respect to time gives (\ref{eq:JG11}).

Another striking result concerns to evolution of several functions as time increases. Let us consider for example the function $\alpha(t)$ for any positive $\pi-$periodic function $f(.)$, and  define its particular value  $\alpha(\pi)$,
\begin{equation} 
\bar{\alpha}=e^{- \int_{0}^{\pi} \mathrm{d}{t}\  f( t)}
\textrm{,}
\label{eq:alphapi}
\end{equation}
From  (\ref{eq:alphat}) we have, by definition:
\begin{equation} 
\alpha(t+\pi)=\bar{\alpha} \alpha(t)
\textrm{.}
\label{eq:alphatpi}
\end{equation}
Moreover, equation(\ref{eq:m(t)}) can be written as
 \begin{equation} 
m(t)= \int_{T} \mathrm{d}{\theta}\ p(\theta,0)\left(-\partial_{t} ( e^{- \int_{0}^{t} \mathrm{d}{t'}\  f(\theta+ t')}) \right)
 \textrm{,}
\label{eq:JG12}
\end{equation}
that gives by integration with respect to time
 \begin{equation} 
M(t)=  1- \int_{T} \mathrm{d}{\theta}\ p(\theta,0) e^{- \int_{0}^{t} \mathrm{d}{t'}\  f(\theta+ t')}  =1-k(t)
 \textrm{.}
\label{eq:JG13}
\end{equation}

The periodicity of $f$ leads to the  relation between $k(t+\pi)$  and $ k(t)$ defined in (\ref{eq:JG13}),
 \begin{equation} 
k(t+\pi)= \bar{\alpha}\; k(t)
 \textrm{.}
\label{eq:JG14}
\end{equation}
Finally an interesting property concerns the evolution of the function $B(t)$. Defining the difference
 \begin{equation} 
\beta(t)= B(t+\pi) - \bar{\alpha}\; B(t) = \int_{t}^{t+\pi} \mathrm{d}{t'}\  b( t') \alpha(t+\pi-t')
 \textrm{.}
\label{eq:JG14b}
\end{equation}

we can show that
 \begin{equation} 
\beta(t)= 1- \bar{\alpha}
\textrm{,}
\label{eq:JG15}
\end{equation}
for $t \geqslant 0$.  Here is the proof, 
 \begin{equation} 
\beta(t)= M(t+\pi) - \bar{\alpha} M(t)= 1-k(t+\pi) -\bar{\alpha}(1-k(t))=1-\bar{\alpha}
\textrm{.}
\label{eq:JG16}
\end{equation}
The relations established above are useful to check the numerical results, in particular we have verified the property (\ref{eq:JG15}). 
Finally we make the conjecture that, as $t$ tends to $+\infty$,  $b(t)$  tends to its  average value, $<b_{st}> $ of $\gamma' \sin^{2}(\theta)$ calculated with the stationary probability given in equation(\ref{eq:pst9}): 
 \begin{equation} 
\lim_{t \to + \infty} b(t)= <b_{st}>
\textrm{.}
\label{eq:JG17}
\end{equation}
This statement is expected intuitively and confirmed by the numerics. We haven't  established it rigorously, but we made a step towards the proof by considering the Fourier transform of $H(t)b(t)$,$H$ being Heaviside function, equal to unity for $t\ge 0$ and zero for $t<0$. Defining the Laplace transform  of $Hb$ by  
$\widehat{Hb}(z)= \int_{0}^{\infty} \mathrm{d}{t}\ b(t) e^{-i z t} $ we show in Appendix B that
 \begin{equation} 
\widehat{Hb}= \frac{<b_{st}>}{iz}  + \frac{\tilde{b}(z)}{\widehat{H_{\pi}\alpha}(z)}
\textrm{,}
\label{eq:JG17b}
\end{equation}
where $H_{\pi}$ is the characteristic function of $[0,\pi]$, and $\tilde{b}(z)$  is an analytical function for any bounded value of $z$. The end of the proof necessitates the knowledge of the zeros of $\widehat{H_{\pi}a}(z)$ in the complex plane.

 \subsection{Steady solution of Kolmogorov equation}
\label{sec:steadysol}

The steady distribution $p_{st}(\theta)$, if it exists,  must satisfy the integro-differential equation
 
\begin{equation}
 \partial_{\theta} p_{st} = \gamma'\left( \delta(\theta)  \left(  \int_{-\pi/2 }^{\pi/2}  \mathrm{d}{\theta}\ p_{st}(\theta)  \sin^2(\theta) \right) - p_{st}((\theta)  \sin^2(\theta) \right)
\textrm{,}
\label{eq:pst}
\end{equation}
which shows that it depends on the single dimensionless parameter $\gamma'= \frac{2\gamma}{\Omega}$. 
It can be checked that the solution of equation(\ref{eq:pst}) is a periodic function of $\theta$ of period $\pi $. By integrating the equation from $-\pi /2 $ to  $+ \pi/2 $ one obtains zero on the right-hand side, whereas the left-hand side is proportional to the difference $p_{st} (\pi/2) - p_{st} (-\pi/2)$ which is zero as well. 

\begin{figure}
\centerline{
(a) \includegraphics[height=1.5in]{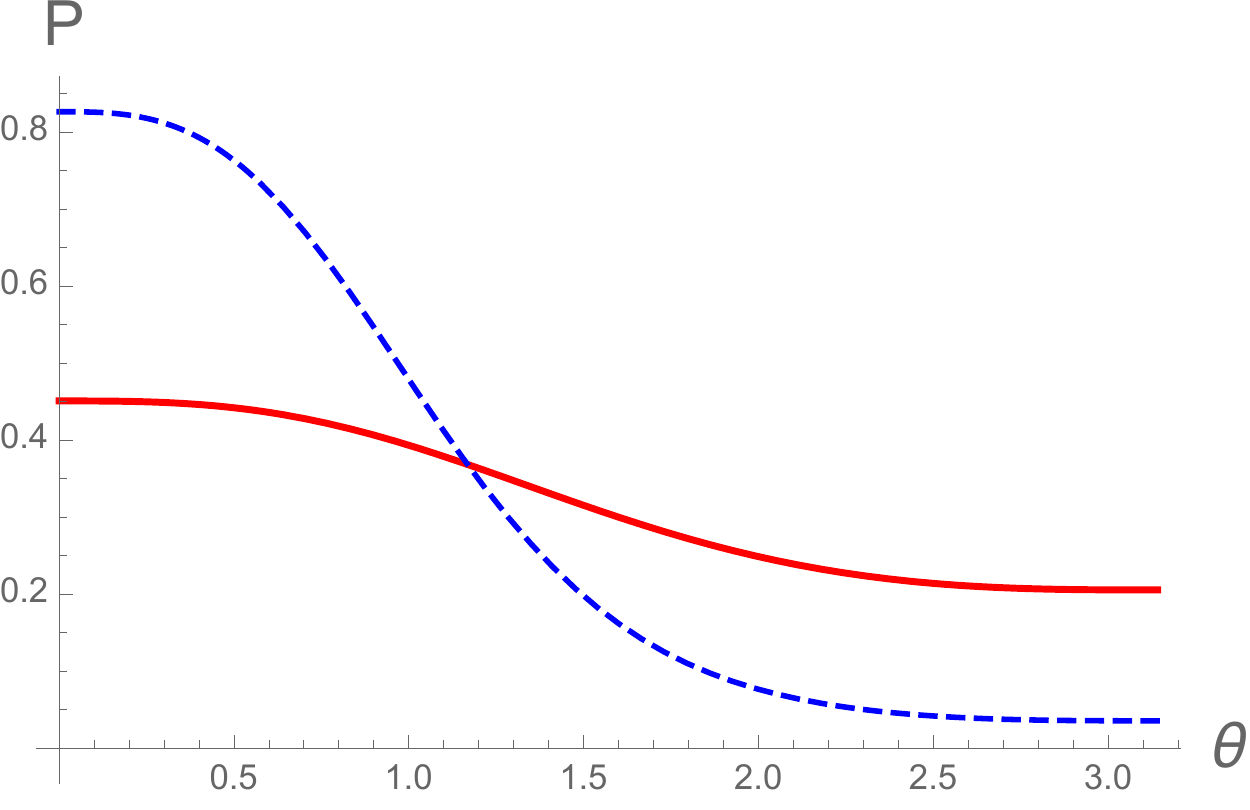}  
(b)\includegraphics[height=1.5in]{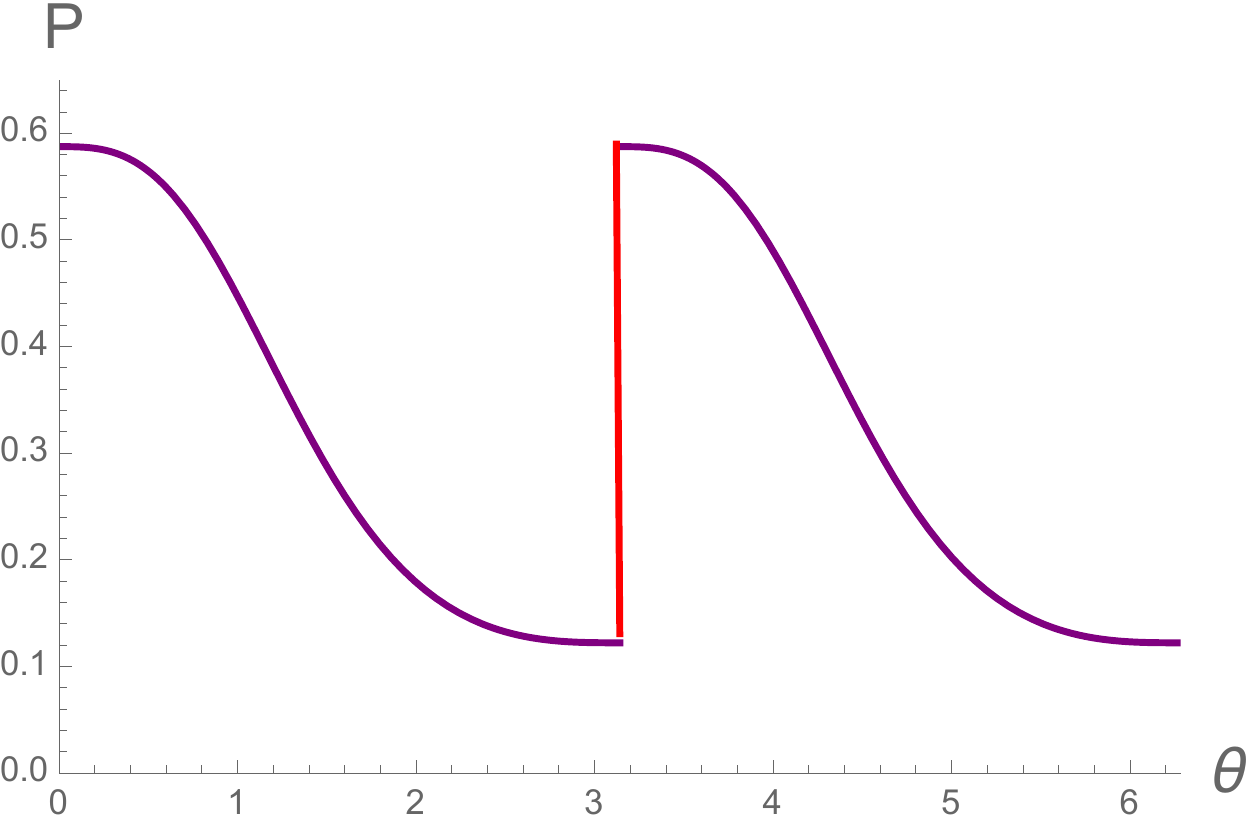}
  }  
\caption{ Stationary probability distribution  $p_{st}(\theta)$,  solution of equation (\ref{eq:pst}). (a) In the interval $[0,\pi]$, solid  line for $\gamma'=0.5$, dashed  line for $\gamma'=2$. The periodicity  of the wrapped distribution is illustrated in (b) for  $\gamma'=1$  on the interval $[0,2\pi]$.
}
\label{fig:Kolst}
\end{figure}

For $\theta$ different of zero  the integro-differential equation (\ref{eq:pst}) reduces to
 \begin{equation}
 \partial_{\theta}\hat{p}_{st}(\theta)= - \gamma' \sin^{2}(\theta) \; \hat{p}_{st} (\theta) 
 \textrm{,}
\label{eq:pst2}
\end{equation}
which can be formally integrated as
 \begin{equation}
\hat{p}_{st}(\theta) = \hat{p}_{st}(0_{+}) \alpha(\theta) 
 \textrm{,}
\label{eq:pst4}
\end{equation}
where, as before, $\alpha(.)$ with only one argument, is the restriction to the line $\theta=t$ of the function $\alpha(\theta,t)$ defined in  equation(\ref{eq:alpha}), it gives $ \alpha(t) =e^{\left(- \int_{0}^{t} \mathrm{d}{t'}\  f(t - t')\right)}$, or
 \begin{equation}
 \alpha(\theta) = e^{-\frac{\gamma'}{4}(2\theta-\sin(2\theta))},
 \label{eq:alpha2}
\end{equation}
 for the particular case of $f$ given in (\ref{eq:f(x)}). This solution is formally not convenient a priori because the exponent of $\alpha$ is not periodic with respect to $\theta$ . However the periodicity is restored by noticing that the solution has a jump at $\theta = 0$. This jump is such that the value of  $p_{st}(\theta)$ for $\theta = 0_{-}$ is just equal to $\hat{p}_{st}(\theta)$ for $\theta = \pi_{_-}$, as given by the solution  of equation(\ref{eq:pst4}). There remains to find the constant of integration $\hat{p}_{st}(0_{+})$. It is derived from the norm constraint (\ref{eq:1}).
Defining $$\mathcal{I}_{x}=  \int_{0}^{x} \mathrm{d}{\theta} \ \alpha (\theta),$$ the
 norm condition  gives 
  \begin{equation}
\hat{p}_{st}(0_{+})= ( \mathcal{I}_{\pi})^{-1},
\label{eq:pst6}
\end{equation}
which depends on the dimensionless parameter $\gamma'$ as illustrated in Fig.\ref{fig:Kolst}-(a).
The  stationary probability distribution, see Fig.(b),  is the wrapped periodic function built by translating  the solution $\hat{p}_{st}(\theta)$ just found, given by equations (\ref{eq:pst4})-(\ref{eq:pst6}) and defined on the interval $ [0_{+}, \pi_{-}]$. This can be written formally as  
 \begin{equation}
p_{st}(\theta)=  \sum_{k=-\infty} ^{\infty} \hat{p}_{st}(\theta-k\pi)
\label{eq:pst7}
\end{equation}
where $k$ is an integer. This function displays discontinuities for any value $\theta=k\pi$. In the interval $[0,\pi]$ the derivative $p_{st,\theta} $ calculated within the meaning of a distribution is equal to $ \partial_{, \theta}p + \mu \delta(\theta)$  where $\mu$ is the size of the jump of $p_{st}$ at $\theta = 0$ , $ \mu= \hat{p}(0_{+})-\hat{p}(\pi_{-})$. This derivative is identical to the right-hand side of (\ref{eq:pst}) and we have  the relation $<b_{st}>= \gamma' <\sin^{2}\theta>  = \hat{p}_{st}(0_{+})-\hat{p}_{st}(\pi_{-})= \hat{p}_{st}(0_{+}) (1-\alpha(\pi))$. From equations (\ref{eq:pst4})-(\ref{eq:pst6}) we derive the two relations
\begin{equation}
\gamma'  \int_{0}^{\pi} \mathrm{d}{\theta}  = 1-\alpha(\pi),
\label{eq:pst8}
\end{equation} 
and
\begin{equation}
<b_{st}>= \frac{1-\alpha (\pi)}{\mathcal{I}_{\pi}},
\label{eq:pst9}
\end{equation} 
where $\alpha$ is given by equation (\ref{eq:alpha2}).

\subsection{Kolmogorov equation in two limits}

Below we look at what kind of approximation can be made in the opposite limits of a large or a small dimensionless ratio $\frac{\gamma}{\Omega}$. 
\subsubsection{ Large damping}
When this ratio is large it means that the damping due to the random emission of photons is large compared to the driving by the external pump field. Therefore the system will be mostly in the fundamental state. it means that the angle $\theta$ remains close to zero. Therefore one can replace in this limit $\sin(\theta)$ by $\theta$. Kolmogorov equation (without introducing any new symbol) becomes: 
\begin{equation}
{\partial_{t}{p}} + \frac{\Omega}{2} {\partial_{\theta}} p = \gamma \left(  \delta(\theta)  \left(  \int_{0_{-}}^{\pi} \mathrm{d}{\theta'}\ p(\theta', t)  (\theta')^2 \right) - p(\theta, t) \theta^2 \right)
\textrm{.}
\label{eq:pgam}
\end{equation}
This can be transformed into a parameterless equation by taking $\left(\frac{\Omega}{\gamma}\right)^{1/3}$ as unit for $\theta$ and $1/2\left({\Omega}^2{\gamma}\right)^{-1/3}$ as unit of time. Because this is a small number in the limit $\gamma$ large we can ignore the condition of periodicity with respect to $\theta$ and take the new transformed scaled variable as going from minus to plus infinity. With this variable Kolmogorov equation becomes:
\begin{equation}
{\partial_{t}{p}} +  {\partial_{\theta}} p = \left(  \delta(\theta)  \left(  \int_{-\infty}^{+\infty} \mathrm{d}{\theta'}\ p(\theta', t)  (\theta')^2 \right) - p(\theta, t) \theta^2 \right)
\textrm{.}
\label{eq:pgam.1}
\end{equation}
The normalization condition becomes:
$$ \int_{-\infty}^{+\infty} \mathrm{d}{\theta'}\ p(\theta', t) = 1\textrm{.}$$
This yields an universal parameterless problem where all the scaling laws are exhibited as a consequence of the transformations made to obtain this parameterless equation. 

\subsubsection{Small damping}
This limit is interesting because it gives an idea of the result of the general solution, which is more involved. In this limit it is easy to check that, if one neglects the right-hand side of equation(\ref{eq:pst}); the solution for the steady distribution $p_{st}(\theta)$ is the constant $1/\pi$.

The case of a small $\gamma$ can be dealt with as follows. The Rabi oscillations make the fast motion. Therefore, during those oscillations, the function $ \sin^2(\theta)$ as it appears in equation(\ref{eq:p}) can be replaced by its average $1/2$. Therefore, in this limit Kolmogorov equation becomes: 
\begin{equation}
{\partial_{t}{p}} + {\partial_{\theta}} p = \gamma' \left( \delta(\sin\theta)  - p(\theta, t) \right)
\textrm{.}
\label{eq:psimpl}
\end{equation}
In the equation above, $2/\Omega$ is taken as time unit and $\gamma' = \frac{\gamma}{\Omega}$ is a small parameter. Let us introduce the function $\hat{p} = e^{-\gamma' t} p( \theta, t)$. It satisfies the equation:
\begin{equation}
{\partial_{t}}{\hat{p}}  +  {\partial_{\theta}}{\hat{p}} = g(\theta, t)
\textrm{.}
\label{eq:psimpl.1}
\end{equation}
where $g(\theta, t) = \gamma' e^{-\gamma' t}  \delta(\sin\theta) =  \gamma' e^{-\gamma' t} \sum_k  \delta(\theta - k \pi)$ where  the index $k$ is any integer, positive, zero or negative. The solution of equation(\ref{eq:psimpl.1}) with initial condition ${\hat{p}}(\theta, t = 0) = p_0(\theta)$ reads: 
\begin{equation}
{\hat{p}}(\theta, t) =  p_0(\theta -  t) + \int_{0}^{t} \mathrm{d}t' g(\theta - t', t - t')
\textrm{.}
\label{eq:psimpl.2}
\end{equation}
The integral over $t'$ can be carried out and yields 
$$\int_{0}^{t} \mathrm{d}t' g(\theta -  t', t - t') = \sum_k e^{-\gamma'(\theta - k\pi)} H(\theta - k\pi) H(t- \theta + k\pi) \textrm{,}$$ where $H(.)$ is Heaviside function. This yields at once the final result for the solution of the initial value problem in the limit $\gamma'$ very small:
\begin{equation}
p(\theta, t) = e^{-\gamma' t}  p(\theta - t, t = 0) + \gamma' \sum_k e^{-\gamma'(\theta - k\pi)} H(\theta - k\pi) H(t- \theta + k\pi)
\textrm{.}
\label{eq:psimpl.3}
\end{equation}
Because of the Heaviside function it is not completely obvious to find the limit of  this expression of $p(\theta, t)$ for $t$ tending to plus infinity.  The sum over $k$ is limited by the condition expressed by the product of Heaviside functions. If one takes $\theta = 0$ to simplify the algebra, $k$ should be such that $ 0 > k > -t/\pi$. In the limit $t$ tending to plus infinity the lower bound on $k'$ can be taken as minus infinity because the sum over $k'$ converges in this limit of large negative values. The result is:
$$  \gamma' \sum_k e^{-\gamma'(- k\pi)} H(- k\pi) H(t + k\pi) \rightarrow \gamma' \sum_{k<0} e^{\gamma' k\pi} =  \gamma'  \frac{\gamma'}{1 - e^{- \gamma'  \pi}}\textrm{.}$$
One can check that, in the limit $\gamma'$ small, this is also the same limit for any value of $\theta$ in the interval $(0, \pi)$ and that this limit is equal to the constant $1/\pi$, the same as the value of $p_{st}(\theta)$ derived before.

\subsection{Between two jumps: statistics of the time intervals}
\label{sec:between}

Here we study the statistics of the point process formed by the quantum jumps.  It has the following properties. $(i)$ The number of jumps in non overlapping intervals are independent variables. $(ii)$ In a small interval $[t, t+dt]$ there is at most one quantum jump. These two properties define a Poisson process. We shall show below that this Poisson process is non stationary, in the sense that its density  (or intensity) is not constant versus  time. 

We shall derive the probability  distribution of the time intervals between two consecutive quantum jumps as follows.  We  consider the dynamics between two successive jumps.  Just after a jump, supposed to occur at time $t=0$, the probability density of the values of the angle, denoted now as  $q(\theta,t)$ has initial condition $q(\theta,0)=\delta(\theta)$. At later times and before the next jump, this conditional probability (the condition being  the event ''there is a jump at $t=0$ ''), follows an equation of motion different of (\ref{eq:p}) because there is no gain term:  this gain term represents the occurrence of quantum jumps, absent if one looks at what happens in a time interval without quantum jumps. 
The equation for $q$ is 
\begin{equation}
{\partial_{t}{q}} + \frac{\Omega}{2} {\partial_{\theta}} q =   - \gamma q(\theta, t)  \sin^2(\theta)
\textrm{.}
\label{eq:qp}
\end{equation}
 The  conditional  probability of the angle  $\theta$,  $q(\theta,t)$,  that we are looking for in this section,  
 is different from $p$
 because there is no gain term in (\ref{eq:qp}).
 The $L^1$-norm of $q(\theta,t)$ depends on time and decays to zero because it is a conditional probability.
 More precisely  during the time interval $(0,t)$ between two successive jumps,  the norm  $\mathcal{N}(t)=\int {q(\theta',t)  \mathrm{d}{\theta'}} $  is expected to vanish for each time $t= k\pi/(\Omega/2)$ in the course of the Rabi oscillations, and its envelope should decrease to zero  in the course of time. Let us also emphasize that we have to consider possibly time intervals longer than a Rabi period because the quantum jumps may be separated by several Rabi cycles, particularly if $\gamma$ is small with respect to $\Omega$. 

With the dimensionless  time $\tilde{t}=\frac{\Omega}{2} t $, the equation for $q(\theta,\tilde{t})$ becomes,
\begin{equation}
{\partial_{\tilde{t}}{q}} + {\partial_{\theta}} q =  - \gamma' q(\theta, \tilde{t})  \sin^2(\theta)
\textrm{,}
\label{eq:q}
\end{equation}
where $\gamma'$  is $\gamma$ divided by  $\Omega/2$, as above. 
Equation(\ref{eq:q}) is a particular case of equation(\ref{eq:JG1}), which is solved in equation(\ref{eq:JG3}). It is the case where $g(.)$ is set to zero. The general solution gives in the present case
 with the initial condition $q(\theta, \tilde{t} = 0) = \delta(\sin\theta)$ , is
\begin{equation}
q({\theta},\tilde{t}) = \alpha(\theta, \tilde{t}) \delta(\sin(\theta - \tilde{t}))
\textrm{,}
\label{eq:qsol}
\end{equation}
where 
\begin{equation}
\alpha(\theta, \tilde{t})) = e^{\left(-\frac{\gamma' \tilde{t}}{2} - \frac{\gamma'}{4}(\sin 2 (\theta - \tilde{t})) -  \sin 2 (\theta))\right)}
\textrm{,}
\label{eq:qsola}
\end{equation}

A first result concerns  the norm of the conditional probability define above, we have
\begin{equation}
\mathcal{N}(\tilde{t}) =e^{-\frac{\gamma'}{4}(2\tilde{ t}  -\sin 2\tilde{t})} 
\textrm{,}
\label{eq:qn}
\end{equation} 
which  displays periodic damped oscillations as expected because the atom has no chance to emit a photon at time $\tilde{t}_{i}=k\pi$, and it has more chance to emit a photon during a time interval of order $\gamma^{-1}$ than much later. Note that $\mathcal{N}(\tilde{t}) = \alpha(t, t) $.
A main result is that  in the interval $[0,\tilde{t}]$ the mean value  of  any function $F(\theta)$, being  an average of $F(\theta)$  weighted by the conditional probability  $q(\theta, \tilde{t}))$ , is such that $ <F(\theta)> = F(\tilde{t})$ because of the initial condition for $q(.)$ includes a Dirac delta function.

As an example,  the probability of transition toward the ground state (or quantum jump) during a small interval of  time $[\tilde{t},t+d\tilde{t}]$ which is equal to  $K( \tilde{t})d\tilde{t} = \gamma' <\sin^{2}(\theta) > $ becomes
\begin{equation}
 K(\tilde{t}) =  \gamma' \alpha(\tilde{t})  \sin^2 \tilde{t} 
 \textrm{,}
 \label{eq:qsol2}
\end{equation}
 where the probability of no jump in the same interval is $(1-K(\tilde{t}) \mathrm{d}\tilde{t}$ because  in small time intervals there is only one or zero quantum jumps (condition $(ii)$ above). 
The function  $K$ in (\ref{eq:qsol2})  was derived as the probability  for the jump to occur at time $\tilde{t}$, conditionally of one jump at time zero and no jump in between because the probability $q(\theta, \tilde{t})$ is derived within the hypothesis of  ''no jump'' and the same initial condition. Finally the probability of no jump in the interval $(t_{0},t)$  if there in one jump at time $t_{0}$ is given by the usual expression for a non stationary Poisson process of time dependent parameter  $\int_{t_{0}}^{t}  dt' f(t')$,
\begin{equation}
P^{(no)}_{t_{0}}(t_{0},t)=\alpha(t - t_{0})
\textrm{.}
\label{eq:nojump0}
\end{equation}

Returning to the unscaled variables and using the relations $K(\tilde{t}) d\tilde{t}= K(t) dt$ 
we get the expression for the probability density of a delay $t$  between two successive jumps
\begin{equation}
K(t) = \gamma  \sin^2 \left( \frac{\Omega}{2} t\right) \ e^{- \frac{\gamma}{2}( t -  \frac {\sin \Omega t} {\Omega} )}
\textrm{.}
\label{eq:newK}
\end{equation}

\begin{figure}
\centerline{
(a) \includegraphics[height=1.5in]{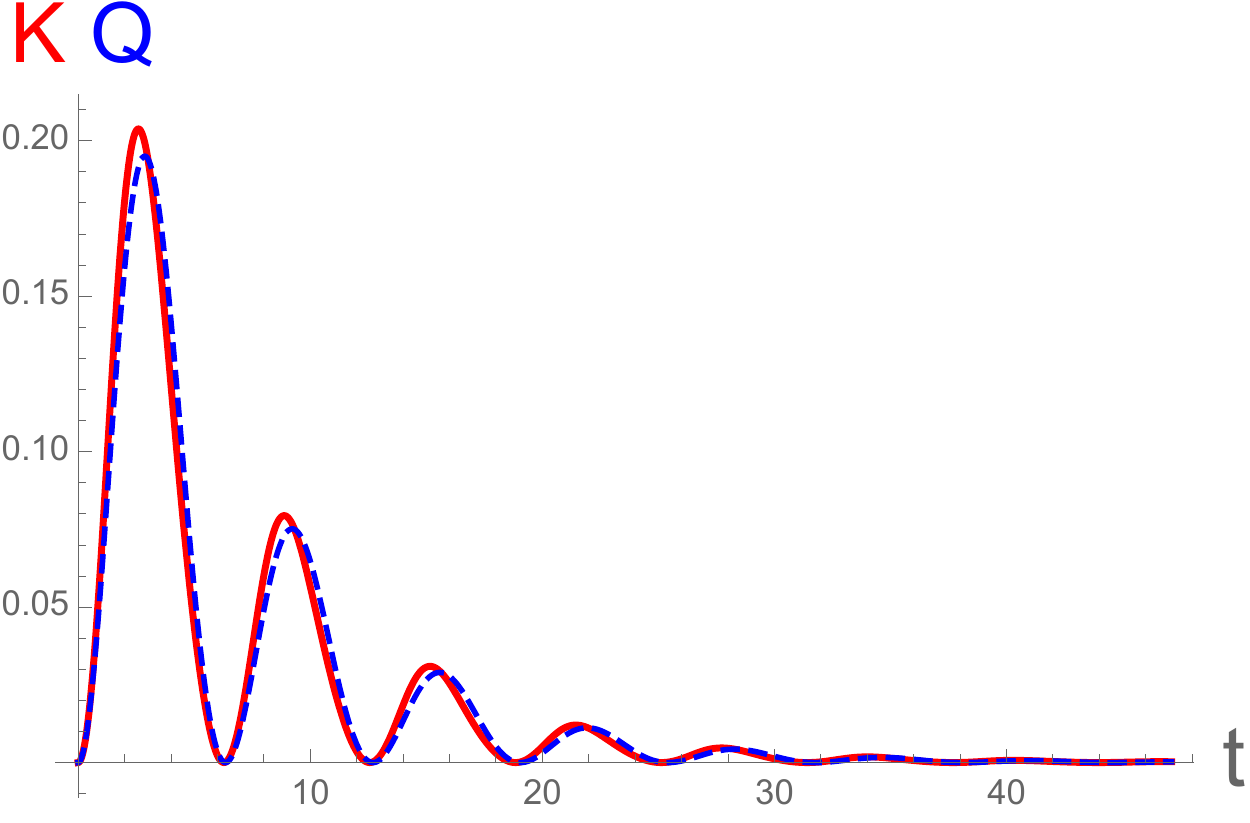}  
(b)\includegraphics[height=1.5in]{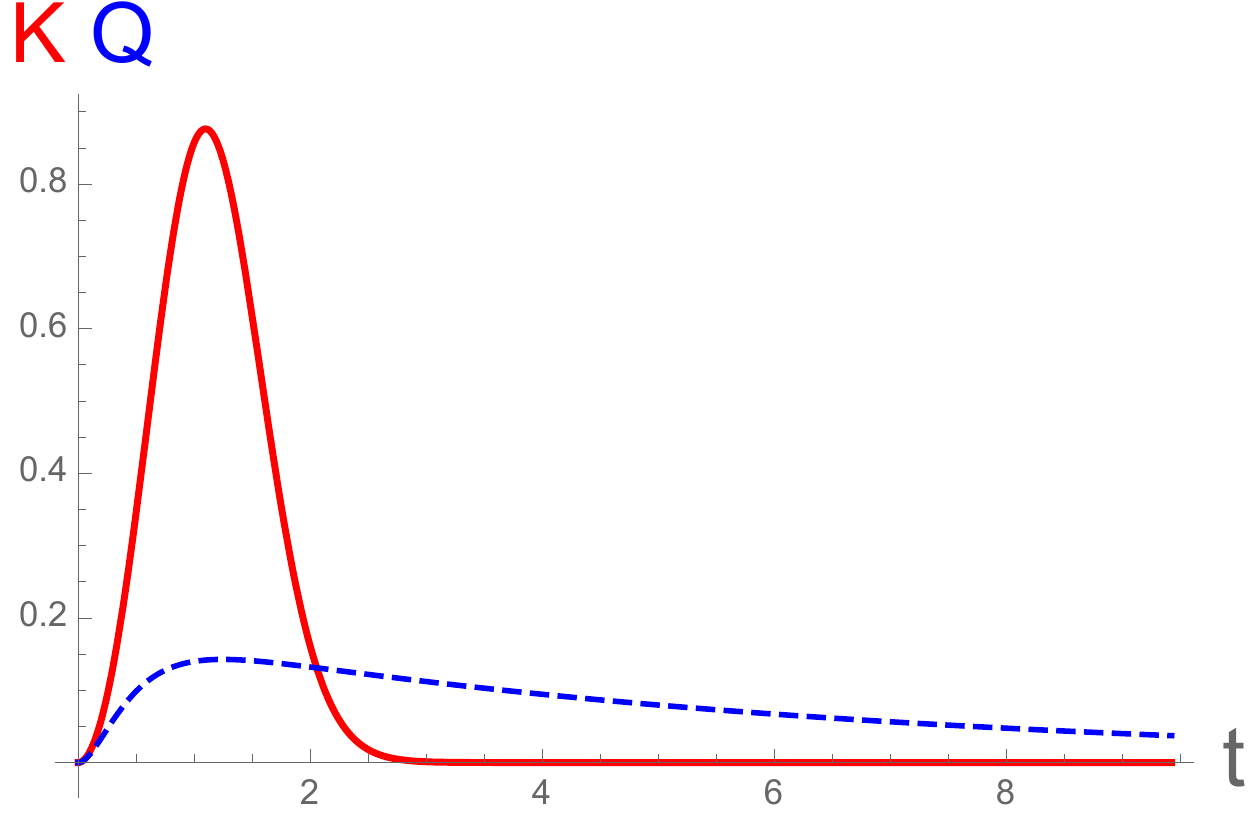}
  }  
\caption{Between two quantum jumps: the probability distribution  $K(t)$ of the delay between two successive quantum jumps, equation (\ref{eq:qsol2})  in solid red line  (a) fort weak damping $\gamma=0.2$, (b) for strong damping $\gamma=6$, both with  $\Omega=1$. The blue dashed lines are  the ''delay functions''  $Q(t)$ of reference \cite{cohen} given here with our notations in equations (\ref{eq:cohen2})-(\ref{eq:cohen}).  
 }
\label{fig:Kol2}
\end{figure}

The expression (\ref{eq:newK}) is of the form $ K(t)= K'(t) exp(-\int_{0}^{t} K'(t'))dt')$ which is the probability of  the first arrival time  for a non stationary Poisson process characterized by the  time dependent intensity 
\begin{equation}
 K'(\tilde{t}) =  \gamma \sin^2 \left( \frac{\Omega}{2} \tilde{t} \right)
 \textrm{,}
 \label{eq:qsol2.1}
\end{equation}
The probability distribution $K(\tilde{t})$ of the delay $\tilde{t}$ between two successive jumps  is drawn in Fig.\ref{fig:Kol2} for weak (a) and strong damping (b)  (resp. strong (a) and weak (b) excitation). As expected, weak damping is associated to long delays on average, or rare quantum jumps, the atom making several Rabi oscillations in between two successive jumps. On the contrary, for strong damping, or large  coupling with the reservoir (or large flux of photons by fluorescence), the delay between two successive jumps is short on average, and the atom has practically no chance to complete a full Rabi nutation in-between two jumps. In summary,  if $\Omega >> \gamma$,  the average delay is of order  $1/\gamma$, with many oscillations at the Rabi frequency during this time.

Note that  the probability distribution  $K(t)$ shows "anti-bunching" of the jumps at different times: the function $K(t)$ vanishes periodically whenever $\Omega t = k \pi $, $k$ positive integer, when the Rabi oscillations bring the atom back to the ground state. This anti-bunching is explained by the property that, starting from the ground state, it takes some time to build-up the amplitude of the excited state by interaction with the pump field.  
Notice that our distribution is "automatically" normalized because integrating over $t$  gives  $\int_{0}^{\infty} \mathrm{d}{t'} K(t')= [ exp^{-\int_{0} ^{t} dt' K'(t')}]_{0}^{\infty}=1$. 

Another limit of interest is when the rate of spontaneous decay is far bigger than the period of the Rabi oscillations, namely the limit $ \Omega << \gamma $.  In this limit one must consider $ \Omega t $ as small in the exponential of equation(\ref{eq:newK}). This yields $$\frac{\gamma}{2}( t -  \frac {\sin \Omega t} {\Omega} ) \approx \frac{\gamma \Omega^2 }{12} t^3 \textrm{,}$$
and 
  \begin{equation}
K(t)  \approx  \gamma \frac{\Omega^2 t^2}{4} e^{-  \frac{\gamma \Omega^2 }{12} t^3}
\textrm{.}
\label{eq:newKas}
\end{equation}
 In the limit $ \Omega << \gamma $ this makes a fair approximation for the bulk of the probability distribution of the intervals between two emitted photons. It shows that this mean time interval is of order 
  \begin{equation}
 \tau_{K}= (\Omega^2 \gamma)^{-1/3}
 \textrm{.}
\label{eq:tauK}
\end{equation}
 It is between the long time $1/ \Omega$ (period of Rabi oscillations)  and the short one $1/  \gamma $ (life time of the excited state). It shows in particular that between two jumps, on average, a complete Rabi oscillation cannot take place because it takes too long.  This time is also longer than $1/ \gamma$, this illustrates the  anti-bunching effect which prevents to have very close quantum jumps, because it takes some time after a jump for the amplitude of the excited state to grow enough until an emitted photon brings back the atom to the ground state.

Let us compare the probability $K(t)$ derived from our statistical analysis  with the  '' delay function'' of Reynaud et al. \cite{cohen}  derived by representing the spontaneous decay by the addition of a (deterministic) linear damping term in the equations for the coupled amplitudes $a_0$ and $a_1$,  omitting the compensating accretive term of the Lindblad equation.
The delay function $Q(t)$ of ref.\cite{cohen} (noted $W(\tau)$ in this reference, with the same meaning as our $K(t)$) is given by the expression, 
\begin{equation}
Q(t)  = \gamma (\frac{\Omega}{\lambda})^{2} \left( \sin(\frac{ \lambda}{2} t)\right)^{2} e^{-\frac{\gamma}{2} t}
\textrm{,}
\label{eq:cohen}
\end{equation}
where $\lambda^{2}=\Omega^{2}-\gamma^{2}/4$.
 In the limit of small damping ($\gamma \ll \Omega$) this becomes
\begin{equation}
Q(t)  = \gamma  \left(\sin(\frac{ \Omega}{2} \tau) \right)^{2} e^{-\frac{\gamma}{2}\tau}
\textrm{.}
\label{eq:cohen2}
\end{equation} 
This delay function in the weak damping limit 
 is close but not identical to our $K(t)$ distribution. The difference is small for strong excitation (or small $\gamma$) as illustrated in our Fig.(\ref{fig:Kol2})-(a)  where $Q(t)$ is plotted in blue dashed line.  
In summary  the derivation of reference \cite{cohen} 
and our theory yield similar but not identical results in the case of  strong excitation (or  small damping).

Let us look at the other limit, namely the  limit of strong damping. In this limit the term $ \sin(\frac{ \lambda}{2} t)$ in equation(\ref{eq:cohen2}) becomes $ \sinh(\frac{ \lambda}{2} \tau)$, that gives
 \begin{equation}
Q(t)  = \gamma (\frac{\Omega}{\lambda})^{2}  \left( \sinh(\frac{ \lambda}{2} t) \right)^{2} e^{-\frac{\gamma}{2} t}
\textrm{.}
\label{eq:cohen}
\end{equation}
 Expanding $\lambda $ in powers of the small parameter $\Omega/\gamma$ shows that $Q(t)$ decreases to zero  asymptotically  over the very long time scale 
 \begin{equation}
  \tau_{Q}=\frac{\gamma }{ \Omega^{2}}
  \textrm{,}
\label{eq:tauQ}
\end{equation}
 a time scale much longer than the inverse of $\gamma$, which is a short time, but also much longer than the period $ 2\pi \Omega^{-1}$ of the Rabi oscillations. This is hard to understand because in the limit of a very strong damping, namely with a short life time of the excited state of order $\gamma^{-1}$, one expects that many photons will be emitted during the (long) period of a Rabi oscillation, if such an oscillation takes place (which is very unlikely). One expects that the delay time decreases when the damping rate increases, although $\tau_{Q}$ increases linearly with $\gamma$.
In this limit of strong damping our delay function  $K(t)$ decreases much more rapidly than $Q(t)$, see Fig.\ref{fig:Kol2}-(b) where the two functions $K(t)$ and $Q(t)$ are plotted for a ratio $\gamma/\Omega=6$.  Our delay function $K(t)$ decreases with the time scale $\tau_{K}$ given in (\ref{eq:tauK}), much shorter than $1/\Omega$, a result opposite to (\ref{eq:tauQ}). Our result is closer to the physical intuition than the one of the theory of reference \cite{cohen} because this strong damping case is associated to strong spontaneous emission, and so to short delays between consecutive quantum jumps. It could be that this limit $\gamma/ \Omega$ very large can hardly be dealt with by models inspired by the Bloch theory for spins, because in this limit the dynamics of the two level system becomes completely dominated by the damping term. 
 Therefore, in this theory, the emission of photons of fluorescence seems to bring back the system to the excited state via the Rabi nutation. This looks somewhat contradictory with the physics:  the emission of photons is a way for the system to leave the excited state and not to reach it.  

We emphasize that the non-stationary character of the Poisson process (made by the set of jumps) is responsible for a true memory effect illustrated by the oscillations of the delay distribution $K(t)$, that is not the case for the standard stationary Poisson process  with constant intensity $\rho$ which has the familiar delay distribution  of the form $K(t)= \rho \exp(-\rho t)$, decreasing monotonically with time.  Nevertheless  because the time intervals between the jumps  are independent from one to the next, the series of time intervals is formally time reversible: it has exactly the same statistical properties when looked at forward and backward in time.

 \subsection{Quantum properties derived from the probability distribution}
 \label{sec:quantum proba}
 
 Below we explain how to get the quantum properties of the fluorescence out of the theory based on Kolmogorov equation. 
  \subsubsection{Quantum state }

First let us describe the quantum state of the two-level atom in the usual sense of quantum mechanics. By quantum state we mean the density matrix $\rho_{ij} (t)$ where $i$ and $j$ have either the value $1$ or $0$, which refer to the ground state (index $0$) or the excited state (index $1$). This two-by-two matrix has two positive diagonal entries, $\rho_{00}$ and  $\rho_{11}$ whose sum is $1$. From the way $p(\theta, t)$ is defined the diagonal entries are just averages of circular functions of $\theta$. The relation is:
 \begin{equation}
\rho_{00} (t) =  \int_{-\pi/2}^{\pi/2} \mathrm{d}{\theta}\ p(\theta, t)  \cos^2(\theta)
\textrm{,}
\label{eq:rho00}
\end{equation}
and
  \begin{equation}
\rho_{11} (t) =  \int_{-\pi/2}^{\pi/2} \mathrm{d}{\theta}\ p(\theta, t)  \sin^2(\theta)
\textrm{,}
\label{eq:rho11}
\end{equation}
Because $p(\theta, t) $ is normalized to one, the trace condition is satisfied. 
The estimate of the off-diagonal entry $\rho_{01} (t)$  implies to return to the original coefficients $c_{0,1}$ introduced in the definition of the wave function, equation (\ref{eq:wavef}). Because the amplitudes $a_0=c_{0}$ and $a_1= c_{1}e^{i\omega_{L}t}$  are derived by getting rid of the time dependence of the pump field, the equations for $c_{0,1}$ are deduced from the coupled equations (\ref{eq:dot0}) and (\ref{eq:dot1})   by multiplying $\Omega$  by the phase factor  $e^{i\omega_{L} t}$. Physically the time-periodic pump field restores continuously the phase relation between state $0$ and $1$ of the two-level system. Thanks to that restoring, the time average of the off-diagonal entries of the density matrix are not zero. In our formalism  this off-diagonal entry is  the average of the product $\cos(\theta)\sin(\theta)$ times the phase factor $e^{\pm i\omega_{L} t}$, the $\pm$ symbol being for  the choice of one of the two off-diagonal components:
  \begin{equation}
\rho_{01} (t) =  \rho_{10}^* (t) = e^{i\omega_{L} t} \int_{-\pi/2}^{\pi/2} \mathrm{d}{\theta}\ p(\theta, t)  \sin(\theta)  \cos(\theta)
\textrm{,}
\label{eq:rho01}
\end{equation}

In next subsection we derive an expression for the spectrum of the emitted light. 

   \subsubsection{Fluctuations of the fluorescence light}
  \label{sec:Fluctlight}  
   
 Experiments give access to the fluorescence light of a single atom or ion and to the statistical properties of this light.  In the case of  a single pumped atom (or ion)  and in the limit $\gamma$ small compared to  $\Omega$ the spectrum of this light has two side peaks at frequencies $\omega_{L} \pm \Omega$,  shifted from a central peak  which is at the frequency of transition (together with the central peak this makes the Mollow triplet). At larger damping this spectrum looses the two side peaks structure  and get the more standard shape of a broad central peak. Below we explain how Kolmogorov approach gives access to statistical properties of the emitted light, including its spectrum and also non trivial correlations exhibiting the loss of time reversal symmetry in the fluctuations of the two-level system. This loss of symmetry is tightly related to the irreversibility of the spontaneous emission process and ultimately of the measurement of the quantum state of the atom, that is linked to the so-called reduction of the wave packet. 

 The electric field emitted by an atom is represented by a quantum operator. This operator acts on quantum states of the whole system, atom and field. We are interested in the correlations of this field at different times. The results of measurement are represented by expressions which are quadratic with respect to the operators (like the one for the electric field) and quadratic with respect to the wave function of the electrons. In the present case, this is the wave-function of the excited state, of amplitude $c_1$. Therefore the radiated electric field is proportional to $c_1$, or  $a_{1}$, an assumption which can be understood as a quantum version of the familiar Hertz formula for the radiation of an oscillating dipole: the radiated electric field is proportional to the strength of the emitting dipole which is proportional to the amplitude $a_1$ because this emission will show-up in a perturbative approach and will give the electric field at the linear order in the amplitude $a_1$.  
 
The  correlation function of the  electric field emitted by the atom at two different times $t_1$ and $t_2$ is proportional to the product of two amplitudes $c_1$ and $c_1^*$ respectively at time $t_1 + t_{D}$ and $t_2 + t_{D}$ where $t_{D}$ is the delay needed for the wave to propagate from the atom to the measuring instrument. Therefore    
 the frequency spectrum $S(\omega)$ of the emitted light is the Fourier transform with respect to the variable $ \tau$ of the function 
 \begin{equation}
  \mathcal{C}( \tau) = e^{-i\omega_{L} \tau } <(a_1(t + \tau) a_1^*(t)> 
\textrm{,}
\label{eq:fluct1}
\end{equation} 
 where the average is taken on all possible histories described by Kolmogorov equation.  Because the amplitude $a_1$ is measured at different times, one has to take into account a possible phase difference between the state at time $t$ and time $(t +  \tau)$, a phase difference irrelevant when estimating single time averages as we did until now. 

\subsubsection{Phase of the wave function}
\label{sec:random phase}
\begin{figure}
\centerline{
 \includegraphics[height=1.5in]{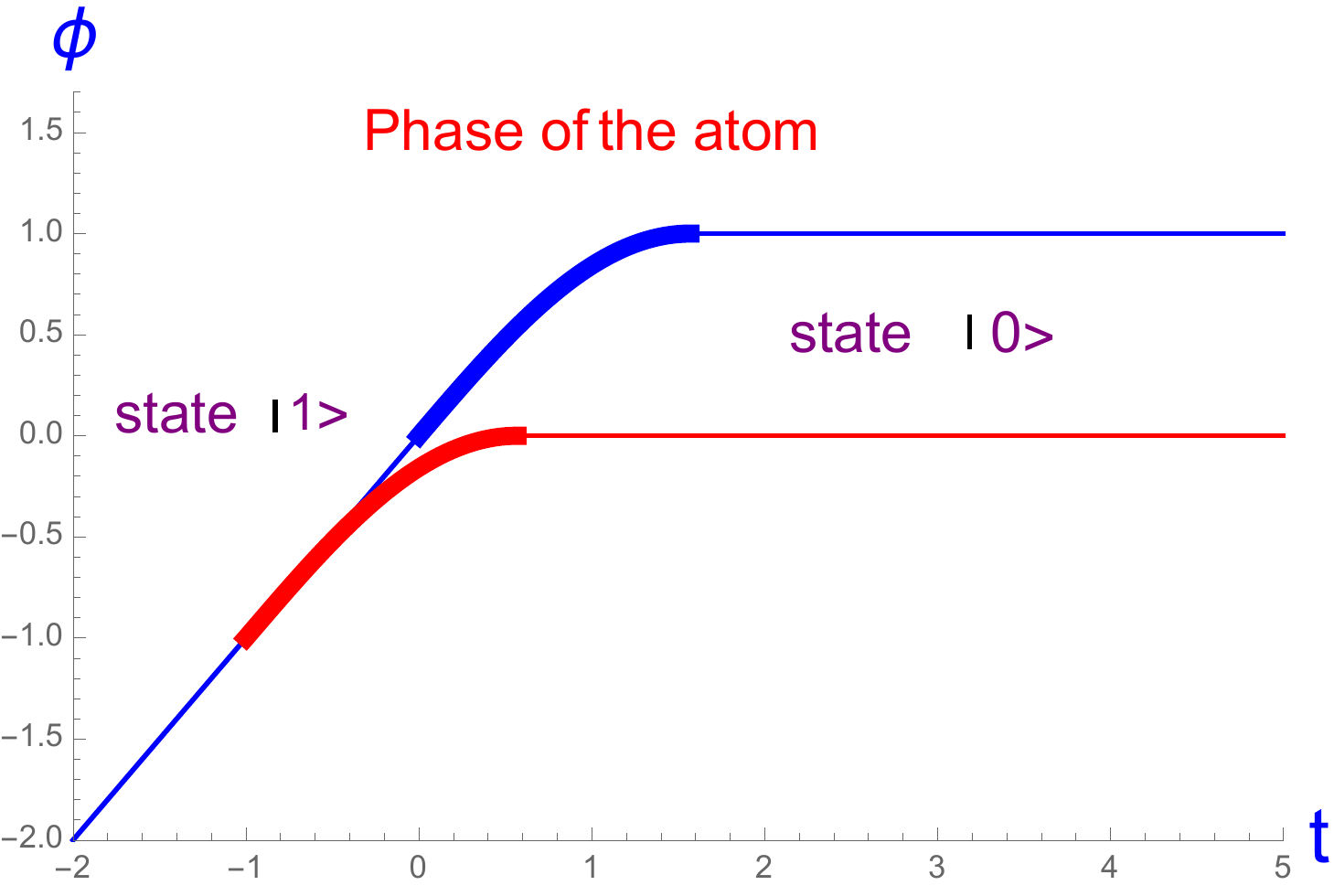}  
  }  
\caption{ Phase change of the atomic state  versus time close to a jump in arbitrary units. The phase increases linearly with time before the jump when the atom is in the state $\vert1>$, the slope being proportional to the energy of the excited state minus the one of the ground state. This slope becomes constant after it, because the atom has decayed to the ground state $\vert0>$. The time of the jump, a quick event, takes place within a window of width $1/\gamma$ much longer than the duration of the jump itself. The two curves mimic the behavior of the phase close to two different jumps sppused to take place at times $t_{1}\sim -1$ (solid line) and  $t_{2}\sim 0$ (dashed line). The widening of the bent part of the curves is to recall that, at the transition, the energy of the atom is not well defined because the atom has left an energy eigenstate and is therefore subject to quantum fluctuations of its energy.} 
\label{fig:phase}
\end{figure}

Let $\phi (t)$ be the phase of the wave-function of state $1$. Recall that we take the light beam exactly at resonance and let $E=E_{1}-E_{0}$. Just before the emission of a photon, at the quantum jump labelled $j$, the time derivative of this  phase is given by the Planck-Einstein relation $$\partial_{t}\phi (t) = E/\hbar = \omega_{L} \textrm{.}$$ 
 As time approaches $t_{j}$,  the emission time  of the photon of energy $E$, the energy of the excited state which is proportional to the time derivative of the phase, decreases abruptly from $E$ to zero.  Therefore  the phase $\phi (t)$ changes from a phase linear with respect to time to a constant as illustrated in Fig.\ref{fig:phase} which shows schematically  a zoom of the evolution of the phase of the wave-function of the atom  during the small time interval  around  quantum jumps. Let us precise that the conservation of energy imposes that the photon gets energy $E$  from the excited state, so that at any time $t$ the sum of the energy of the photon and of the excited state remain constant and equal to $E$ (this neglects the effect of the energy input from the pump field, which makes the system non stationary and so a priori not at constant energy, this yields negligible effects during the jump because of the weakness of the interaction with the pump compared to $E$). Because the instant $t_{j}$ is  random, it takes place in a window of time of width $1/\gamma$. It follows that the  change of behavior of the phase also takes place in this window of time. Therefore, after the transition $j$, assumed to behave similarly  for all jumps, 
the final  phase varies at random  within a very large range of order $ \delta \phi  =E (t_{j}-t_{j-1})/(\hbar)= \omega_{L}(t_{j}-t_{j-1})$, where $(t_{j}-t_{j-1})$ is the  duration of the deterministic evolution before the event $j$ .
 In Fig.\ref{fig:phase}  the final phase of the atomic state (this is a well defined quantity because one can aways measure the energy of this atomic state, which is the time derivative of this phase)  is the ordinate of the horizontal portion of the curve, a random variable which depends on the random time where the jump occurs, see captions. 
 It remains constant during the time intervals of deterministic dynamics and jumps randomly at the time of quantum jumps. This randomness is a consequence of the randomness of the time of jump. After such a jump, the oscillations of the wave functions start again but at a random time at the time scale of period of the laser field, which changes very rapidly with respect to the amplitudes $a_0$ and $a_1$.  Notice also that during the jump the atom  is not  in an energy eigenstate, so that its energy shows quantum fluctuations. Therefore we did widen the trajectory of the phase as a function of time near the bent of this curve.  

In summary, the equations of deterministic motion which are in the form of equations(\ref{eq:dot0}) and (\ref{eq:dot1}) are derived by getting rid of the rapid phase factor $e^{-i\omega_{L} t}$, but the phase change necessary to do this operation induces a  random phase  difference between the two amplitudes $a_0$ and $a_1$ after the jump.
 In principle this random phase should have been already included into the probability distribution as one of its argument, making this probability a function $p(\theta, \phi, t)$ instead of a function of  $\theta$ and $t$ only, as was done before. This imposes also to make the transition probabilities $\Gamma$ dependent on $(\theta, \phi \; ;\theta',\phi')$. Because of the randomness  of $ \phi $ after a jump, the dependence of the transition probability $\Gamma$ with respect to $ \phi'$ is just a constant function of this phase defined on the interval $(0, 2\pi)$. 

The Kolmogorov equation for the two-level case including the phase angle $ \phi $ is derived from equation(\ref{eq:p}) which becomes 
\begin{equation}
\partial_{t}{p} + \frac{\Omega}{2} p_{\theta} = \gamma \left(  \delta(\theta)  \left( \frac{1}{2\pi}\int_{0 }^{2\pi} \mathrm{d}{\phi'} \int_{-\pi/2 }^{\pi/2} \mathrm{d}{\theta'}\ p(\theta', \phi',t)  \sin^2(\theta') \right) - p(\theta, \phi ,t)  \sin^2(\theta) \right)
\textrm{.}
\label{eq:p-phi}
\end{equation}
By integrating both sides of this equation with respect to $\phi$ one recovers Kolmogorov equation (\ref{eq:p}) for the function  $\frac{1}{2\pi}\int_{0 }^{2\pi} \mathrm{d}{\phi} \ p(\theta, \phi ,t) $.  This allows to calculate the integrated gain term in equation(\ref{eq:p-phi}). Once this is done, one can solve this equation for each value of $ \phi $ which becomes a parameter, because changing it does not change the equation at all. Therefore this equation can be solved, at least in principle, for arbitrary initial conditions by using the general method of section  \ref{sec:exactsol}.

 The explicit calculation of the two-time correlation function $ \mathcal{C}( \tau)$  is done by using equation(\ref{eq:dotprob1}). We take
  the initial condition
  \begin{equation}
  p (\theta, \phi, 0) = \delta(\sin(\theta - \theta_0)) \delta(\phi - \phi_0) 
 \textrm{,}
\label{eq:fluct2}
\end{equation} 
  where  $(\theta_0, \phi_0) $ are arbitrary values of $(\theta, \phi)$,  and introduce $p (\theta, \phi, t ) = p(\theta, \phi, t \vert  \theta_{0}, \phi_{0}, 0)  p (\theta_{0}, \phi_{0}, 0)$ in the integrand of equation(\ref{eq:p-phi}). In the following we write the conditional probability as $p_{\delta} (\theta, \phi, t )=p(\theta, \phi, t \vert \theta_{0}, \phi_{0}, 0) $, in order to lighten the notation, having in mind that $p_{\delta} (\theta, \phi, t )$ depends on  $(\theta, \phi)$ and on $ (\theta_0, \phi_0)$.  
  The important point is the following. Because the initial condition (\ref{eq:fluct2}) for  $p_{\delta} (\theta, \phi, t)$ is factorized into a function of $\theta$ and a function of $\phi$, the solution of equation(\ref{eq:p-phi}) remains factorized at later times because this equation allows separation of variables.  This makes the solution fairly simple. 

\subsubsection{Correlation function and spectrum}
\label{sec:corell-spectrum}
  
   The correlation function defining the spectrum, as given in equation(\ref{eq:fluct1}), is given by the expression
 \begin{equation}
 \mathcal{C}( \tau) =  \frac{e^{-i\omega_{L} \tau }}{4\pi^2}\int_{0 }^{2\pi} \mathrm{d}{\phi_0} \int_{-\pi/2}^{\pi/2} \mathrm{d}{\theta_0}  \ p_{st} (\theta_0, \phi_0)  \sin(\theta_0)\int_{0 }^{2\pi} \mathrm{d}{\phi} \int_{-\pi/2}^{\pi/2} \mathrm{d}{\theta}  \ p_{\delta} (\theta,  \phi, \tau )\sin(\theta) e^{i(\phi_0 - \phi)}
\textrm{.}
\label{eq:fluct3}
\end{equation} 
Taking into account the factorization of the probability  $p(\theta, \phi, t)=p_{\theta} (\theta, t) p_{\phi} (\phi, t)$ equation (\ref{eq:fluct3}) becomes
 \begin{equation}
 \mathcal{C}( \tau) =  e^{-i\omega_{L} \tau } <e^{i(\phi_0 - \phi)}> \mathcal{C}_{1}(\tau)
\textrm{,}
\label{eq:fluct4}
\end{equation} 
where $T$ is any interval of length $\pi$, the period of the functions $f(\theta)$ and $p_{\delta}(\theta,t)$ at any time, and $ \mathcal{C}_{1}(\tau)$ is the double integral,
 \begin{equation}
 \mathcal{C}_{1}(\tau)=  \int_{T} \mathrm{d}{\theta_0}  \ p_{st} (\theta_0)  \sin(\theta_0)  \int_{T} \mathrm{d}{\theta}  \ p_{\delta} (\theta, \tau )\sin(\theta) 
\textrm{.}
\label{eq:fluct4b}
\end{equation}

The  average value of the phase difference in equation (\ref{eq:fluct4}) is easy to derive. We have
\begin{equation}
<e^{i(\phi_0 - \phi)}>= P^{(no)}(0, \tau)
\textrm{,}
\label{eq:fluct5}
\end{equation} 
 where $P^{(no)}(0, \tau)$ is the probability that there is no jump in the time interval $(0,\tau)$. The relation (\ref{eq:fluct5})  follows from the fact that  $\phi_0 = \phi$  if there is no jump in the time interval $(0,t)$,  and   $<e^{i(\phi_0 - \phi)}>= <e^{i\phi_0 }><e^{-i \phi} >=0$ otherwise, because the distribution of the phase is uniform in $(0,\pi)$.  The probability  of no jump in $(t, t+\tau)$ is linked to $P_{t_{0}}^{(no)}$  derived in section \ref{sec:between}  but is not identical to it, because the latter was a conditional probability, with the condition that a jump exist at time zero. The probability 
 of no jump in the interval $(t,t+\tau)$ is given by the integral
\begin{equation}
P^{(no)}(t, t+\tau)= \int_{-\infty}^{t}  \mathrm{d}{u} P_{u}^{(no)}(u,t+\tau)  K(u)
\textrm{,}
\label{eq:nojump}
\end{equation} 
where $K(u) \mathrm{d}u$ is the probability of one jump in the time interval $(u,u+ \mathrm{d}u)$. Introducing the relations (\ref{eq:nojump0}) and (\ref{eq:qsol2}), we obtain
\begin{equation}
P^{(no )}(t, t+\tau)= \int_{-\infty}^{t}  \mathrm{d}{t'} \alpha(t+\tau-t')  \alpha(t') f(t')
\textrm{,}
\label{eq:nojump2}
\end{equation} 
or
\begin{equation}
P^{(no)}(0, \tau)= e^{-\frac{\gamma' \tau}{2}}  \int_{0}^{\infty}  \mathrm{d}{t'}  f(t') e^{-\frac{\gamma' }{4}(\sin(2\tau+2t')-\sin(2t'))} 
\textrm{,}
\label{eq:nojump3}
\end{equation}

\begin{figure}
\centerline{
(a)\includegraphics[height=1.5in]{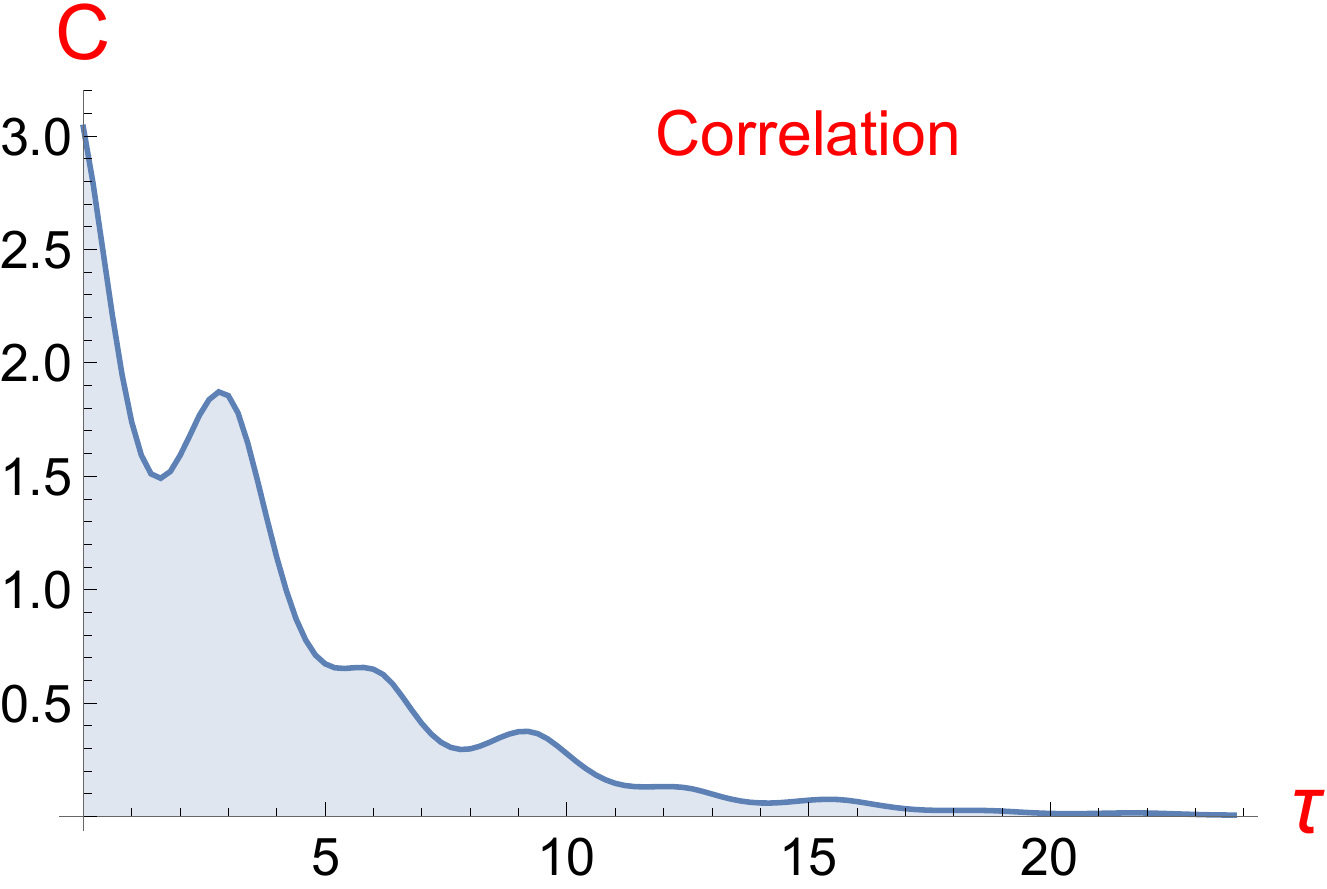}  
(b)\includegraphics[height=1.5in]{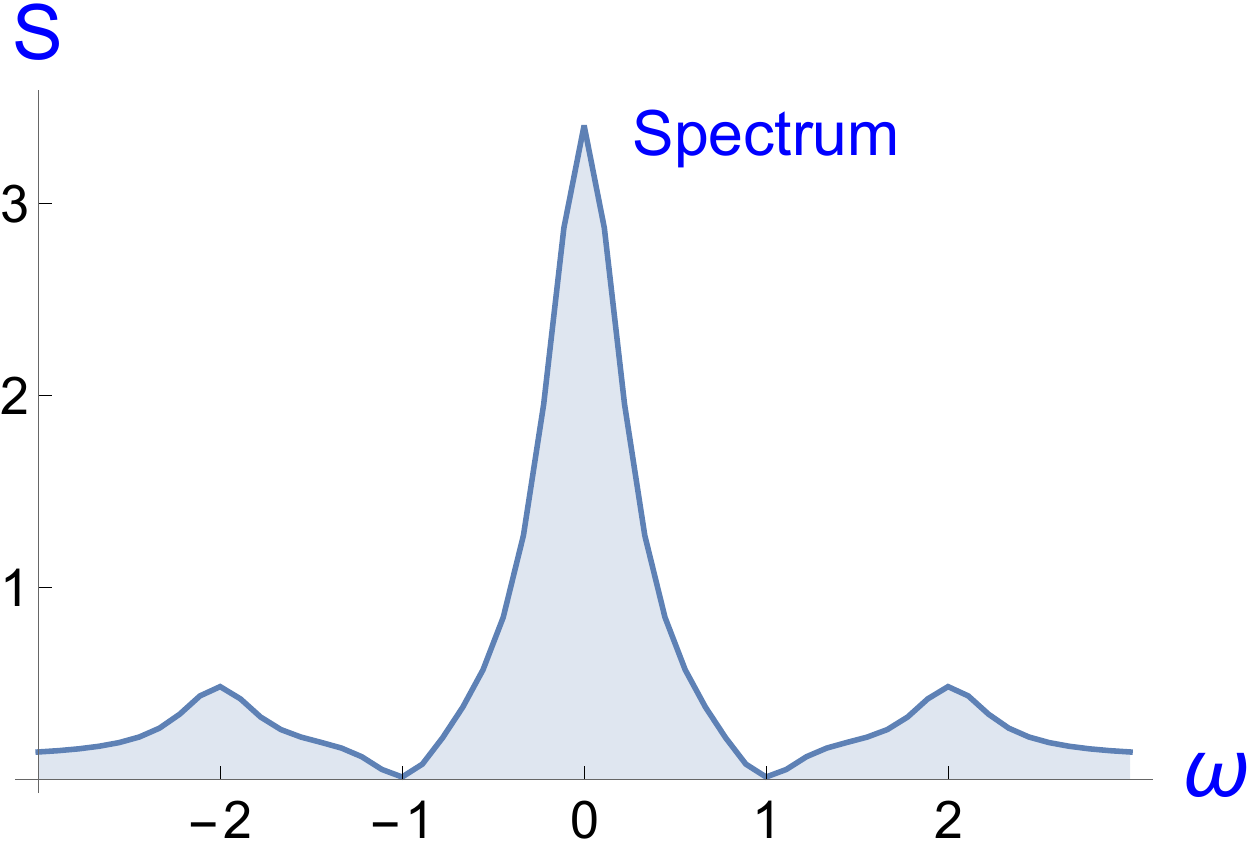}
 }  
\caption{ Correlation function  and spectrum for small damping, $\gamma'=1/7.8$. (a) $C(\tau) -C(\infty)$ defined in equation (\ref{eq:fluct4}), (b) $S(\omega)$  defined in (\ref{eq:spect}), centered on the laser frequency $\omega_{L}$.   Time is in units of $ \frac{2}{\Omega}$ and frequency in units of $\Omega/2$.
}
\label{fig:correll}
\end{figure}

 The double integral (\ref{eq:fluct4b}) is easily calculated after the integral equation for $b(t,\theta_{0})$ has been solved.  
In view of the next subsection, we  give  the expression of the double integral for more general cases involving the correlation between two different functions $F(\theta) $ and $G(\theta)$ at different times, whereas we had $F(\theta)=G(\theta)= \sin(\theta)$ in the case of the auto-correlation function $C(\tau)$. Let us define the intercorrelation function
 \begin{equation}
 \mathcal{C}_{FG}(\tau)=  \int_{T} \mathrm{d}{\theta_0}  \ p_{st} (\theta_0)  F(\theta_0)  \int_{T} \mathrm{d}{\theta}  \ p_{\delta} (\theta, \tau )G(\theta) 
\textrm{.}
\label{eq:fluct4FG}
\end{equation} 
Defining $ \mathcal{B}_{G}(\theta_{0} ,\tau) =  \int_{T} \mathrm{d}{\theta}  \ p_{\delta} (\theta,\tau) G(\theta) $, we get

 \begin{equation}
 \mathcal{C}_{FG}( \tau) =  \int_{T} \mathrm{d}{\theta_0}  \ p_{st} (\theta_0) F(\theta_0)  \mathcal{B}_{G}(\theta_{0},\tau) 
\textrm{,}
\label{eq:cor1}
\end{equation} 
where $ \mathcal{B}_{G}( \theta_{0}, \tau) $ takes a form similar to (\ref{eq:b(t)}) but with $b(t)$ already known,
 \begin{equation}
 \mathcal{B}_{G}( \theta_{0}, \tau) =  \mathcal{M}_{G}( \theta_{0}, \tau)  + \int_{0}^{t} \mathrm{d}{t'} \mathcal{L}_{G}(t') b(\theta_{0},t-t')
 \textrm{,}
\label{eq:BG}
\end{equation}  
where
  \begin{equation}
 \mathcal{M}_{G}( \theta_{0}, \tau) =   \int_{T} \mathrm{d}{\theta}   \alpha(\theta,t) G(\theta) p(\theta-t,0)
 \textrm{,}
\label{eq:MG}
\end{equation}  
and 
 \begin{equation}
 \mathcal{L}_{G}( t) = G(t)  \alpha(t)
 \textrm{.}
\label{eq:LG}
\end{equation}   
The self-correlation  $ \mathcal{C}(\tau)$  defined in (\ref{eq:fluct4b}) is given by  equations (\ref{eq:fluct5}) and (\ref{eq:cor1})-(\ref{eq:LG}), with $F=G=\sin\theta$.
We show in Fig.\ref{fig:correll}-(a) the numerical result for a case of small damping. The curve $C(\tau)$ displays several oscillations before getting  the stationary value $C(\infty)$.

The spectrum of the field emitted  by the atom is given by the expression
 \begin{equation}
S(\omega) =2 \Re{\left(\int _{0}^{\infty} \mathrm{d}{\tau} e^{i\omega \tau}  C(\tau)\right)}
 \textrm{,}
\label{eq:spect}
\end{equation}  
where $\Re{(f)}$ means real part of $f$.
As shown in Fig.\ref{fig:correll}-(b), the spectrum is centered on the laser frequency $\omega_{L}$. It displays two side peaks at frequencies $\omega_{L} \pm \Omega$ which are broader than the central peak and have smaller intensities.  These characteristics also occurs  with the Bloch equations description. More precisely the Bloch equations lead to  sidebands of width larger than the one of the central peak by a factor $3/2$, and to heights three times smaller in the limit of large ratio $\Omega/\gamma$. By comparison Kolmogorov description  gives a spectrum in qualitative agreement with the quantum Bloch equations description, the width of the three peaks are similar, but the height of the sidebands is smaller in Kolmogorov theory.

\begin{figure}
\centerline{
(a) \includegraphics[height=1.5in]{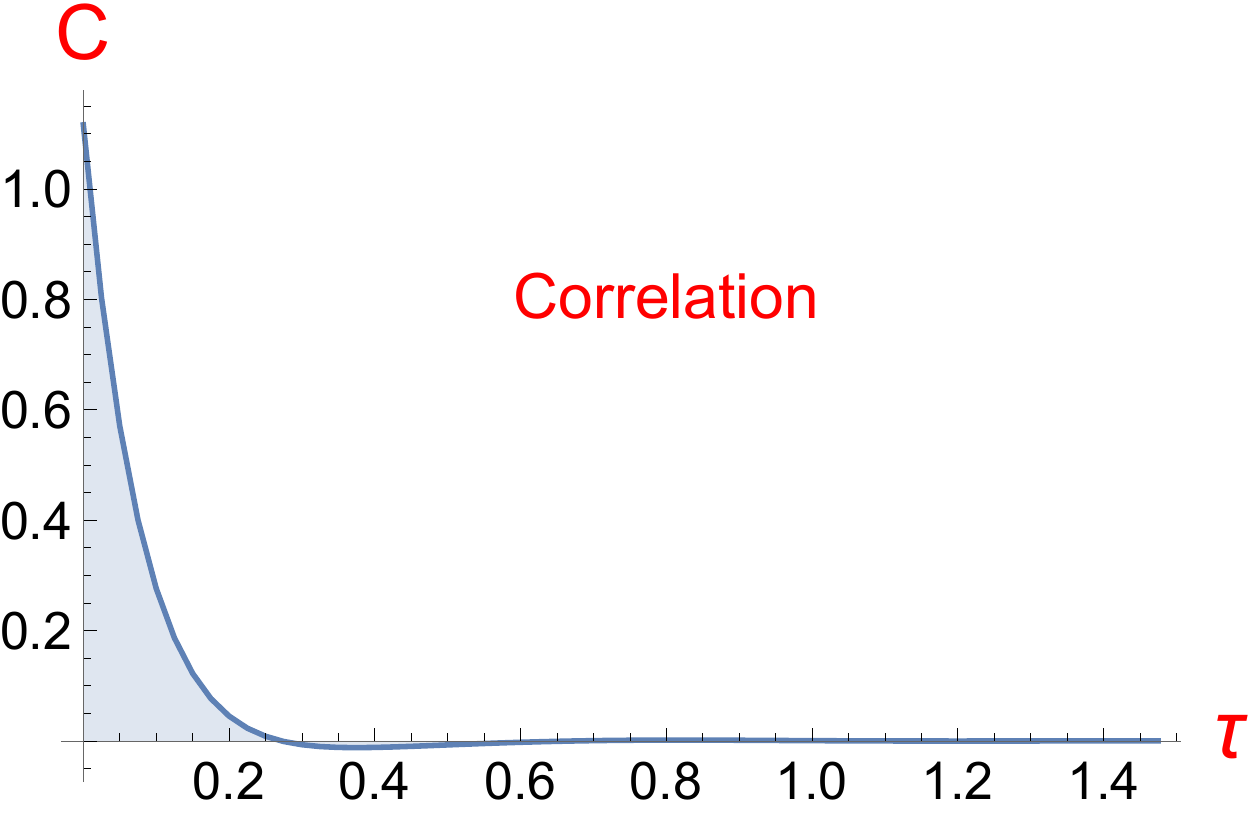}  
(b)\includegraphics[height=1.5in]{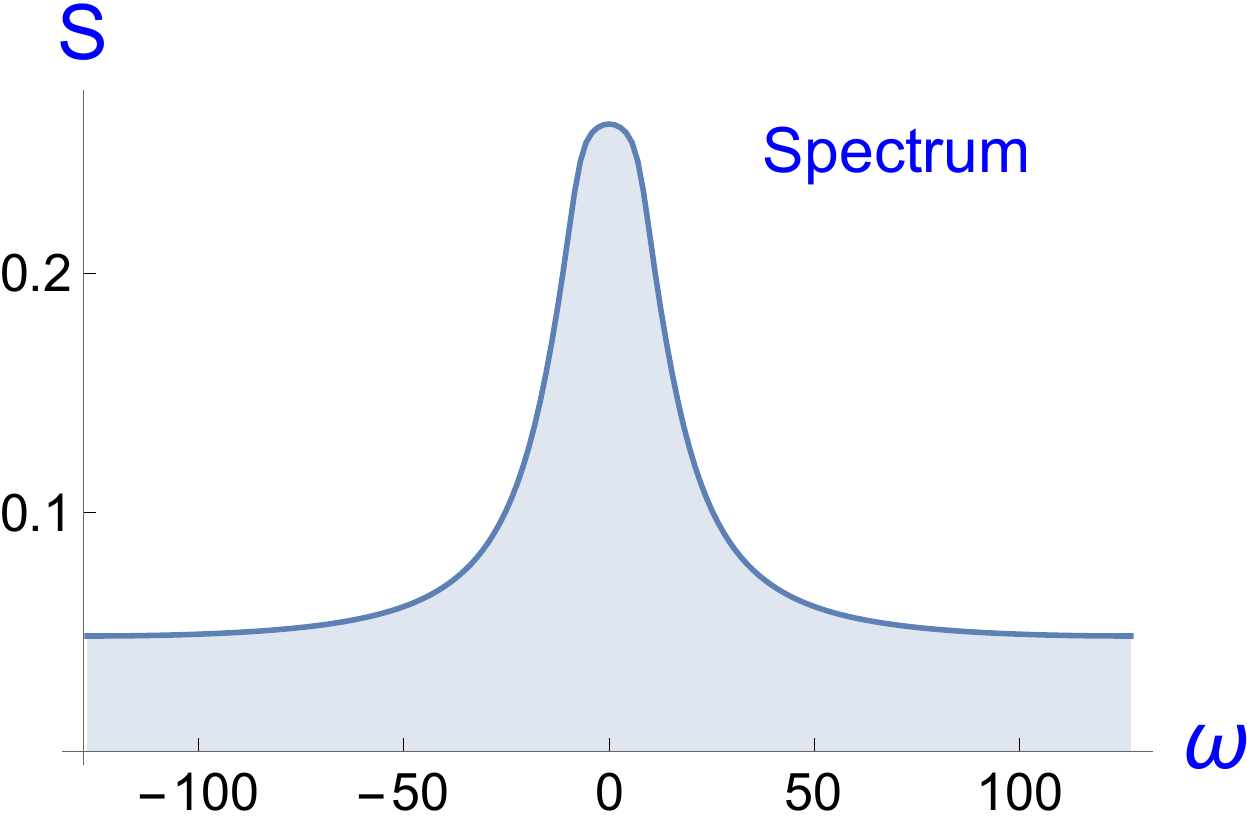}
  }  
\caption{ Correlation function  and spectrum for large damping, $\gamma'=10$. (a) $C(\tau) -C(\infty)$;  (b) $S(\omega)$  Same units as in Fig.\ref{fig:correll}.
}
\label{fig:correl2}
\end{figure}
For large damping, the correlation function is narrow in time and the spectrum displays a single peak centered on the laser frequency, as illustrated in Fig.\ref{fig:correl2}.

   \subsection{Test function(s) of irreversibility by using fluctuations of the fluorescence light}
  \label{sec.irr}
  
In mathematical literature, irreversibility and reversibility are defined by reference to the equations of motion. For instance Newtonian mechanics is well-known to be reversible because, by inverting velocities and keeping the same positions, the trajectories of a set of interacting particles will trace back exactly their history. Seemingly this property was already known to Newton himself: he spared computing work by calculating trajectories of a mass around a center by inverting the speed  at the apex to get the next part of the orbit. This definition of reversibility does not help much in real life because it cannot be used "practically". Consider for instance fluctuations in a turbulent fluid. One cannot reverse the speeds of all molecules at some time to check if the turbulent flow is in a state of reversible dynamics or not. Therefore another definition of reversibility should be used to have measurable consequences. This was done by one of us in  \cite{ref.1}. There the idea was introduced that by analyzing some time correlation functions one can decide if a fluctuating signal is invariant or not under time reversal. 

This breaking of time-reversal invariance of a time dependent random signal can be measured in many different ways. The idea is to compare time dependent correlations which are different if one reverses the direction of time. This excludes pair correlations of the same observable, like $<A(t) A(t + \tau)>$, a function of $\tau$ that is invariant under the exchange of $\tau$ and $-\tau$, for a random signal stationary on average.  Indeed this is because the same observable $A$ has been taken for the measurement  at time $t$ and time $(t +\tau)$. When picking-up {\emph{different}} functions of the fluctuations at time $t$ and  $(t + \tau)$, this symmetry under the exchange of $\tau$ and $-\tau$ is not guaranteed anymore in general and becomes a property shared or not by the system under consideration. Various examples of breaking of this invariance under time reversal are given in ref.\cite{ref.1}. As pointed out there, generally speaking this invariance is absent  in out-of-equilibrium systems  like a model of steady shear flow or turbulent flows which are then irreversible.

Suppose a signal $x(t)$ fluctuating in the course of time, $x$ real quantity. We shall assume that this signal is statistically stationary, so that averages like $C_{FG}(t_{1},t_{2})= <F(x(t_{1}) G(x(t_2))>$  depend only on the difference $(t_1- t_2)$, $F$ and $G$ being smooth functions of $x$. Practically such averages are given by time integrals like 
$$  <F(x(t)) G(x(t +\tau))> =  \lim_{T \rightarrow \infty } \frac{1}{T}\int_0^{T} \mathrm{d}t   \ F(x(t)) G(x(\tau +t)) \textrm{.} $$
As shown in  \cite{ref.1} one may extract from such a signal what was called test functions for irreversibility. This is done by subtracting from  $<F(x(t)) G(x(t +\tau))>$ the function found  by exchanging $F$ for $G$, which is equivalent to reverse the sense of time. Such a test function is 

\begin{equation}
 \Psi_{1,2}(\tau)  = < F(x(t)) G(x(t +\tau)) - G(x(t)) F(x(t +\tau))>
\textrm{.}
\label{eq:function.3t.1}
\end{equation}
  
If such a function is not zero for $F$ different of $G$ and $\tau$ different of $0$, the signal $x(t)$ is not time-reversible. Various theoretical examples of this asymmetry are given in  \cite{ref.1}, and it was also recalled that equilibrium fluctuations have the very special property of time-reversal symmetry, as had been shown by Onsager.  Another function for testing the symmetry of fluctuations under time reversal can be 
\begin{equation}
 \Psi^1(\tau) = < x(t) (x(t+2\tau) - x(t+\tau))x(t + 3\tau)>
 \textrm{,}
\label{eq:function.3t}
\end{equation}
which involves multi-time correlations of the same function  $x(t)$,  instead of  the two-time correlation functions with different $F$ and $G$ defined in (\ref{eq:function.3t.1}). 
 By reversing time, the correlation $ \Psi^1(\tau) $ changes sign. 
 Therefore it is a test-function for irreversibility. This is an odd function of $\tau$ and its Taylor expansion near $\tau = 0$ begins with a cubic term.  
 
 \begin{figure}
\centerline{
(a) \includegraphics[height=1.5in]{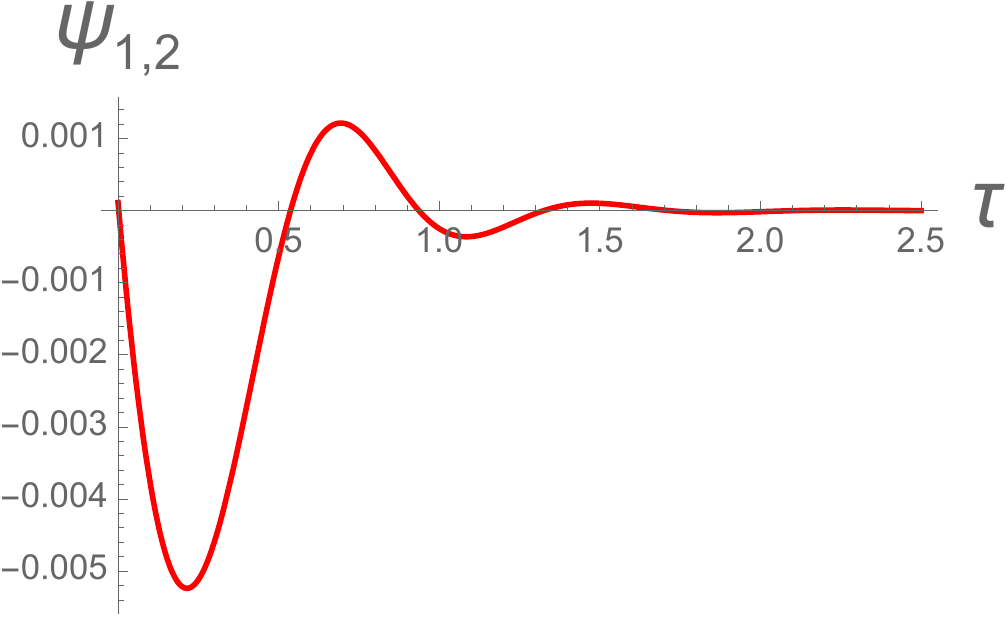}
(b)   \includegraphics[height=1.5in]{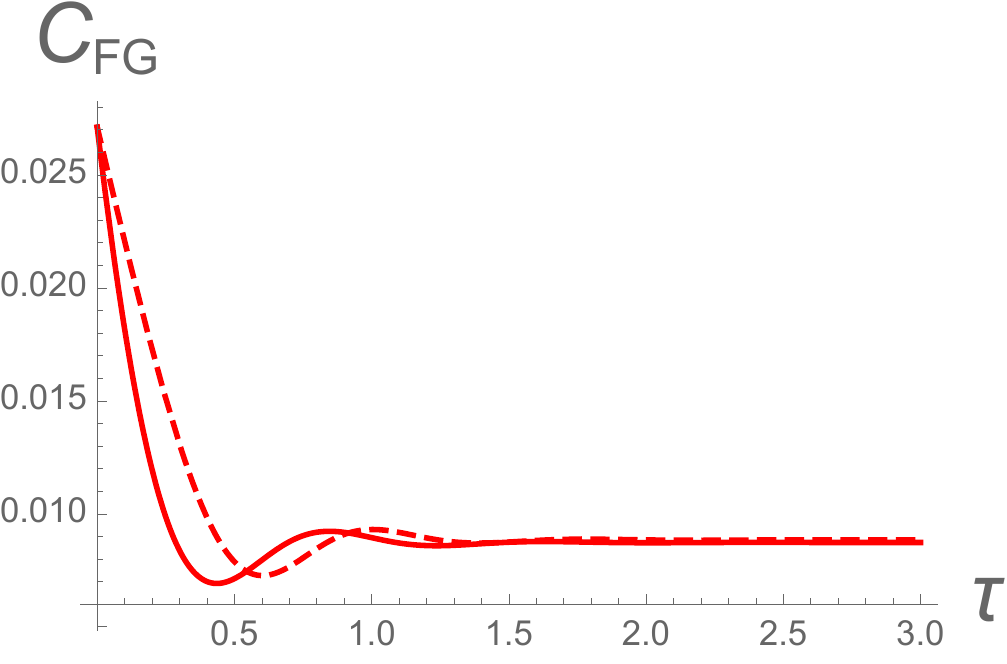}
}
\centerline{
 \includegraphics[height=1.5in]{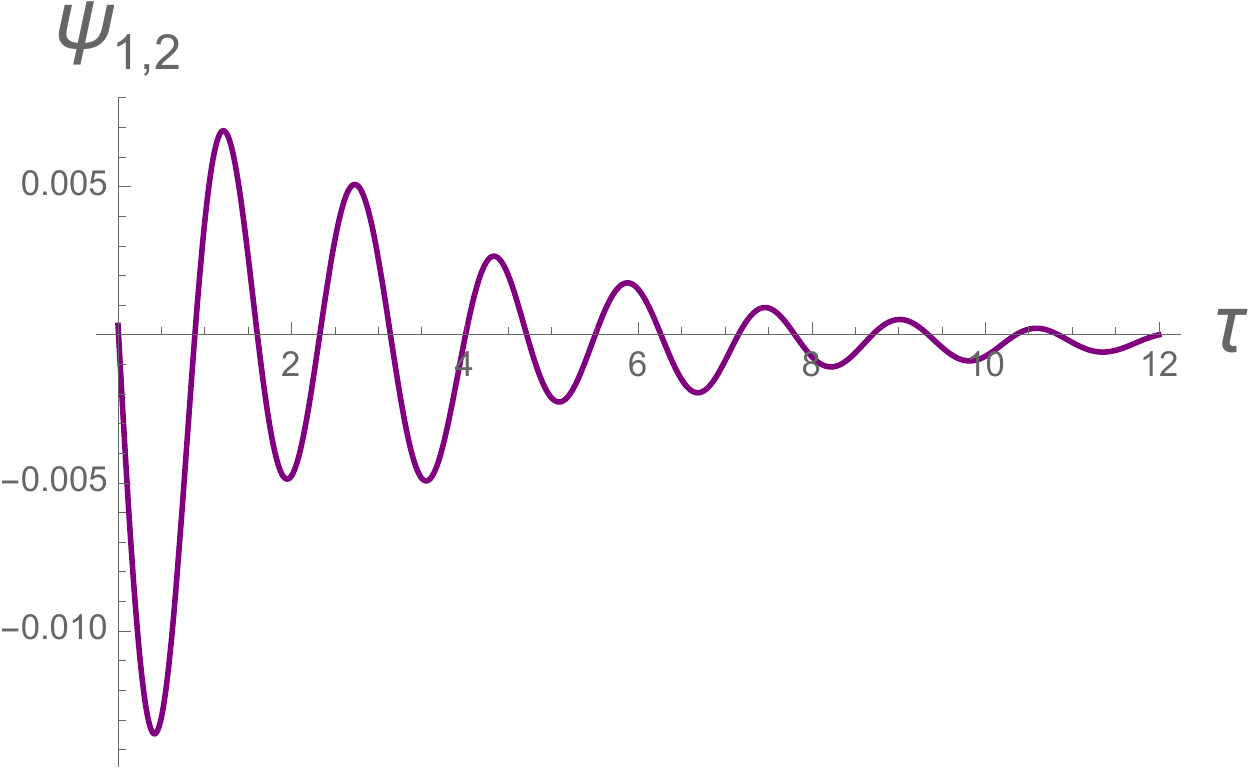}  
  }  
\caption{ Test function $\Psi_{1,2}(\tau)$ defined in equation (\ref{eq:function.3t.2}). (a) for  $\gamma'=10$ and (c) for  $\gamma'=0.6$.  For strong damping the  irreversible character  is clearly visible  on the  curves $C_{FG}=< I(t) I^{2}(t +\tau)>$ (solid line) and $C_{GF}=< I^{2}(t) I(t +\tau)>$ (dashed line) shown in (b) which are well separated.
}
\label{fig:test}
\end{figure}

 In the case of the fluorescence emitted by a two-level system illuminated by a resonant laser beam, one may use various functions of the (fluctuating) intensity emitted by the atom. We shall choose one of the simplest, namely the one where $F = I^2 (t)$ is the square of the emitted intensity at time $t$ and $G = I(t)$ is this intensity itself.  Therefore we shall consider the following correlation:
 \begin{equation}
 \Psi_{1,2}(\tau)  = < I(t) I^{2}(t +\tau) -  I^{2}(t) I(t +\tau) >
\textrm{.}
\label{eq:function.3t.2}
\end{equation}
Up to a constant multiplying factor, the intensity $I(t)$ is given by $\sin^2(\theta)$ in our representation of the state  of the two-level atom (or ion).  Using the general expression for two-time correlations given in equation(\ref{eq:dotprob1}) one obtains 
\begin{equation}
 \Psi_{1,2}(\tau)  =  \int_{-\pi/2}^{\pi/2} \mathrm{d}{\theta_0}  \ p_{st} (\theta_0)  \int_{-\pi/2}^{\pi/2} \mathrm{d}{\theta}  \ p_{\delta} (\theta, \tau ) \left( \sin^2(\theta_0)\sin^4(\theta) - \sin^4(\theta_0)\sin^2(\theta)\right)
\textrm{.}
\label{eq:function.3t.3}
\end{equation}
Notice that the phase angle $\phi$ does not appear in the expression above which involves product of intensities at a given time,  insensitive to phase differences of the quantum states. In equation (\ref{eq:function.3t.3}) the functions  $\ p_{st} (\theta_0)$ and  $p_{\delta} (\theta, \tau )$ have the same definition as  before, $p_{\delta} (\theta, \tau )$ is the solution of equation (\ref{eq:p}) with $\tau$ as time,  the initial condition is $p_{\delta} (\theta, \tau = 0 ) =  \delta(\sin(\theta - \theta_0)) $ and $\ p_{st} (\theta_0)$ is the steady solution of Kolmogorov equation computed in section \ref{sec:steadysol}. 

Numerically one finds a clear proof  of the irreversible character of the two-level atom fluorescence. The test function is shown in figs.\ref{fig:test} (a) and (c) for large and small damping rates respectiveley. In these figures the  difference $\Psi_{1,2}=C_{FG}-C_{GF}$ displays oscillations which have noticeable amplitudes in a time interval of order of the  correlation time (the width of the correlation function $\mathcal{C}(\tau)$ of the fluorescence field).  For small damping the  difference $\Psi_{1,2}$ is small (it is of order one per cent of the amplitude of  $C_{FG}= < I(t) I^{2}(t +\tau)>$ for the curve (c)). The important result is that the  ratio $\Psi_{1,2}/C_{FG} $ increases  with the damping rate $\gamma$. This ratio becomes about $25$ per cent for $\gamma'=10$, as illustrated in figs (a)-(b). This proves that \textit{ quantum jumps are responsible for the irreversible character of the spontaneous emission}.

 \section{Kolmogorov equation for a three-level atom.} 
 \label{3levelJG}

\begin{figure}
\centerline{
\includegraphics[height=1.5in]{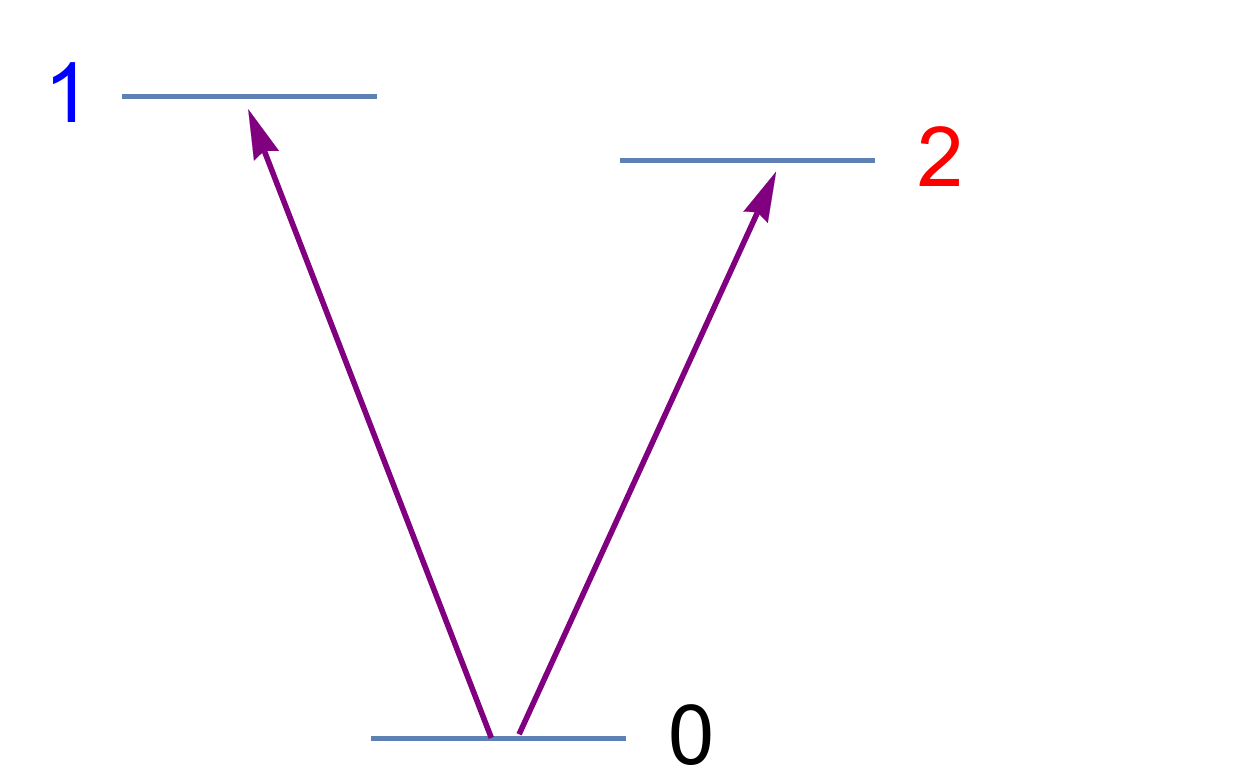}
  }  
\caption{ Scheme of the three level  atom in V-configuration.
}
\label{fig:scheme}
\end{figure}

Let us consider a three-level system illuminated by two laser beams, each one at the frequency of transition between the ground state and one of the two excited states, the configuration imagined by Dehmelt \cite{ref.2}, named V-configuration.  In the absence of  spontaneous decay of the excited states, the equation of motion connects three complex amplitudes, $a_0 (t) $ for the ground and $a_1(t), a_2(t)$ for the two excited states, see Fig.\ref{fig:scheme}. Those equations are\cite{ref.3}

\begin{equation}
\partial_{t}{a}_0 = - i \omega_{1}  a_1 - i \omega_{2}  a_2
\textrm{,}
\label{eq:dot0b}
\end{equation} 
\begin{equation}
\partial_{t}{a}_1 = - i \omega_{1} a_0
 \textrm{,}
\label{eq:dot1b}
\end{equation}
and 
\begin{equation}
\partial_{t}{a}_2 = - i \omega_{2} a_0
 \textrm{.}
\label{eq:dot2b}
\end{equation}
In those equations, $2\omega_{1}$ and $2\omega_{2}$ are Rabi frequencies associated to the transitions from level $1$ to zero (index $1$) and from level $2$ to zero (index $2$), each one being proportional to the amplitude of an electromagnetic wave at the frequency of the atomic transition between level $1$ or $2$ and the ground state. 
In the following we shall assume, as in Dehmelt's proposal, that the transition $\vert 0> \to \vert 1>$ is saturated, contrary to the transition $\vert 0> \to \vert 2>$,  and that the level $2$ has a very long lifetime (such that spontaneous emission  from level  $1$ is frequent while it is very rare from level $2$, and such that  stimulated emission  from $2$ is still more rare), namely we shall consider the situation where the parameters fulfill the conditions,
\begin{equation}
\omega_{1} \gg  \gamma_{1} \gg \gamma_{2} \gg \omega_{2}
 \textrm{.}
\label{eq:dehmelt-cond}
\end{equation}
Within this frame we shall try to describe the intermittent fluorescence observed in the experiments \cite{nagourney} where clear period of darkness were  appearing in the fluorescent signal.

\subsection{Solution in the deterministic regime}
\label{amplitude sol}
Setting
\begin{equation}
A(t)=
 \left( \begin{array}{ccc}
a_{0} \\
a_{1} \\
a_{2}
\end{array}
\right)
\textrm{,}
\label{eq:vect3}
\end{equation}
equations (\ref{eq:dot0b})-(\ref{eq:dot2b}) can be written as

\begin{equation}
i \partial_{t}A(t)= 
 \left( \begin{array}{ccc}
0 & \omega_{1} & \omega_{2} \\
\omega_{1}  & 0 & 0 \\
\omega_{1}  & 0 & 0
\end{array}
\right)
A(t)
\textrm{.}
\label{eq:mat3}
\end{equation}
The eigenvalues of the matrix in (\ref{eq:mat3}) are $(0, \pm \omega)$ with $\omega^2= \omega_{1}^2 + \omega_{1}^2 $.  We  introduce the parameter  $\epsilon$ defined by the relations $\cos \epsilon= \omega_{1}/ \omega$  and $\sin \epsilon= \omega_{2}/ \omega$ ($\epsilon$  will be small in the situation considered below).  It is interesting to introduce two new amplitudes which are linear combination of $a_{1}$ and $a_{2}$,

\begin{equation}
 \begin{array}{c}
a_{3}= \cos\epsilon \; a_{1} +  \sin\epsilon \; a_{2} \\
a_{4}= \sin\epsilon \; a_{1} -  \cos\epsilon\;  a_{2}
\end{array}
\textrm{,}
\label{eq:34}
\end{equation}
and fulfill the  condition $\vert a_{1}^{2} \vert + \vert a_{2}^{2} \vert = \vert a_{3}^{2} \vert + \vert a_{4}^{2} \vert $. The system (\ref{eq:mat3}) becomes
\begin{equation}
i\partial_{t}
\left(\begin{array}{c}
 a_{0} \\
 a_{3}\\
a_{4}
\end{array}
\right)
=\omega
\left( \begin{array}{c}
 a_{3} \\
 a_{0} \\
0
\end{array}
\right)
\textrm{.}
\label{eq:sys34}
\end{equation}
In this system the amplitude $a_{4}$ remains constant, and 
the first and second equation in (\ref{eq:sys34})  give $(\partial_{t}^{2} + \omega^{2})a_{0}=0$ and  $(\partial_{t}^{2} + \omega^{2})a_{3}=0$. Therefore the dynamics for  the original three amplitudes (\ref{eq:vect3}) reduce to the evolution of  only two. We have 
\begin{equation}
 \begin{array}{c}
 a_{0}(t) = \cos \omega t \;  a_{0}(0) -i \sin \omega t  \; a_{3}(0)\\
 a_{3}(t) = \cos \omega t \;  a_{3}(0) -i \sin \omega t  \; a_{0}(0)\\
 a_{4}(t)=a_{4}(0).
\end{array}
\label{eq:sys03}
\end{equation}
Note that the amplitude $a_{4}$ enters in the norm condition but plays no role in the dynamics.

Let us return to the original amplitudes and look at their evolution, assuming that at time $t=0$ the atom emits a photon by  the transition $\vert 1> \to \vert 0>$ . 
Because this transition does not concern the state $\vert 2>$, the amplitude $a_{2}$ should be the same just before and just after this quantum jump. Differently the amplitudes $a_{0}$ and $a_{1}$ should change at t=0, in particular we  have to set
\begin{equation}
a_{1}(0_{+})=0
\textrm{,}
\label{eq:a1zero}
\end{equation} 
because the emitted photon comes from the state $\vert 1>$ which is empty at $t=0+$. Using equations (\ref{eq:34}) and  (\ref{eq:sys03}) and setting $a_{i\pm} =a_{i}(0\pm)$,   the  solution in terms of $a_{2+}=a_{2-}=a_{20}$ and $a_{0+}$ is given by the equations
\begin{equation}
a_0 (t)=  \cos \omega t \;  a_{0+}  - i  \sin \epsilon \sin \omega t  \; a_{20}
\textrm{,}
\label{eq:sol0}
\end{equation} 

\begin{equation}
a_1(t)=  \sin \epsilon \cos \epsilon (\cos \omega t-1)\; a_{20}  - i  \cos \epsilon \sin  \omega t \;   a_{0+}
\textrm{,}
\label{eq:sol1}
\end{equation} 
and 
\begin{equation}
a_2(t)=  (\sin^{2} \epsilon\cos \omega t +\cos ^{2}\epsilon) \;a_{20}  - i  \sin \epsilon \sin  \omega t  \;  a_{0+}
\textrm{.}
\label{eq:sol2}
\end{equation} 
The relation (\ref{eq:a1zero}) is satisfied, as well as the norm constraint $\vert a_{0}^{2} \vert + \vert a_{1}^{2} \vert + \vert a_{2}^{2} \vert =1$.
At time $0+$  the two complex amplitudes $a_{0+}$  and $a_{20}$ are linked by a single relation $ \vert a_{0+} \vert ^{2} + \vert a_{20} \vert ^{2} =1$ , that allows to write 
\begin{equation}
\left(
 \begin{array}{c}
 a_{0+} \\
 a_{20}
\end{array}
\right)
=e^{i \phi}
\left(
 \begin{array}{c}
\cos \varphi\\
\sin \varphi e^{i \xi}
\end{array}
\right)
\textrm{.}
\label{eq:ci}
\end{equation} 
Finally we have three unknown quantities, the angular variable $\varphi$ (which describes the amplitudes of levels $\vert 2>$ and $\vert 0>$ after the jump), plus the two phases $\phi$ (in factor) and $\xi$, the  phase difference between the amplitudes $a_{20}$ and $a_{0+}$. Note that  only $\xi$  is useful  if we want to derive the dynamics of squared amplitudes.  Contrary to the two-level atom where we pointed out that the two amplitudes evolve in quadrature, here  we have no reason to assume this and must consider  the relative phase $\xi$ as a random variable (with uniform distribution for instance). Moreover the  phase $\phi$  which appears in factor in the r.h.s. of (\ref{eq:ci}), enters into play for the calculation of the correlation functions of the complex amplitudes,  as for the two-level atom case.

 \subsection{Change of coordinates for the Schr\"{o}dinger flow.}
\label{3level param}
Let us consider the solution (\ref{eq:sys03}) of the pseudo-two-level atom  and try to find the pertinent phase space to describe this deterministic dynamics. We can introduce first the variable 
\begin{equation}
s=\vert a_{4} \vert^{2},
\label{eq:s}
\end{equation}  
which fulfills the relation $ \vert a_{0} \vert^{2}+\vert a_{3} \vert^{2}=1-s$. This relation differs from the  true two-level case where one should have $s=0$. We  look for a description of the dynamics which takes the form of  a rotation on a circle with angular velocity proportional to $\omega$. 
Let us us consider the relation
\begin{equation}
i \partial_{t} (a_{0}^{*}a_{3})= \omega(2\vert a_{0} \vert^{2}-1+s)
\textrm{,}
\label{eq:par1}
\end{equation} 
and define the functions 
\begin{equation}
u=2\vert a_{0} \vert^{2}-1+s
\textrm{,}
\label{eq:u}
\end{equation} 
 and 
\begin{equation} 
 v=2 \Im (a_{0}a_{3}^{*})
 \textrm{.}
\label{eq:v}
\end{equation} 

  They obey the system
\begin{equation}
 \begin{array}{c}
\partial_{t} u =- 2 \omega \; v \\
\partial_{t} v = 2 \omega \; u 
\end{array}
\textrm{,}
\label{eq:duv}
\end{equation}
The  time dependent variables  ($u(t),v(t)$) may be associated to  the  variables ($r, \theta(t)$) defined by
\begin{equation}
 \begin{array}{c}
u =r\; \cos \theta \\
v =r\; \sin \theta
\end{array}
\textrm{,}
\label{eq:ruv}
\end{equation} 
The radius $r=\sqrt{\vert u\vert^{2}+\vert v\vert^{2}}$ is constant during the deterministic motion, and the angular variable $\theta(t)$ obeys the equation
\begin{equation}
\partial_{t}\theta= 2 \omega
\textrm{.}
\label{eq:angvel}
\end{equation}

\begin{figure}
\centerline{
(a) \includegraphics[height=1.5in]{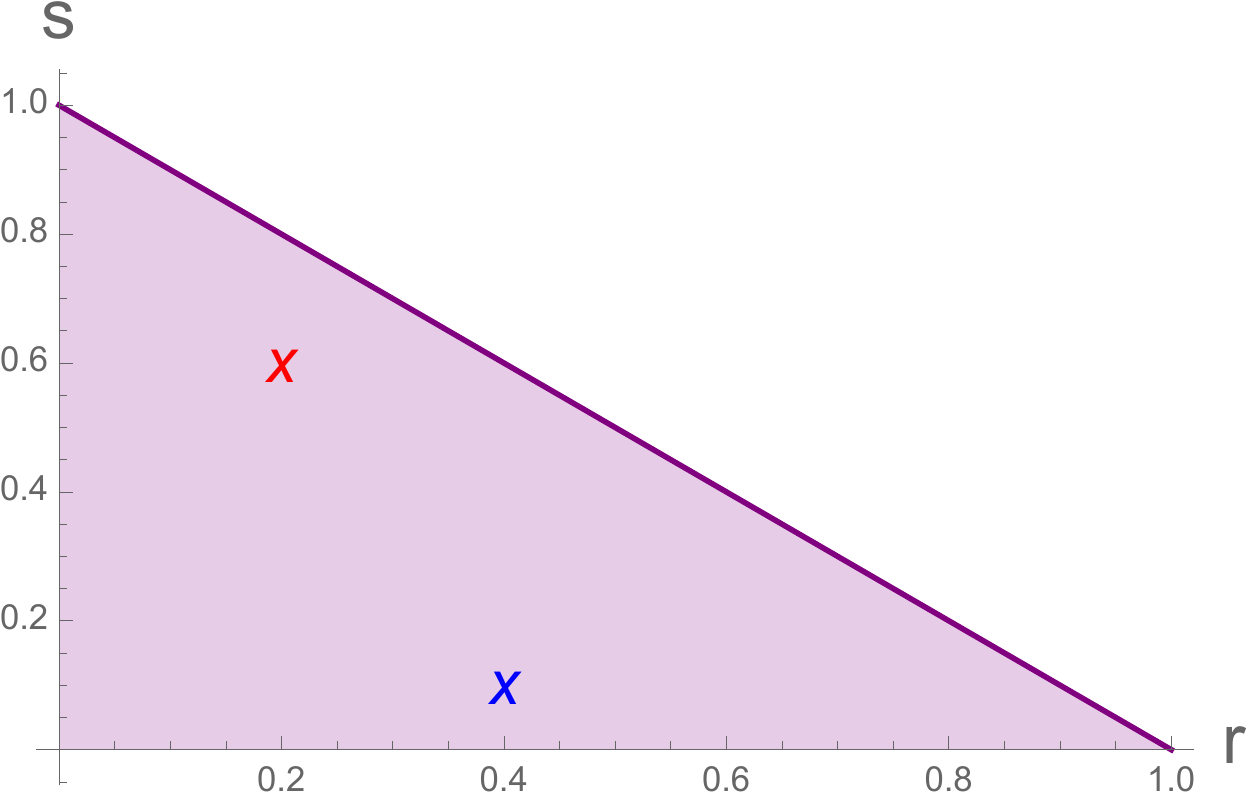}  
(b)\includegraphics[height=1.5in]{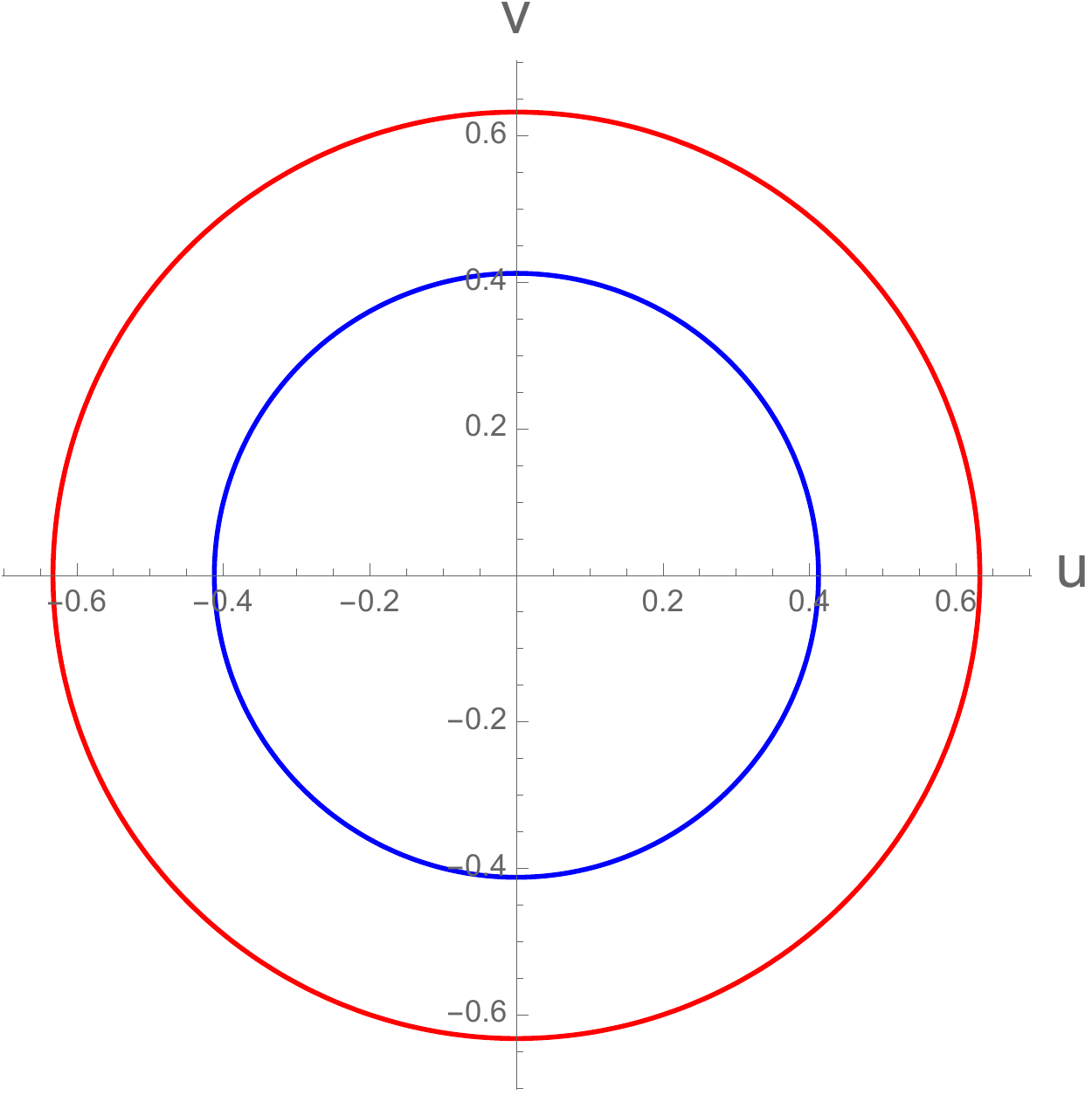}
  }  
\caption{  (a) Domain of the phase space $r,s$  bounded by the relation (\ref{eq:rs}). (b) Trajectory of the coupled $u,v$ or $r,\theta$ functions for the two  different initial conditions corresponding to the red and blue points in (a).
}
\label{fig:frontier}
\end{figure}

Recall that ($u,v$) are defined in terms of the amplitudes ($a_{0}, a_{3}$) by equations (\ref{eq:u})-(\ref{eq:v}).  In this picture the variable $r$ is the radius of a circle along which  the coordinates ($u,v$),  or ($r,\theta$) evolve with the angular velocity $2 \omega$, see Figure \ref{fig:frontier}-(b).

 The phase space ($s,r$) is bounded by the frontier 
\begin{equation}
r+s \le 1
\textrm{,}
\label{eq:rs}
\end{equation} 
because we have $v^{2} = 4 \vert \Im(a_{0} a_{3}^{*} ) \vert ^{2} \le 4  \vert a_{0} \vert ^{2}  \vert a_{3} \vert ^{2} $ which gives $u^{2} +v^{2} \le (1-s)^{2}$.
At time zero after a given jump, the trajectory  starts from one particular point $r,s$  of the 2D-phase space shown  inside the filled region of Fig.\ref{fig:frontier}-(a). The motion  is known as soon as the initial value of ($r,s$) is given, this pair of variables changing after each quantum jump.

In summary one may consider a probability distribution $p(r,s,\theta; t)$ taken inside the domain ($0 \le r \le r+s \le 1$ , $0 \le \theta \le 2 \pi$). Doing this one has the transport equation
\begin{equation}
\partial_{t} p + 2\omega \partial_{\theta} p =0
\textrm{,}
\label{eq:rs}
\end{equation} 
because the two variables ($r,s$) are constant during the deterministic dynamics which depends on the initial conditions. The initial conditions in the phase space ($r,s$) can be expressed in terms of the initial values (\ref{eq:ci}). Using (\ref{eq:sys03}) with $a_{1+}=0$, and the definitions of ($u,v$) with (\ref{eq:u})-(\ref{eq:v}) we get $a_{3+}=a_{20} \sin \epsilon$ and $a_{4+}= - a_{20} \cos \epsilon$, that gives 
\begin{equation}
s_{+}= \vert  a_{4+} \vert^{2}=  \vert  a_{20} \vert^{2} \cos^{2}\epsilon
\textrm{.}
\label{eq:s+}
\end{equation} 
 Using the relations (\ref{eq:ci}) deduced from the norm constraint, we get
\begin{equation}
u_{+}=1 - \vert a_{20} \vert ^{2}(1+ \sin^{2} \epsilon) = \sin^{2} \varphi (1+ \sin^{2} \epsilon)
\textrm{,}
\label{eq:u+}
\end{equation} 
and 
\begin{equation}
v_{+}=2 \sin \epsilon\; \Im(a_{0+} a_{20}^{*}) = -\cos \varphi \sin \varphi \sin\xi
\textrm{.}
\label{eq:v+}
\end{equation} 

 \subsection{Shelving}
\label{sec:shelving}
To figure out how the shelving process occurs, we can imagine to begin an experiment with a single laser, as done in ref. \cite{nagourney}, the intense one of frequency $\omega_{1}$ induces the transitions between the states $\vert 0>$ and $\vert 1>$. At this stage the atom makes Rabi nutations between these two states and emits photons  $''1''$ by stimulated and spontaneous emission. At a given time taken as the origin  and coinciding with a quantum jump from $\vert 1>$ and $\vert 0>$, the second laser is switched on. At this instant we have the initial conditions
\begin{equation}
a_{0+}=1;  \qquad   a_{1+}= a_{2+}=0
\textrm{,}
\label{eq:cis}
\end{equation} 
or $r_{+}=1$ and $s_{+}=0$ which  corresponds to  a rotation on the circle of unit radius in the phase plane ($u,v$), the situation of the true two-level atom. For positive time, the trajectory in the plane ($u,v$) follows the circle of radius unity until the next quantum jump. At this time the initial conditions change from the coordinates ($r_{0}=1, s_{0}=0$) in the plane $(r,s)$ towards  new values ($r_{1}, s_{1}$). We show below that the distance between $(1,0)$ and the new initial conditions is of order $\epsilon^{2}$ in each direction. Therefore the new trajectory is a circle of radius slightly smaller  than unity. 

To be more precise let us consider the change of  the amplitudes $a_{i}$.  With initial conditions (\ref{eq:cis}), the solution (\ref{eq:sol0})-(\ref{eq:sol2}) shows that  the amplitude $a_{2}$  evolves (before the next emission of a photon ) as $a_{2}(t)= -i \sin\epsilon \sin\omega t$ (the two other ones as $a_{0}=\cos\omega t$ and $a_{1}= -i \sin\epsilon \sin\omega t$). Recall that we have assumed that $\epsilon$ is much smaller than unity, because $\omega_{1}  \ll \omega_{2}$.  
Let $t_{1}>0$ be  the emission time of the next photon $``1``$. We have $a_{2}(t_{1})= -i \sin\epsilon \sin\omega t_{1}$ which gives the initial condition for the next deterministic stage,
\begin{equation}
a_{0+}=\cos\varphi^{(1)}e^{i \phi} ;  \qquad   a_{1+}= 0; \qquad a_{2+}=-i  \sin\varphi^{(1)}
\textrm{.}
\label{eq:cis1}
\end{equation} 
where $ \sin\varphi^{(1)} =  \sin\epsilon \sin\omega t_{1}$ is of order $\epsilon$.
By comparing the initial conditions (\ref{eq:cis}) and (\ref{eq:cis1}) we see that the population of the level $2$  increases during the first deterministic stage. 

Let describe now the shift of the initial coordinates in the phase space ($(r,s)$). Using expressions (\ref{eq:cis1}) and relations (\ref{eq:u+})-(\ref{eq:v+}), the initial conditions after the time $t_{1}$ are $u_{+}^{(1)}=1-\sin^{2}\varphi^{(1)}(1+\sin^{2}\epsilon)$ and  $v+^{(1)}=  \sin \epsilon \cos\phi \sin(2 \varphi^{(1)})$.
At first order with respect to the parameter $\epsilon$, it gives $u_{+}^{(1)}=1-(\phi^{(1)})^{2}$ and  $v_{+}^{(1)}=  \sin \epsilon \cos\xi_{0}\varphi^{(1)}$. The radius of the next deterministic stage becomes equal to $1-(\varphi^{(1)})^{2}$. Moreover from (\ref{eq:s+}) we have  $s_{+}= \cos^{2 \epsilon}(\sin\varphi^{(1)})^{2}$, that gives 
\begin{equation}
r^{(1)}\sim 1-(\varphi^{(1)})^{2};  \qquad   s^{(1)}\sim (\varphi^{(1)})^{2}
\textrm{,}
\label{eq:cis2}
\end{equation} 
at first order with respect to $\phi^{(1)} \sim \epsilon \sin\omega t_{1}$.
This relation describes a small shift of the initial conditions along the frontier in Fig.\ref{fig:frontier}-(a), of order $\epsilon$. The deterministic motion between $t_{1}$ and $t_{2}$ (emission time of the next photon from level $\vert 1>$), leads to more complicated expressions,  that we shall not write. It appears that 
 setting $\sin\omega t_{1}/ \sin \omega t_{2}=1$ and $cos \phi=0$,
 the modulus  of $a_{2}$ gets a new increment equal to $\varphi^{(1)}$ at first order. This gives a new shift  of the initial conditions  $r,s$ along the frontier. We infer that step by step the modulus of the amplitude $a_{2}$ will grow with time sufficiently to allow the transition of the atom from the state $\vert 0>$ to $\vert 2>$ after a given number of successive deterministic stages interrupted randomly by emissions of photons $1$.

 \section{ Summary and conclusion}
 
 We intended to show how helpful are some concepts of non-equilibrium statistical mechanics in the understanding of the phenomenon of  fluorescence of an atom or ion submitted to an electromagnetic wave at the frequency of resonance between two or three quantum levels. The coherent part of the dynamics of this system is well understood and is standard quantum physics. The spontaneous decay brings randomness into this system, a phenomenon requiring statistical methods. The constraint of conservation of probability led us quite naturally to a Kolmogorov equation where the randomness inherent to the decay process and determinism linked to the interaction with the light beam are put together in a coherent picture. Thanks to this we have been able to recover the physically reasonable result that, if the natural lifetime of the excited state decreases, the time lag-between two emissions of photons by the atom (or ion) decreases  also. Other properties of fluorescence have been derived also. Among them we did show that the fluorescence process is objectively irreversible in the sense that two-time correlations of the emitted intensity are not symmetric under time reversal. This irreversibility is the result of the randomness of the time of the quantum jumps, and so it is related to one of the fundamental aspects of quantum mechanics. This remark could have a relevance in other situations, like perhaps the cosmological background, because this method gives a precise test of the fact that a signal comes from an equilibrium or from a non-equilibrium system.  Another instance where this type of analysis could be done is what is called sonoluminescence \cite{ref.4} which, despite many efforts, remains poorly understood : it could allow to know if this radiation is some sort of black-body radiation or not.

\appendix
\section{Kolmogorov equation for  a detuned  atom-laser transition: the 2-level case}
\label{app:A}
For simplicity of the presentation let us set $\xi=0$ (see below the solution for $\xi \ne 0$).  With $A=(a_{0},a_{1})^{\dagger}$, equations (\ref{eq:dot0})-(\ref{eq:dot1}) can be written as
 \begin{equation}
 2i\partial _{t} A = \Omega
 \left(
 \begin{array}{cc}
0 &e^{i(t \delta -\xi)}\\
e^{-i(t \delta -\xi)} &0
\end{array}
\right)
A
\textrm{.}
\label{eq:sysa}
\end{equation}
in the general case of non zero detuning $\delta$. The solution corresponding to the initial condition $a_{0}(0)=1$ and $a_{1}(0)=0$ is given by the expressions
\begin{equation}
a_{0} (t)= e^{i\frac{t \delta }{2}} \left( \cos \omega t - i \frac{\delta}{\Omega_{\delta}}\sin\omega t \right)
\textrm{,}
\label{eq:apa0}
\end{equation} 
and 
\begin{equation}
a_{1} (t)= - i e^{-i(\frac{t \delta}{2}- \xi)}  \frac{\Omega}{\Omega_{\delta}} \sin \omega t 
\textrm{,}
\label{eq:apa1}
\end{equation} 
where $\omega=\Omega_{\delta}/2$ with $\Omega_{\delta}$ the effective Rabi frequency in presence of detuning,
\begin{equation}
\Omega_{\delta}=\sqrt{\Omega^{2} + \delta^{2}}
\textrm{.}
\label{eq:omegdet}
\end{equation} 
Using these variables $a_{0,1}$  the  evolution equation  for the probability (or any other function of the amplitudes $a_{0,1}$)  should have partial derivatives not only with respect to $t$ and $\theta=\omega t$ but also with respect to $\psi=\delta t $ and $\xi$ that leads to  awful expressions for Kolmogorov equation. We are going to show that a fair change of variables allows to obtain an unexpected result, namely a Kolmogorov equation of the same form as the one for the resonant case. The choice of appropriate phase space comes from the one used for three-level case treated in section \ref{3levelJG}.
In the next subsection we assume that the atomic dipole moment is real, or $\xi=0$,  in order to simplify the presentation.  The solution for $\xi \ne 0$ is given at the end of this appendix.  

\subsubsection{New amplitudes}
 We can define a parameter $\eta$  which obeys the relations 
\begin{equation}
\sin 2\eta=\frac{\delta}{\Omega_{\delta}} \; \;      \; \; \cos 2\eta=\frac{\Omega}{\Omega_{\delta}}
\textrm{.}
\label{eq:eta}
\end{equation}   
To get rid of the phases factors, let us define the vector $B$ (which has nothing to do with the function B(t) defined in equation (\ref{eq:JG8})),
 \begin{equation}
 B=
 \left(
 \begin{array}{c}
b_{0}\\
b_{1}
\end{array}
\right)
=
 \left(
 \begin{array}{c}
e^{-it \delta /2} a_{0}\\
e^{it \delta/2} a_{1}
\end{array}
\right)
\textrm{.}
\label{eq:vectb}
\end{equation}

The system (\ref{eq:sysa}) becomes
 \begin{equation}
 i\partial _{t} B = \frac{\Omega_{\delta}}{2}
 \left(
 \begin{array}{cc}
\sin 2\eta & \cos 2 \eta\\
\cos 2\eta & -\sin 2 \eta
\end{array}
\right)
B
\textrm{,}
\label{eq:matdb}
\end{equation}
with the initial condition $B(0)=A(0)$.
Introducing the two matrices 
 \begin{equation}
\sigma_{1}=
 \left(
 \begin{array}{cc}
0 & 1\\
1&0
\end{array}
\right)
  \qquad \sigma_{3}=
 \left(
 \begin{array}{cc}
 1& 0\\
0 &-1
\end{array}
\right)
\textrm{,}
\label{eq:sigma13}
\end{equation}
the system (\ref{eq:matdb}) becomes
 \begin{equation}
 i\partial _{t} B =\frac{\Omega_{\delta}}{2} \left(\sigma_{3} \sin 2 \eta +\sigma_{1} \cos 2 \eta \right) B
\textrm{.}
\label{eq:matdbb}
\end{equation}
 Using the relation
 \begin{equation}
\left(\sigma_{3} \sin 2 \eta +\sigma_{1} \cos 2 \eta \right)^{2}=1
\textrm{,}
\label{eq:relb}
\end{equation}
 the general solution of (\ref{eq:matdbb}) is
 \begin{equation}
B(t)= \left(\cos\frac{\Omega_{\delta}}{2} t- (\sigma_{3} \sin 2 \eta +\sigma_{1} \cos 2 \eta)\sin \frac{\Omega_{\delta}}{2} t \right) B(0)
\textrm{.}
\label{eq:solbt}
\end{equation}
Using (\ref{eq:vectb}) we get the general solution for $A$, which gives (\ref{eq:apa0})-(\ref{eq:apa1}) for $a_{1}(0)=0$.  

Let us introduce the rotation matrix, 
 \begin{equation}
R(\eta) = 
 \left(
 \begin{array}{cc}
\cos \eta & -\sin \eta\\
\sin \eta & \cos \eta
\end{array}
\right)
\textrm{.}
\label{eq:rotR}
\end{equation}
It is interesting to notice that the right-hand side of equation (\ref{eq:matdb}) can be written as a product of matrix, equation (\ref{eq:matdbb}) becomes
 \begin{equation}
i\partial_{t}B(t)= \frac{\Omega_{\delta}}{2} R^{-1} \sigma_{1}R B
\textrm{.}
\label{eq:dtB}
\end{equation}
Using the spinor $C=(c_{0},c_{1})^{\dagger}$ defined by the relation,
 \begin{equation}
C(t)=R(\eta) B
\textrm{,}
\label{eq:RB}
\end{equation}
we get  the system
 \begin{equation}
i\partial_{t}C(t)= \frac{\Omega_{\delta}}{2}  \sigma_{1}C
\textrm{.}
\label{eq:dtC}
\end{equation}
Lastly the equations for the amplitudes ($c_{0}, c_{1}$) are identical to those of the amplitudes ($a_{0}, a_{1}$) in the resonant case. They are $$ i\partial_{t}c_{0}=\frac{\Omega_{\delta}}{2} c_{1}$$ and  $$i\partial_{t}c_{1}=\frac{\Omega_{\delta}}{2} c_{0}.$$ 

\subsubsection{Change of variables}
To find a  pertinent  phase space we adapt the procedure detailed in section \ref{3levelJG} for the three level atom case.  We define the  new variables,
 \begin{equation}
u=2\vert c_{0}\vert^{2} -1=1-2\vert c_{1}\vert^{2}
\textrm{,}
\label{eq:uap}
\end{equation}
and
 \begin{equation}
v=2\Im( c_{0}c_{1}^{*})
\textrm{,}
\label{eq:vap}
\end{equation}
where $\Im{(f)}$ is for the imaginary part of $f$.
They obey the same differential equations as (\ref{eq:duv}) with $2\omega=\Omega_{\delta}$.  Using the polar coordinates ($r,\theta$) associated to  the variables ($u,v$) by the relation  $u+iv=r\exp(2i\theta)$, we are able to describe the motion in a 2D phase space (in the three level case the phase space is $3D$, with variables $r,s,\theta$). In this phase space the dynamics is governed by the equations 
 \begin{equation}
\partial_{t}r=0
\textrm{,}
\label{eq:dtr}
\end{equation}
and
 \begin{equation}
\partial_{t}\theta=\frac{\Omega_{\delta}}{2}
\textrm{.}
\label{eq:dttheta}
\end{equation}
It follows that
the  deterministic motion 
is a rotation  along a circle of constant radius $r=\sqrt{u^{2}+v^{2}}$, with angular velocity $\Omega_{\delta}/2$. Note that the angular variable $\theta$ has not the same meaning as the one used to describe the motion in the phase space of the amplitudes $a_{0,1}$. Here $\theta$ is twice the one of the resonant case, because  we consider in this subsection the motion of the functions $u(t),v(t)$ which are quadratic with respect to the amplitudes ($a_{0}, a_{1}$). 

\subsubsection{Motion after an emission}
Just after the emission of a photon, the initial conditions for the deterministic motion is $b_{1+}=a_{1+}=0$ and $\vert b_{0+1}\vert =\vert a_{0+}\vert=1$. In terms of  the amplitudes $c_{0,1}$ it gives
  $$c_{0+}=b_{0+} \cos\eta= a_{0+}\cos\eta$$ and  $$c_{1+}=b_{0+} \sin\eta= a_{0+}\sin\eta.$$ The two amplitudes $c_{0,1}$  are then in phase, it follows that $v_{+}=\theta_{+}=0$, $\Re(c_{0}c_{1}^{*})_{+}=\sin\eta \cos\eta$ and  $r_{+}= 2\cos^{2}\eta-1$. In summary the motion start from the same initial point $$\theta=0$$ on the trajectory defined by the circle of radius
 \begin{equation}  
r_{\delta}= \cos2\eta=\Omega/(2\omega)
\textrm{.}
\label{eq:rapa}
\end{equation}  

\subsubsection{Kolmogorov equation}
We have proved that the phase space reduces to a single trajectory. On this trajectory one may put a probability $p(\theta,t)d\theta $. The left hand side of the kinetic equation is the one of equation (\ref{eq:pdet}). The right hand side is built as in the resonant case, help to the transition function
 \begin{equation}
\Gamma(\theta,\theta')=\gamma \delta_{p}(\theta') \vert a_{1}(\theta)\vert^{2}
\textrm{,}
\label{eq:apGamma}
\end{equation}
where $\delta_{p}$ stands for the wrapped Dirac distribution of period $p$. To calculate the amplitude $a_{1}(\theta)$ we use equations (\ref{eq:uap})-(\ref{eq:vap}) together with the relation
 \begin{equation}  
\Re{(c_{0}c_{1}^{*})}= k
\textrm{,}
\label{eq:k}
\end{equation}  
where $k$ is a constant. This constant can be deduced from the initial conditions $a_{1}=b_{1}=0$ and the expressions $c_{0}=b_{0} \cos\eta -b_{1} \sin\eta$, $c_{1}=b_{0}\sin\eta -b_{1} \cos\eta$ taken from (\ref{eq:RB}). It gives $$k=\vert b_{0}\vert^{2}= \sin\eta \cos\eta.$$ 
Using (\ref{eq:vectb}), we have $ \vert a_{1}\vert^{2}= \vert b_{1}\vert^{2}$. In terms of $c_{0,1}$ it gives $\vert a_{1}\vert^{2}=\vert -c_{0}\sin\eta+c_{1}\cos\eta\vert^{2}$. Introducing equation (\ref{eq:uap})  and the relation $u=r\cos2\theta$ with $r=r_{\delta}$ given in (\ref{eq:rapa}), we get  
\begin{equation} 
\vert a_{1}\vert^{2}=\frac{1}{2} (1+u)\sin^{2}\eta+ \frac{1}{2}(1-u)\cos^{2}\eta-2\sin^{2}\eta \cos^{2}\eta =  \cos^{2}(2\eta) \sin^{2}\theta
\textrm{.}
\label{eq:a1b}
\end{equation}  
Finally the transition function becomes
 \begin{equation}  
\Gamma(\theta,\theta')= \gamma (\frac{\Omega}{\Omega_{\delta}})^{2}\delta_{p}(\theta')\sin^{2}\theta
\textrm{.}
\label{eq:Gammadet}
\end{equation}  
The transport equation,  or Kolmogorov equation is given in (\ref{eq:Kolmdet}). It  takes  the same form as in the resonant case when using the same variables, but with  different coefficients.  In the left hand side the  Rabi frequency is given in (\ref{eq:omegdet}) and  in the right hand side the  function $f(\theta)$ is given in (\ref{eq:fdet}).

\section{asymptotic value of $b(t)$}
\label{app:B}
We try to solve equation (\ref{eq:b(t)}) by Fourier Transform. Using relations (\ref{eq:JG11}) and (\ref{eq:JG9}) we get
\begin{equation}
Hb \;\star\;H\alpha = HM
\textrm{.}
\label{eq:ap1}
\end{equation}
where $H(x)$ is Heaviside function defined in section (\ref{sec:propK}), and $\star$ is for convolution. The function $b(t)$ is positive and bounded, $0 \le b \le \Vert f\Vert _{\infty}$, and $0\le H\alpha \le 1$, that gives the solution in the Fourier space
\begin{equation}
\widehat{Hb } = \frac {\widehat{HM}} { \widehat{H\alpha}} 
\textrm{.}
\label{eq:ap2}
\end{equation}
By definition, the Fourier transform of $H\alpha$ as a function of $z=x+i y$  is given by the integral,
\begin{equation}
\widehat{H\alpha } (z)= \int_{0}^{\infty} \mathrm{d}{t}\  \alpha(t) e^{-itz}
\textrm{.}
\label{eq:ap2}
\end{equation}
which can be written as
\begin{equation}
\widehat{H\alpha } (z)= \sum_{n\ge 0} \int_{n\pi}^{(n+1)\pi} \mathrm{d}{t}\  \alpha(t) e^{-itz}
\textrm{.}
\label{eq:ap3}
\end{equation}

Putting relation (\ref{eq:alphapi}) into (\ref{eq:ap3}) we get

\begin{equation}
\widehat{H\alpha} (z)= \sum_{n\ge 0} \int_{0}^{\pi} \mathrm{d}{t}\  \alpha(t) e^{-itz} e^{-(\bar{f}+iz)n\pi}
\textrm{,}
\label{eq:ap4}
\end{equation}
where $$\bar{f}=(1/\pi) \int_{0}^{\pi} \mathrm{d}{t}\ f(t)$$  is equal to $\gamma'/2$ in the particular case treated here ($f(x)=\sin(x)$) but we introduce the parameter $\bar{f}$ to be more general . For $y<\bar{f}$, equation (\ref{eq:ap4}) becomes 

\begin{equation}
\widehat{H\alpha} (z)=\frac{1}{1-e^{-(\bar{f}+i z)\pi}} \int_{0}^{\pi} \mathrm{d}{t}\  \alpha(t) e^{-itz} 
\textrm{,}
\label{eq:ap5}
\end{equation}

where the first r.h.s. factor is a meromorphe function of $z$ having simple poles at $z=2k+i\bar{f}$, $k$ integer. The latter term is holomorphe for $y < \bar{f}$ and never vanishes.
The second factor in equation (\ref{eq:ap5}) is an entire function of $z$. Its imaginary part 
\begin{equation}
\Im{ \int_{0}^{\pi} \mathrm{d}{t}\  \alpha(t) e^{-itz}}= - \int_{0}^{\pi} \mathrm{d}{t}\  \alpha(t) e^{ty} \sin{tx}
\textrm{,}
\label{eq:ap6}
\end{equation}
is negative and have no zero for $y<0$ because in this domain $\alpha(t) e^{y t}$  is decreasing.  The question is: does this factor in (\ref{eq:ap6}) vanish in the domain  $0<y< \bar{f}$ ?
Let us return to the original problem, namely the asymptotic solution of $b(t)$ from equation (\ref{eq:ap1}). The function $M$  defined by
\begin{equation}
M(t)= 1-\int_{0}^{\pi} \mathrm{d}{\theta}\  p(\theta,0) e^{-\int_{0}^{t}  \mathrm{d}{t'}\ f(\theta+t')}=1-k(t)
\textrm{,}
\label{eq:ap7}
\end{equation}
has a Fourier transform given by
\begin{equation}
\widehat{HM} (z)= \widehat{H} - \widehat{Hk}
\textrm{.}
\label{eq:ap8}
\end{equation}
Treating  the term $\widehat{Hk}$ in the same way as we did for $\widehat{H\alpha}$, we get
\begin{equation}
\widehat{H k} (z)= \int_{0}^{\infty}  \mathrm{d}{t}\ k(t) e^{-itz}=\frac{1}{1-e^{-(\bar{f} + iz)\pi} } \int_{0}^{\pi}  \mathrm{d}{t}\ k(t) e^{-itz}
\textrm{,}
\label{eq:ap9}
\end{equation}
where the second factor 
\begin{equation}
 \int_{0}^{\pi}  \mathrm{d}{t}\ k(t) e^{-itz}=  \int_{0}^{\pi}  \mathrm{d}{\theta}\  p(\theta,0)  \int_{0}^{\pi}  \mathrm{d}{t}\ e^{( -\int_{0}^{t}  \mathrm{d}{t'}\  f(\theta+t') - i tz)  }
\textrm{,}
\label{eq:ap10}
\end{equation}
is still an entire function of $z$. Defining $H_{\pi}(t)$ as the characteristic function on the domain $0\le t \le \pi$ (equal to unity in this domain and null outside), we finally get
\begin{equation}
 \widehat{Hb} (z)=\left( \frac{1-e^{-(\bar{f} + iz)\pi}}{i z }  -  \widehat{H_{\pi}k} (z) \right)\frac{1}{ \widehat{H_{\pi}a}(z)}
\textrm{.}
\label{eq:ap11}
\end{equation}

Finally let us write $<b_{st}>$ in terms of the above functions. 
The stationary distribution is given by $p_{st}(\theta)=p_{st}(0) \alpha(\theta)$ in the domain $0<\theta <\pi$. The norm condition 
\begin{equation}
p_{st}(0) \int_{0}^{\pi} \mathrm{d}{\theta}\ \alpha(\theta)   =1
\textrm{,}
\label{eq:ap20}
\end{equation}
gives
\begin{equation}
p_{st}(\theta)  =\frac{\alpha(\theta)}{ \widehat{H_{\pi}\alpha} (0) }
\textrm{,}
\label{eq:ap21}
\end{equation}
with $\widehat{H_{\pi}\alpha} (0) = \int_{0}^{\pi} \mathrm{d}{t}\ \alpha(t)$.The average stationary value of $b$ is defined by
\begin{equation}
<b_{st}>  = \int_{0}^{\pi} \mathrm{d}{\theta}\ f(\theta) p_{st}(\theta)=- \frac {1}{ \widehat{H_{\pi}\alpha} (0)} \int_{0}^{\pi} \mathrm{d}{\theta}\ \alpha_{\theta}(\theta)
\textrm{,}
\label{eq:ap22}
\end{equation}
that gives the relation
\begin{equation}
<b_{st}>  =  \frac {1-\bar{\alpha}}{ \widehat{H_{\pi}\alpha} (0)} 
\textrm{.}
\label{eq:ap23}
\end{equation}
Returning to equation (\ref{eq:ap11})  we get
\begin{equation}
 \widehat{Hb} (z)=\frac{<b_{st}>}{i z }  +  \frac{\tilde{b}(z)}{ \widehat{H_{\pi}a}(z)}
\textrm{,}
\label{eq:ap12}
\end{equation}
where $\tilde{b}(z)$ is an entire function of $z$, that allows to make the following conjecture
\begin{equation}
 Hb (t)=H<b_{st}>+  F(t)
\textrm{,}
\label{eq:ap13}
\end{equation}
where $F(t)$ tends to zero as time tends to infinity. To conclude with more arguments we have to control the zeros of $\widehat{H_{\pi}a}(z)$,  although we have only stated that  $\widehat{H_{\pi}a}(z)$ doesn't vanish  for $y\le 0$.

\section{Toward the writing of the Kolmogorov equation in the three-level case}
\label{app:C}
This Appendix is only to give an idea of the way the Kolmogorov equation can be derived (hopefully to be solved) in the three-level case. 
This writing is not only a matter of dealing with cumbersome algebra but it has also to do with more fundamental issues. Mathematically speaking the problem under consideration has to deal with two seemingly opposite constraints: ({\it{i}}) the equations of motion are linear with respect to the amplitudes, ({\it{ii}}) the total probability should be conserved in the course of time. 

Those two constraints are not easy to reconcile. To see it, consider the following simple looking problem: let $a_{0,-}$ and  $a_{1,-}$ be the amplitudes of state $0$ and $1$ just before a jump from $1$ to $0$ and let us try to find a linear map of $(a_{0,-}, a_{1,-})$ to $(a_{0,+}, a_{1,+})$ just after the jump. Of course one has $a_{1,+} = 0$ and therefore all the information is stored into the linear map
$$ a_{0,+} = \alpha a_{0,-} +  \beta a_{1,-} \textrm{,} $$ 
where $\alpha$ and $\beta$ are given complex coefficients, independent (because of the constraint of linearity) of the amplitudes. The conservation of probability imposes that $\vert a_{0,+} \vert^2 = \vert a_{0, -} \vert^2 + \vert a_{1,-} \vert^2$ for any choice of $a_{0, -} $ and $a_{1, -}$. Because of the cross term $(\alpha \beta^*a_{0,-} a_{1,-}^* + \alpha^* \beta a_{0,-}^*  a_{1,-})$  in  $\vert a_{0,+} \vert^2$ this constraint cannot be satisfied in general unless $a_{1,-} = 0$, a trivial situation. The way out of this dilemma is well known: one assumes that $\beta$ has a random phase and one carries in the expression of $\vert a_{0,+} \vert^2$ an average over this random phase in order to get rid of the cross terms. Thanks to this average, one finds that $\vert a_{0,+} \vert^2 = \vert a_{0, -} \vert^2 + \vert a_{1,-} \vert^2$ is satisfied with $\vert  \alpha \vert^2 = \vert \beta \vert^2 = 1$. Thanks to this the conservation of probability and the linearity can be imposed together. Let us explain how this can by used to write the Kolmogorov equation for the three level system.  To cancel the cross term by averaging on a random phase, one has to assume that the final result of the quantum jump is not a wave-function in the usual sense but instead a linear superposition of such wave-functions with random phases, namely that it can be described only by a density matrix built out of the two states $0$ and $1$. In this schema, the effect of the quantum jump is a linear transformation of this density matrix from its values before to its one after jump. This should be the valid in the two-   level case as well as in the three-level case. But the two-level case is somewhat trivial in this respect because, after the jump, the amplitude of the excited state is brought to zero and the conservation of probability imposes that the amplitude of the ground state $a_{0,+}$ is only a complex number $e^{i  \phi}$ where as we showed, the angle $\phi$ is random. The three-level case is far more complex and we shall sketch its analysis from the present point of view.   

Let us find the relationship between the density matrices of the three-level system just after jump number $j$ from state $1$ to state zero and emission of a photon. The next jump, number $(j + 1)$, is also from from state $1$ to zero. We are looking for the linear map between the density matrix just after jump $j$ an the density matrix just after jump $(j +1)$. After each jump the amplitude $a_1$ is set to zero. In terms of density matrix this is equivalent to say that all entries if the density matrix with the subscript $1$ are zero. Therefore, after emission of a photon by a jump $1$ to $0$, the only surviving terms of the density matrix are $ \rho_{00, +}$,  $ \rho_{22, +}$, $ \rho_{20, +}$ and $ \rho_{02, +} = \rho_{20, +}^*$. The subscript $+$ is to  recall that this is the density matrix just after a jump. We shall add also a superscript, either $j$ or $(j + 1)$ to recall that the jump under consideration carries either the number $j$ or $(j + 1)$. Therefore we are looking for the linear map from the density matrix of entries $ \rho_{00, +}^j$,  $ \rho_{22, +}^j$, $ \rho_{20, +}^j$ and $ \rho_{02, +}^j$ to the matrix of entries $ \rho_{00, +}^{(j+1)}$,  $ \rho_{22, +}^{(j+1)}$, $ \rho_{20, +}^{(j+1)}$ and $ \rho_{02, +}^{(j+1)}$. 

This linear map splits into two pieces. First, knowing the matrix $( \rho_{00, +}^j ,  \rho_{22, +}^j , \rho_{20, +}^j , \rho_{02, +}^j )$ one has to find the full three by three density matrix  $ \rho_{mn,  -}^{(j+1)}$ with $m, n = 1, 2 , 3$, which is the density matrix just before the jump $(j +1)$ and then the matrix (not a density matrix) relating the density matrix before and after jump by a simple product of matrices. The relationship between $ \rho_{mn,  +}^{(j)}$ and $ \rho_{mn,  -}^{(j+1)}$ is, in principle straightforward to derive from the general expressions given in section \ref{amplitude sol}. For instance, from equation (\ref{eq:sol0}) one obtains: 
\begin{equation}
 \rho_{00,  -}^{(j+1)} =  \cos^2(\theta)  \rho_{00,  +}^{(j)} + \sin^2(\theta)  \sin^2(\epsilon) \rho_{22,  +}^{(j)} + i \cos(\theta) \sin(\theta)  \sin(\epsilon) ( \rho_{02,  +}^{(j)} - \rho_{20,  +}^{(j)}) 
\textrm{,}
\label{eq:rho00}
\end{equation}
where $\theta = \omega (t_{(j+1)} - t_j)$, $t_j$ being the time of jump number $j$.  This equation is derived by replacing $ \rho_{00,  -}^{(j+1)}$ by the product $a_0 (t_{(j+1)} 
a_0^* (t_{(j+1)}$, inserting the expression of $a_0(t)$ given in equation (\ref{eq:sol0})  and substituting into the result the entry $\rho_{mn} (t_j)$ of the density matrix at time $t_j$ for each product $a_m(t_j) a_n^*(t_j)$. Similar expressions hold for the other entries of the three by three density matrix at time $t_{(j+1), -}$. To get the non-zero entries of the density matrix just after the jump, one needs to find the linear relationship between the density matrix before and after a jump. 

This is done as follows. As said before, after the jump, all the norm of the state $1$ is collapsed to state $0$. This is represented very simply by the condition $$ 
 \rho_{00, +}^{(j+1)} =  \rho_{00, -}^{(j+1)}+  \rho_{11, -}^{(j+1)} \textrm{.}$$
The amplitude of level $2$ remains continuous at jump from $1$ to $0$. Therefore the matrix element $ \rho_{22}$ is continuous across this jump, so that:
$$  \rho_{22, +}^{(j+1)} =  \rho_{22, -}^{(j+1)} \textrm{.}$$
Lastly we have to consider the non-zero off-diagonal elements of the density matrix after the jump, namely $ \rho_{20, +}^{(j+1)}$ and  $ \rho_{02, +}^{(j+1)}$. 
Such matrix elements behave like products $a_0 a_2^*$ and $a_2 a_0^*$.  In this product, $a_2$ is continuous across the jump. As said before, the amplitude $a_{0, +}$ is the sum $a_{0, -} + e^{i  \phi} a_{1, -}$ where $\phi$ is a random phase. To get the density matrix, one has to make an average over this random phase. therefore the added term proportional to $a_1$ is eliminated by this average, so that the off-diagonal entries $ \rho_{20}$ and $ \rho_{20}$ of the density matrix are continuous across the jump: 
$$  \rho_{20, +}^{(j+1)} =  \rho_{20, -}^{(j+1)} \textrm{,}$$
and 
$$  \rho_{02, +}^{(j+1)} =  \rho_{02, -}^{(j+1)} \textrm{.}$$
This completes the (sketchy) derivation of the linear relation between the density matrices just after two consecutive jumps.  Notice that the transformation of the density matrix from one jump to the next is a linear map but has no reason to be a multiplication of the density matrix by another matrix of the same size. 

The linear map relating $  \rho_{mn, +}^{(j+1)} $, ($m, n = 0, 2$) to $  \rho_{mn, +}^{(j)} $ can be written thanks to a function $Y(m, n; m' , n'\vert \theta_{j, j+1})$ like: 
\begin{equation}
 \rho_{mn, +}^{(j+1)} = \sum_{m', n'} Y(m, n; m' , n'\vert \theta_{j, j+1}) \rho_{m'n', +}^{(j)}
\textrm{,}
\label{eq:rhomn}
\end{equation}
 where $ \theta_{j, j+1} =  \Omega (t_{j+1} - t_j$. this equation looks a bit like the equation (\ref{eq:dotprobM1.1}) for a discrete Markov process. The functions $Y(m, n; m' , n'\vert \theta_{j, j+1})$ are constrained by the conservation of probability. As the total probability is givenby the trace of the density matrix, the conservation of probability imposes:  
 $$  Y(m, m; m' , n'\vert \theta_{j, j+1})  =  \delta_{m', n'} Y(m, m; m' , m'\vert \theta_{j, j+1})$$ 
 and 
 $$  \sum_{m} Y(m, m; m' , m'\vert \theta_{j, j+1})  = 1 \textrm{,}$$
 Moreover the iteration given by equation (\ref{eq:rhomn}) must keep the property that the density matrix is hermitian. This amounts to impose that 
 $$  Y(m, n; m' , n'\vert \theta_{j, j+1}) = Y^*(n, m; n' , m'\vert \theta_{j, j+1}) \textrm{.}$$

In the significant case of a nutation frequency much bigger than the damping rate by emission of photons of fluorescence, one can replace the angle $\theta$ by a random angle so that the linear mapping from one jump to the next one become a linear random map, where the element of randomness is in this angle $\theta$, which appears in the matrix mapping the non-zero elements of the density matrix after a jump. 

{}
\end{document}